\documentclass[12pt]{scrartcl}    
\usepackage{a4} 
\usepackage{amsmath}    
\usepackage{paralist}
\usepackage{latexsym} 
\usepackage{amssymb}  
\usepackage{amsfonts}  
\usepackage{mathrsfs}  
\usepackage{dsfont}
\usepackage{xcolor}
\usepackage{bbm,exscale}
\definecolor{Myblue}{rgb}{0,0,0.6}  
\usepackage[colorlinks,citecolor=Myblue,linkcolor=Myblue,urlcolor=Myblue,pdfpagemode=None]{hyperref}
\usepackage{amsthm}
\usepackage{accents}
\usepackage[square,numbers,sort&compress]{natbib} 
\usepackage[all,cmtip]{xy}
\usepackage{ifthen} 
\usepackage{bbding}
\usepackage{stmaryrd}  
\usepackage{verbatim}
\usepackage{bbding} 
\usepackage{wasysym}  
\usepackage{soul}  %allow linebreak for underlined with \ul
	\setuldepth{Berlin}
\usepackage[yyyymmdd,hhmmss]{datetime}
\usepackage{booktabs}
\usepackage{enumitem}
\usepackage{color}
\usepackage{textcomp}
\usepackage{gensymb}
\usepackage{pdfpages}
\usepackage{tcolorbox}
\usepackage[bb=boondox]{mathalfa} % for bordism double categories
\usepackage{tikz}
\usepackage{tikz-cd}
\usetikzlibrary{calc}
\usetikzlibrary{decorations.markings}
\usetikzlibrary{fadings,decorations.pathreplacing}
\usetikzlibrary{matrix,arrows}
\usetikzlibrary{patterns}
\usetikzlibrary{arrows,calc,decorations.pathreplacing,decorations.markings,shapes.geometric,shadows}

\DeclareSymbolFont{bbold}{U}{bbold}{m}{n}
\DeclareSymbolFontAlphabet{\mathbbold}{bbold}

\colorlet{SurfaceColor}{orange!30!white}
\colorlet{BackSurfaceColor}{orange!20!white}
\colorlet{IdentityColor}{red!80!black}
\colorlet{IdentityLineColor}{gray}
\colorlet{IdentityColorMid}{red!50!white}
\colorlet{DefectColor}{blue!50!black}
\colorlet{GreenColor}{green!80!black}
\colorlet{SidelineColor}{black}
\tikzset{
	Surface/.style={ SurfaceColor, opacity=0.8 },
	BackSurface/.style={ BackSurfaceColor, opacity=0.8 },
	LineDefect/.style = {ultra thick, DefectColor },
	Sideline/.style = {	very thin , black },
	IdentitySlice/.style = { very thick , IdentityColor },
	IdentitySliceMid/.style = { very thick , IdentityColorMid },
	Multiplication/.style = {thick, BlueColor },
	IdentityLine/.style = {thin, IdentityLineColor,dashed},
	DefaultSettings/.style = {very thick,scale=0.7,color=blue!50!black, baseline=0.5cm}
}
\tikzset{
	string/.style={draw=#1, postaction={decorate}, decoration={markings,mark=at position .51 with {\arrow[draw=#1]{>}}}},
	costring/.style={draw=#1, postaction={decorate}, decoration={markings,mark=at position .51 with {\arrow[draw=#1]{<}}}},
	ostring/.style={draw=#1, postaction={decorate}, decoration={markings,mark=at position .47 with {\arrow[draw=#1]{>}}}},
	ustring/.style={draw=#1, postaction={decorate}, decoration={markings,mark=at position .56 with {\arrow[draw=#1]{>}}}},
	oostring/.style={draw=#1, postaction={decorate}, decoration={markings,mark=at position .43 with {\arrow[draw=#1]{>}}}},
	uustring/.style={draw=#1, postaction={decorate}, decoration={markings,mark=at position .59 with {\arrow[draw=#1]{>}}}},
	directed/.style={string=blue!50!black}, 
	odirected/.style={ostring=blue!50!black}, 
	udirected/.style={ustring=blue!50!black}, 
	oodirected/.style={oostring=blue!50!black}, 
	uudirected/.style={uustring=blue!50!black},     
	redirected/.style={costring= blue!50!black},
	redirectedgreen/.style={costring= green!50!black},
	directedgreen/.style={string= green!50!black},
}

\def\nicedashedpalecolourscheme{\shadedraw[top color=orange!22, bottom color=orange!22, draw=gray, dashed]}
\def\nicedashedpalecolourschemegreenedition{\shadedraw[top color=green!22, bottom color=green!22, draw=gray, dashed]}

\tikzset{-dot-/.style={decoration={
			markings,
			mark=at position 0.5 with {\fill circle (2pt);}},postaction={decorate}}}

\tikzset{
	Fdot/.style={circle, draw, fill, inner sep=0pt}, 
	Odot/.style={circle, draw, inner sep=0.1pt, minimum size=0.1cm}
}

\newcommand\tikzzbox[1]
{#1}

\vfuzz \hfuzz
\makeatletter
\newcommand{\raisemath}[1]{\mathpalette{\raisem@th{#1}}}
\newcommand{\raisem@th}[3]{\raisebox{#1}{$#2#3$}}
\makeatother

\newcommand{\A}{\mathcal{A}}

\newcommand*{\longhookrightarrow}{\ensuremath{\lhook\joinrel\relbar\joinrel\rightarrow}}

\newcommand{\B}{\mathcal{B}}

\newcommand{\Borb}{\B_{\mathrm{orb}}}

\newcommand{\C}{\mathds{C}}

\newcommand{\R}{\mathds{R}}
\newcommand{\Z}{\mathds{Z}}

\def\1{\ifmmode\mathrm{1\!l}\else\mbox{\(\mathrm{1\!l}\)}\fi}
\newcommand{\one}{\mathbbm{1}}
\newcommand{\be}{\begin{equation}}
  \newcommand{\ee}{\end{equation}}
\newcommand{\bes}{\begin{equation*}}
  \newcommand{\ees}{\end{equation*}}

\let\Bar\undefined
\newcommand{\Bar}{\textrm{B}}

\newcommand{\End}{\operatorname{End}}

\def\LG{\mathcal{LG}}

\newcommand{\ev}{\operatorname{ev}}

\newcommand{\tev}{\widetilde{\operatorname{ev}}}
\newcommand{\coev}{\operatorname{coev}}

\def\lra{\longrightarrow}

\def\lmt{\longmapsto}

\DeclareMathOperator{\Jac}{Jac}

\DeclareMathOperator{\colim}{colim}

\newcommand{\Bord}{\textrm{Bord}}

\newcommand{\Bordor}{\Bord^{\textrm{or}}}

\newcommand{\zz}{\mathcal{Z}}

\newcommand{\zzss}{\mathcal{Z}^{\textrm{ss}}}
\newcommand{\zztriv}{\mathcal{Z}^{\textrm{{\tiny triv}}}}

\newcommand{\Vectk}{\operatorname{Vect}_\Bbbk}

\newcommand{\vectk}{\operatorname{vect}_\Bbbk}
\newcommand{\sVect}{\operatorname{sVect}}
\newcommand{\sVectk}{\operatorname{sVect}_\Bbbk}

\newcommand{\ssFrob}{\operatorname{ssFrob}}

\newcommand{\X}{\mathcal{X}}

\newcommand{\dX}{{}^\dagger\hspace{-1.8pt}X}
\newcommand{\Xd}{X^\dagger}

\newcommand{\Borddef}{\Bord_{n,n-1}^{\textrm{def}}(\mathds{D})}
\newcommand{\Borddefblank}{\Bord^{\textrm{def}}}

\newcommand\arxiv[2]      {\href{https://arXiv.org/abs/#1}{#2}}
\newcommand\doi[2]        {\href{https://dx.doi.org/#1}{#2}}

\allowdisplaybreaks

\deffootnote[1em]{1em}{1em}{\textsuperscript{\thefootnotemark}}

\theoremstyle{definition} 
\newtheorem{definition}{Definition}
\newtheorem{proposition}[definition]{Proposition}
\newtheorem{theorem}[definition]{Theorem}
\newtheorem{definitiontheorem}[definition]{Definition and Theorem}

\newtheorem{remark}[definition]{Remark}

\newtheorem{example}[definition]{Example}

\numberwithin{equation}{section}
\numberwithin{definition}{section}
\numberwithin{figure}{section}

\newcommand\void[1]{}

\begin{document}

\title{%
Orbifolds of topological quantum field theories%
}

\author{%
	Nils Carqueville %\quad 
	\\[0.5cm]
	\normalsize{\texttt{\href{mailto:nils.carqueville@univie.ac.at}{nils.carqueville@univie.ac.at}}} 
	\\[0.5cm]  %
	{\normalsize\slshape Universit\"at Wien, Fakult\"at f\"ur Physik,}
	\\[-0.1cm] 
	{\normalsize\slshape Boltzmanngasse 5, 1090 Wien, \"{O}sterreich}\\
}

\date{}
\maketitle

\begin{abstract} 
	The orbifold construction via topological defects in quantum field theory can either be understood as a state sum construction internal to a given ambient theory, or as the procedure of (identifying and) gauging ordinary and ``non-invertible'' symmetries. 
	Here we explain how this is rigorously understood in the case of topological QFTs. 
	We provide various examples and outline general features, also of relevance for full QFT.  
	
	\medskip 
	
	\noindent
	{\scriptsize 
	This is a contribution to the Encyclopedia of Mathematical Physics (editors-in-chief: M.~Bojowald and R.~Szabo), for the section edited by C.~Meusburger.}

%\medskip 
%
%\noindent 
%{\scriptsize Keywords: 
%	orbifold construction, 
%	topological quantum field theory, 
%	gauging non-invertible symmetries, 
%	state sum models, 
%	higher categories with adjoints, 
%	higher algebras, 
%	stratified bordisms}
\end{abstract}

\hfill 

%{\footnotesize 
\tableofcontents
%}

\newpage 

\section{Introduction and overview}
\label{sec:Introduction}

The generalised orbifold construction takes as input a quantum field theory~$\zz$ and a collection of \textsl{topological} defects~$\A$ of~$\zz$, to produce a new quantum field theory~$\zz_\A$. 
In particular, 
%arXiv_v2: 
	%the partition function which~$\zz_\A$ associates to 
	 correlation functions which~$\zz_\A$ associates to
a closed spacetime~$M$ 
%arXiv_v2: 
	%is 
	 are 
computed by filling~$M$ with a 
%arXiv_v2: 
	network or 
``foam'' of defects of type~$\A$ (of all codimensions), and then evaluating with~$\zz$, e.g.\
%arXiv_v2: 
	%(suppressing orientations)
	 schematically 

%arXiv_v2: 
	\vspace{-0.3cm}
	
\be 
\mathcal Z_{\mathcal A}(S^2) 
\;
= 
\;
\mathcal Z_{\mathcal A}\Biggl(
\tikzzbox{%
	%%%%%%%%%%%%%%%%%%%%%% 
	\begin{tikzpicture}[thick,scale=0.9, color=green!50!black, baseline=0cm]
	\fill[ball color=orange!40!white] (0,0) circle (1.55);
	\fill[orange!30!white, opacity=0.7] (0,0) circle (1.55);
	\end{tikzpicture}
	%%%%%%%%%%%%%%%%%%%%%% 
}
\Biggr)
\;
:= 
\;
\mathcal Z\Bigg(
\tikzzbox{%
	%%%%%%%%%%%%%%%%%%%%%% 
	\begin{tikzpicture}[thick,scale=0.9, color=green!50!black, baseline=0cm]
	%
	%%%%%%%%%%%%%%%%%%%%%%
	\coordinate (v1) at ($(-0.75,-0.75)$);
	\coordinate (v2) at ($(+0.75,-0.75)$);
	\coordinate (v3) at ($(+0,0.85)$);
	\coordinate (v4) at ($(0,0)$);
	\coordinate (f1) at (190:1.55);
	\coordinate (f2) at (-10:1.55);
	\coordinate (f3) at (80:1.55);
	%%%%%%%%%%%%%%%%%%%%%%
	%
	%%%%%%%%%%%%%%%%%%%%%%
	% "ball": 
	\fill[ball color=orange!40!white] (0,0) circle (1.55);
	%%%%%%%%%%%%%%%%%%%%%%
	%
	%%%%%%%%%%%%%%%%%%%%%%
	% defects BACK: 
	% 0-strata: 
	\fill (v4) circle (2.5pt) node[left] (0up) { };
	%
	% 1-strata: 
	\draw[very thick] (f1) .. controls +(0,0.25) and +(-0.25,0) .. (v4);
	\draw[very thick] (f2) .. controls +(0,0.25) and +(+0.25,0) .. (v4);
	\draw[very thick] (f3) .. controls +(0.5,0) and +(0,+0.25) .. (v4);
	% 
	% labels: 
	\fill ($(v4)+(-0.15,0.15)$) circle (0pt) node {\scalebox{.5}{$\mathcal A_0$}};
	%%%%%%%%%%%%%%%%%%%%%%
	%
	%%%%%%%%%%%%%%%%%%%%%%
	% sphere front: 
	\fill[orange!30!white, opacity=0.7] (0,0) circle (1.55);
	%%%%%%%%%%%%%%%%%%%%%%
	%
	%%%%%%%%%%%%%%%%%%%%%%
	% defects FRONT: 
	% 0-strata: 
	\fill (v1) circle (2.5pt) node[left] (0up) { };
	\fill (v2) circle (2.5pt) node[left] (0up) { };
	\fill (v3) circle (2.5pt) node[left] (0up) { };
	%
	% 1-strata: 
	\draw[very thick] (v1) .. controls +(0.5,-0.25) and +(-0.5,-0.25) .. (v2);
	\draw[very thick] (v1) .. controls +(0,0.75) and +(-0.25,0) .. (v3);
	\draw[very thick] (v2) .. controls +(0,0.75) and +(+0.25,0) .. (v3);
	\draw[very thick] (v1) .. controls +(-0.25,0) and +(0,-0.25) .. (f1);
	\draw[very thick] (v2) .. controls +(+0.25,0) and +(0,-0.25) .. (f2);
	\draw[very thick] (v3) .. controls +(0,+0.25) and +(-0.25,0) .. (f3);
	%%%%%%%%%%%%%%%%%%%%%%
	%
	%%%%%%%%%%%%%%%%%%%%%%
	% labels: 
	\fill (40:1.2) circle (0pt) node {\scalebox{.5}{$\mathcal A_2$}};
	\fill (140:1.2) circle (0pt) node {\scalebox{.5}{$\mathcal A_2$}};
	\fill (270:1.3) circle (0pt) node {\scalebox{.5}{$\mathcal A_2$}};
	\fill (270:0.4) circle (0pt) node {\scalebox{.5}{$\mathcal A_2$}};
	\fill (270:0.82) circle (0pt) node {\scalebox{.5}{$\mathcal A_1$}};
	\fill (-1.2,-0.45) circle (0pt) node {\scalebox{.5}{$\mathcal A_1$}};
	\fill (+1.2,-0.45) circle (0pt) node {\scalebox{.5}{$\mathcal A_1$}};
	\fill (-0.68,+0.3) circle (0pt) node {\scalebox{.5}{$\mathcal A_1$}};
	\fill (+0.68,+0.3) circle (0pt) node {\scalebox{.5}{$\mathcal A_1$}};
	\fill (-0.1,+1.30) circle (0pt) node {\scalebox{.5}{$\mathcal A_1$}};
	\fill ($(v1)+(0,-0.2)$) circle (0pt) node {\scalebox{.5}{$\mathcal A_0$}};
	\fill ($(v2)+(0,-0.2)$) circle (0pt) node {\scalebox{.5}{$\mathcal A_0$}};
	\fill ($(v3)+(0,-0.2)$) circle (0pt) node {\scalebox{.5}{$\mathcal A_0$}};
	%%%%%%%%%%%%%%%%%%%%%%
	%
	\end{tikzpicture}
	%%%%%%%%%%%%%%%%%%%%%% 
}
\Bigg)
. 
\ee 

The construction should not depend on the choice of defect foam, and this condition imposes constraints on the 
%arXiv_v2: 
	%$j$-dimensional defects~$\A_j$. 
	 labels~$\A_j$ for defects supported on $j$-dimensional strata (of which~$\zz$ detects only isotopy classes due to the topological nature of the defects). 
If the foam is taken to be Poincar\'{e} dual to a triangulation of~$M$, triangulation invariance imposes finitely many constraints. 
%arXiv_v2: 
	%(coming from Pachner moves, see Section~\ref{sec:Orbifolds})
A collection of defects~$\A$ satisfying these constraints is called an \textsl{orbifold datum}. 

The eponymous source of orbifold data~$\A$ are (gaugeable) \textsl{symmetries} of the theory~$\zz$, in which case all $\A$-defects of non-zero dimension are invertible with respect to their fusion. 
%arXiv_v2: 
	%product. 
%arXiv_v2: 
	%Then the associated gauged theory, or \textsl{orbifold theory}, is~$\zz_\A$. 
	 Then~$\zz_\A$ is the associated gauged theory, or \textsl{orbifold theory}, obtained by averaging over all gauge connections. 
Examples~\ref{exa:OrbifoldDataFromGroupActions}, \ref{exa:OrbifoldSigmaModels} and~\ref{exa:Gequiv} 
%arXiv_v2: 
	%make the connection 
	 give the relation 
to orbifold stacks in the context of sigma models. 

The defects in an orbifold datum however need not come from a group action, and they need not be invertible. 
Indeed, (\textsl{lattice} or) \textsl{state sum models} are other special cases of the
%arXiv_v2: 
	%(generalised) 
	 generalised 
orbifold construction, namely when~$\zz$ is the ``trivial'' theory as explained in Examples~\ref{exa:OrbForSSM} and \ref{exa:SSMTQFT}, see also Remark~\ref{rem:HigherSSMs}. 
It is thus natural to think of the orbifold construction as a state sum construction ``internal''  to any given theory~$\zz$, and we drop the attribute ``generalised''. 

Alternatively, even if~$\A$ does not arise from a group action one may still think of it as encoding a ``generalised'' symmetry of~$\zz$. 
Other common designations are ``non-invertible'', ``topological'', or ``categorical'' symmetries. 
A foundational result in \cite{ffrs0909.5013} is that a huge class of 2-dimensional conformal field theories are orbifolds of one another, but that those from group actions do not suffice. 
More recently, $n$-dimensional orbifold data have appeared (at least behind the scenes) as \textsl{non-invertible symmetries} of $(n-1)$-dimensional quantum field theories. 

\medskip 

A mathematically rigorous account of the orbifold construction is lacking for general quantum field theories~$\zz$. 
However, much about (the gauging of) non-invertible symmetries can be separated from~$\zz$ and discussed only with respect to a higher-dimensional \textsl{topological} quantum field theory, cf.\ 
%arXiv_v2: 
	e.g.\ 
\cite{FreedMooreTeleman2022}. 

Over the last decade or so, the orbifold construction has been developed rigorously for TQFTs of arbitrary dimension~$n$. 
The purpose of this chapter is to give an overview of this theory, as well as some of its applications for $n\leqslant 4$. 
Further motivations include the phenomenon of ``gauging topological phases of matter'' (via the relation between the latter and invertible TQFTs, cf.\ \cite{FreedHopkinsRefelctionPositivity2016, Yonekura2018}) and the higher representation theory of orbifold data mentioned below. 

More precisely, we will work with (non-extended) \textsl{defect TQFTs} (reviewed in Section~\ref{sec:DefectTQFTs}, after the less structure-rich \textsl{closed} TQFTs in Section~\ref{sec:ClosedTQFTs}), formalised as symmetric monoidal functors on stratified and labelled bordism categories. 
Such TQFTs are expected (in low dimension: known) to give rise to higher categories, and one finds that orbifold data are naturally certain types of algebras in this context. 
This is explained in Section~\ref{subsec:OrbifoldData}, along with the basics of Pacher moves, their algebraic incarnations in orbifold data, and several examples. 

The construction of the (closed) \textsl{orbifold TQFT}~$\zz_\A$ from a defect TQFT~$\zz$ and an orbifold datum~$\A$ is described in Section~\ref{subsec:OrbifoldConstruction}. 
This is then illustrated by identifying state sum models as well as gaugings of (higher) symmetry group actions as special cases of the construction, and we give several examples that are of neither extreme type. 
We stress that a higher-categorical approach is often convenient, but it is not necessary to construct~$\zz_\A$. 

Since orbifold data can be viewed as (higher) algebras, it is natural to consider their higher representation category, called ``orbifold completion''. 
In Section~\ref{subsec:OrbifoldCompletion} we describe how this allows to further generalise the orbifold construction to output \textsl{defect} TQFTs (understood in detail for $n\leqslant3$), which in turn provides a 
%arXiv_v2: 
	%much 
more powerful technology for applications, and to develop the theory conceptually. 

\medskip 

It is worth noting that the orbifold construction has so far mostly been developed for oriented TQFTs, i.e.\ the bordisms on which they evaluate come with the structure of an orientation. 
It is expected that analogous constructions can be worked out for other tangential structures -- a promising avenue for future research. 
For example, the ``condensation monads'' of \cite{GaiottoJohnsonFreyd} are thought to be ``framed'' variants of orbifold data, and one can build spin quantum field theories from oriented ones in an orbifold-esque way, see \cite{NovakRunkel, RunkelSzegedyWatts}. 

Relatedly but separately, it would be interesting to build a general theory of orbifolds for extended TQFTs, defined as symmetric monoidal functors on higher bordism categories. 
At least in the fully extended case, the setting of $(\infty,n)$-categories and the ideas described in \cite[Sect.\,4.3]{l0905.0465} seem to make this a technical, but non-conceptual challenge, given the simplicial origins of orbifold data. 
The case of orbifolds from group actions of once-extended TQFTs is described in \cite{SchweigertWoike}. 

\medskip 

\noindent
\textbf{Acknowledgements. }
I am grateful to all the colleagues with whom I have collaborated and learned from in connection with the topic of orbifold TQFT, especially to 
	V.~Mulevi\v{c}ius, 
	I.~Runkel, 
	and 
	G.~Schaumann, 
who also contributed insightful comments on an earlier version of this manuscript. 
Moreover, I thank 
	B.~Bartlett, 
	C.~Lieberum, 
	C.~Meusburger, 
	L.~Müller, 
	and 
	C.~Schweigert 
for further helpful comments, and I acknowledge support from the DFG Heisenberg Programme.

\section{Reminder on closed TQFTs}
\label{sec:ClosedTQFTs}

A quantum field theory can be thought of as a structure-preserving map from spacetime to the algebraic description of physical processes therein. 
By discarding most of the geometry while retaining all of the topology of spacetime, one can define an \textsl{$n$-dimensional closed topological quantum field theory (TQFT) for topological structure~$X$ and with values in~$\mathcal C$} to be a symmetric monoidal functor
\be 
\label{eq:Zclosed}
\zz \colon \Bord_{n,n-1}^X \lra \mathcal C \, . 
\ee 
The codomain~$\mathcal C$ is often taken to be the category of (super) vector spaces, while the domain of~$\zz$ has 
%arXiv_v2: 
	smooth 
$(n-1)$-dimensional closed $X$-manifolds as objects and (equivalence classes of)
%arXiv_v2: 
	smooth 
$n$-dimensional $X$-bordisms as morphisms. 
The composition of bordisms $M\colon E\lra E'$ and $N\colon E' \lra E''$ is the glueing $N \sqcup_{E'} M$, the monoidal structure on $\Bord_{n,n-1}^X$ is given by 
%arXiv_v2: 
	the 
disjoint union, and the symmetric braiding 
%arXiv_v2: 
	%$E\sqcup E' \lra E'\sqcup E$ is the mapping cylinder over the swap map. 
	 is obtained from its universal property. 
We refer to \cite{AtiyahTQFT, Kockbook, TVireBook} for details, and illustrate the structure of the bordism category with an example for $n=2$: 
\be 
\label{eq:ClosedBordism}
\tikzzbox{
	%%%%%%%%%%%%%%%%%%%%%% 
	\begin{tikzpicture}[thick,scale=2.83, color=black, baseline=1.14cm]
	\coordinate (p1) at (-0.55,0);
	\coordinate (p2) at (-0.2,0);
	\coordinate (p3) at (0.2,0);
	\coordinate (p4) at (0.55,0);
	\coordinate (p5) at (0.175,0.8);
	\coordinate (p6) at (-0.175,0.8);
	\coordinate (s) at (0.175,0.07);
	\coordinate (d) at (0,0.5);
	%
	%
	% Back surface colouring
	\fill [BackSurface] 
	(p1) .. controls +(0,0.1) and +(0,0.1) ..  (p2)
	-- (p2) .. controls +(0,0.35) and +(0,0.35) ..  (p3)
	-- (p3) .. controls +(0,0.1) and +(0,0.1) ..  (p4)
	-- (p4) .. controls +(0,0.5) and +(0,-0.5) ..  (p5)
	-- (p5) .. controls +(0,0.1) and +(0,0.1) ..  (p6)
	-- (p6) .. controls +(0,-0.5) and +(0,0.5) ..  (p1)
	;
	%
	%: Front surface colouring: 
	\fill [Surface] 
	(p1) .. controls +(0,-0.1) and +(0,-0.1) ..  (p2)
	-- (p2) .. controls +(0,0.35) and +(0,0.35) ..  (p3)
	-- (p3) .. controls +(0,-0.1) and +(0,-0.1) ..  (p4)
	-- (p4) .. controls +(0,0.5) and +(0,-0.5) ..  (p5)
	-- (p5) .. controls +(0,-0.1) and +(0,-0.1) ..  (p6)
	-- (p6) .. controls +(0,-0.5) and +(0,0.5) ..  (p1)
	;
	\draw[thick] (p2) .. controls +(0,0.35) and +(0,0.35) ..  (p3); 
	\draw[thick] (p4) .. controls +(0,0.5) and +(0,-0.5) ..  (p5); 
	\draw[thick] (p6) .. controls +(0,-0.5) and +(0,0.5) ..  (p1);
	\draw[very thick, red!80!black] (p1) .. controls +(0,-0.1) and +(0,-0.1) ..  (p2); 
	\draw[very thick, red!80!black] (p3) .. controls +(0,-0.1) and +(0,-0.1) ..  (p4); 
	\draw[very thick, red!80!black] (p5) .. controls +(0,0.1) and +(0,0.1) ..  (p6); 
	\draw[very thick, red!80!black] (p5) .. controls +(0,-0.1) and +(0,-0.1) ..  (p6); 
	\draw[very thick, red!80!black] (p1) .. controls +(0,0.1) and +(0,0.1) ..  (p2); 
	\draw[very thick, red!80!black] (p3) .. controls +(0,0.1) and +(0,0.1) ..  (p4); 
	\fill[red!80!black] (0,0.07) circle (0pt) node[below] {{\tiny$E'$}};
	\fill[red!80!black] (0.26,0.88) circle (0pt) node[below] {{\tiny$E''$}};
	\fill[gray] (0.35,0.56) circle (0pt) node[below] {{\tiny$N$}};
	\end{tikzpicture}
	%%%%%%%%%%%%%%%%%%%%%%
}
\circ 
\tikzzbox{
	%%%%%%%%%%%%%%%%%%%%%% 
	\begin{tikzpicture}[thick,scale=2.83,color=black, baseline=-1.14cm]
	\coordinate (p1) at (-0.55,0);
	\coordinate (p2) at (-0.2,0);
	\coordinate (p3) at (0.2,0);
	\coordinate (p4) at (0.55,0);
	\coordinate (p5) at (0.175,-0.8);
	\coordinate (p6) at (-0.175,-0.8);
	\coordinate (s) at (0.175,0.07);
	\coordinate (d) at (0,-0.5);
	%
	%
	% Back surface colouring
	\fill [BackSurface] 
	(p1) .. controls +(0,0.1) and +(0,0.1) ..  (p2)
	-- (p2) .. controls +(0,-0.35) and +(0,-0.35) ..  (p3)
	-- (p3) .. controls +(0,0.1) and +(0,0.1) ..  (p4)
	-- (p4) .. controls +(0,-0.5) and +(0,0.5) ..  (p5)
	-- (p5) .. controls +(0,0.1) and +(0,0.1) ..  (p6)
	-- (p6) .. controls +(0,0.5) and +(0,-0.5) ..  (p1)
	;
	%
	% Front surface colouring: 
	\fill [Surface] 
	(p1) .. controls +(0,-0.1) and +(0,-0.1) ..  (p2)
	-- (p2) .. controls +(0,-0.35) and +(0,-0.35) ..  (p3)
	-- (p3) .. controls +(0,-0.1) and +(0,-0.1) ..  (p4)
	-- (p4) .. controls +(0,-0.5) and +(0,0.5) ..  (p5)
	-- (p5) .. controls +(0,-0.1) and +(0,-0.1) ..  (p6)
	-- (p6) .. controls +(0,0.5) and +(0,-0.5) ..  (p1)
	;
	\draw[thick] (p2) .. controls +(0,-0.35) and +(0,-0.35) ..  (p3); 
	\draw[thick] (p4) .. controls +(0,-0.5) and +(0,0.5) ..  (p5); 
	\draw[thick] (p6) .. controls +(0,0.5) and +(0,-0.5) ..  (p1); 
	\draw[very thick, red!80!black] (p1) .. controls +(0,-0.1) and +(0,-0.1) ..  (p2); 
	\draw[very thick, red!80!black] (p3) .. controls +(0,-0.1) and +(0,-0.1) ..  (p4) ; 
	\draw[very thick, red!80!black] (p5) .. controls +(0,-0.1) and +(0,-0.1) ..  (p6); 
	\draw[very thick, red!80!black] (p1) .. controls +(0,0.1) and +(0,0.1) ..  (p2); 
	\draw[very thick, red!80!black] (p3) .. controls +(0,0.1) and +(0,0.1) ..  (p4); 
	\draw[very thick, red!80!black] (p5) .. controls +(0,0.1) and +(0,0.1) ..  (p6); 
	\fill[red!80!black] (0,0.07) circle (0pt) node[below] {{\tiny$E'$}};
	\fill[red!80!black] (0.23,-0.73) circle (0pt) node[below] {{\tiny$E$}};
	\fill[gray] (0.33,-0.45) circle (0pt) node[below] {{\tiny$M$}};
	\end{tikzpicture}
	%%%%%%%%%%%%%%%%%%%%%%
}
= 
\;
\tikzzbox{%
	%%%%%%%%%%%%%%%%%%%%%% 
	\begin{tikzpicture}[color=blue!50!black, scale = 0.4, >=stealth, baseline = 0cm]
	%% Coordinates
	\coordinate (l1) at (0,0);
	\coordinate (l2) at (2,0);
	\coordinate (r2) at (6,0);
	\coordinate (r1) at (8,0);
	\coordinate (ld1) at (1,0);
	\coordinate (ld2) at (7,0);
	\coordinate (t1) at (2.5,-1.25);
	\coordinate (t2) at (5.5,-1.25);
	\coordinate (c1) at (3,-3);
	\coordinate (c2) at (5,-3);
	\coordinate (2t1) at (2.5,1.25);
	\coordinate (2t2) at (5.5,1.25);
	\coordinate (2c1) at (3.25,3);
	\coordinate (2c2) at (4.75,3);
	\coordinate (d1) at (4,-2.63);
	\coordinate (d2) at (4.25,-3.36);
	\coordinate (d3) at (2.1,1.38);
	\coordinate (s1) at (3.5,-2.25);
	\coordinate (s2) at (4.5,-2.25);
	\coordinate (s3) at (3.5,-1.5);
	\coordinate (s4) at (4.5,-1);
	\coordinate (s5) at (1,0.5);
	\coordinate (s6) at (7.5,0);
	\coordinate (s7) at (2.3,1.8);
	\coordinate (s8) at (5,1.5);
	\coordinate (s9) at (3.5,-2.7);
	\coordinate (s10) at (4.5,-2.7);
	\coordinate (b1) at (3.2,-2.75);
	\coordinate (b2) at (3.8,-2.65);
	\coordinate (b3) at (4.67,-2.7);
	\coordinate (b4) at (3.5,2.7);
	\coordinate (b5) at (4.2,2.64);
	\coordinate (phi) at (4,-1.5);
	%
	% Filling (back)
	\fill[orange!30!white, opacity=0.8] (l1) .. controls +(0,3) and +(0,3) .. (r1) .. controls +(0,-3) and +(0,-3) .. (l1) ;
	%
	% Filling front
	\fill[orange!30!white, opacity=0.8] (l1) .. controls +(0,3) and +(0,3) .. (r1) .. controls +(0,-3) and +(0,-3) .. (l1) ;
	%
	%%%%
	% Outer curve
	\draw[black, thick] (l1) .. controls +(0,3) and +(0,3) .. (r1) .. controls +(0,-3) and +(0,-3) .. (l1) ;
	% Inner curve
	\fill [white] (l2) .. controls +(1,1) and +(-1,1) .. (r2) .. controls +(-1,-1) and +(1,-1) .. (l2) ;
	\draw[black, thick] (l2) .. controls +(1,1) and +(-1,1) .. (r2) .. controls +(-1,-1) and +(1,-1) .. (l2) ;
	\draw[black, thick] (l2) -- ($(l2) + (-0.25, 0.25)$) ;
	\draw[black, thick] (r2) -- ($(r2) + (0.25, 0.25)$) ;
	%
	%%%%%%%%%%%%%%%%%%%%%%%%%%%%%%%%%
	% "trunk1":
	\fill[white] (c1) -- (c2) .. controls +(0,1) and +(0,0) .. (t2) -- (t1) .. controls +(0,0) and +(0,1) .. (c1) ;
	\fill[orange!30!white, opacity=1] (c1) .. controls +(0,-0.5) and +(0,-0.5) .. (c2) .. controls +(0,1) and +(0,0) .. (t2) -- (t1) .. controls +(0,0) and +(0,1) .. (c1) ;
	\fill[orange!15!white, opacity=1] (c1) .. controls +(0,0.5) and +(0,0.5) ..  (c2) -- (c2) .. controls +(0,-0.5) and +(0,-0.5) .. (c1);
	\draw[black, thick] (c1) .. controls +(0,1) and +(0,0) .. (t1);
	\draw[black, thick] (c2) .. controls +(0,1) and +(0,0) .. (t2);
	%%%%%%%%%%%%%%%%%%%%%%%%%%%%%%%%%
	%
	%%%%%%%%%%%%%%%%%%%%%%%%%%%%%%%%%
	% "trunk2":
	\fill[white] (2c1) -- (2c2) .. controls +(0,-1) and +(0,0) .. (2t2) -- (2t1) .. controls +(0,0) and +(0,-1) .. (2c1) ;
	\fill[orange!30!white, opacity=1] (2c1) .. controls +(0,0.5) and +(0,0.5) .. (2c2) .. controls +(0,-1) and +(0,0) .. (2t2) -- (2t1) .. controls +(0,0) and +(0,-1) .. (2c1) ;
	\fill[orange!15!white, opacity=1] (2c1) .. controls +(0,-0.5) and +(0,-0.5) ..  (2c2) -- (2c2) .. controls +(0,0.5) and +(0,0.5) .. (2c1);
	\draw[black, thick] (2c1) .. controls +(0,-1) and +(0,0) .. (2t1);
	\draw[black, thick] (2c2) .. controls +(0,-1) and +(0,0) .. (2t2);
	%%%%%%%%%%%%%%%%%%%%%%%%%%%%%%%%%
	%
	% boundary circles: 
	\draw[very thick, red!80!black] (c1) .. controls +(0,0.5) and +(0,0.5) ..  (c2); 
	\draw[very thick, red!80!black] (c1) .. controls +(0,-0.5) and +(0,-0.5) ..  (c2);
	\draw[very thick, red!80!black] (2c1) .. controls +(0,0.5) and +(0,0.5) ..  (2c2); 
	\draw[very thick, red!80!black] (2c1) .. controls +(0,-0.5) and +(0,-0.5) ..  (2c2); 
	\fill[red!80!black] (5.4,-2.55) circle (0pt) node[below] {{\tiny$E$}};
	\fill[red!80!black] (5.4,+3.6) circle (0pt) node[below] {{\tiny$E''$}};
	\fill[gray] (7.3,-1.8) circle (0pt) node[below] {{\tiny$N\!\sqcup_{E'}\!M$}};
	\end{tikzpicture}
	%%%%%
}
\ee 
The topological structure~$X$ may e.g.\ be chosen to be a spin or string structure, a framing, or a principal bundle for some group. 
Here we will exclusively consider orientations. 
In this case classification results are available for $n\leqslant 3$: 
for $n=1$ the groupoid of symmetric monoidal functors $\Bordor_{n,n-1} \lra \mathcal C$ is equivalent to that of dualisable objects in~$\mathcal C$, 
for $n=2$ the classification is via commutative Frobenius algebras in~$\mathcal C$, while for $n=3$ via ``$J$-algebras''; 
%arXiv_v2: (added semicolon)
	%, as explained in \cite{Juhasz2014}. 
	 see \cite{Juhasz2014}. 
	 
\medskip 

A certain type of algebra internal to any given rigid monoidal category~$\mathcal C$ plays an important role for many examples below, namely \textsl{$\Delta$-separable symmetric Frobenius algebras}. 
They consist of an object~$A$ together with (co)multiplication maps $\mu\colon A\otimes A \lra A$ and $\Delta\colon A \lra A\otimes A$ as well as (co)units $\one\lra A$ and $A\lra \one$, subject to the defining relations (see e.g.\ \cite{FuchsStignerFrobeniusAlgebras} for a review on such algebras and their modules, and \cite{2dDefectTQFTLectureNotes} for the string diagrammatic calculus)
\be 
\tikzzbox{%
	%%%%%%%%%%%%%%%%%%%%%%%%%%%%
	\begin{tikzpicture}[very thick,scale=0.33,color=green!50!black, baseline=0.29cm]
	\draw[-dot-] (3,0) .. controls +(0,1) and +(0,1) .. (2,0);
	\draw[-dot-] (2.5,0.75) .. controls +(0,1) and +(0,1) .. (3.5,0.75);
	\draw (3.5,0.75) -- (3.5,0); 
	\draw (3,1.5) -- (3,2.25); 
	\end{tikzpicture} 
	%%%%%%%%%%%%%%%%%%%%%%%%%%%% 
}%
=
\tikzzbox{%
	%%%%%%%%%%%%%%%%%%%%%%%%%%%%
	\begin{tikzpicture}[very thick,scale=0.33,color=green!50!black, baseline=0.29cm]
	\draw[-dot-] (3,0) .. controls +(0,1) and +(0,1) .. (2,0);
	\draw[-dot-] (2.5,0.75) .. controls +(0,1) and +(0,1) .. (1.5,0.75);
	\draw (1.5,0.75) -- (1.5,0); 
	\draw (2,1.5) -- (2,2.25); 
	\end{tikzpicture} 
	%%%%%%%%%%%%%%%%%%%%%%%%%%%% 
}%
\,,\;\;
\tikzzbox{%
	%%%%%%%%%%%%%%%%%%%%%%
	\begin{tikzpicture}[very thick,scale=0.25,color=green!50!black, baseline=-0.07cm]
	\draw (-0.5,-0.5) node[Odot] (unit) {}; 
	\fill (0,0.6) circle (7.0pt) node (meet) {};
	\draw (unit) .. controls +(0,0.5) and +(-0.5,-0.5) .. (0,0.6);
	\draw (0,-1.5) -- (0,1.5); 
	\end{tikzpicture} 
	%%%%%%%%%%%%%%%%%%%%%% 
}%
=
\tikzzbox{%
	%%%%%%%%%%%%%%%%%%%%%%%%%%%%
	\begin{tikzpicture}[very thick,scale=0.25,color=green!50!black, baseline=-0.07cm]
	\draw (0,-1.5) -- (0,1.5); 
	\end{tikzpicture} 
	%%%%%%%%%%%%%%%%%%%%%%%%%%%% 
}%
=
\tikzzbox{%
	%%%%%%%%%%%%%%%%%%%%%%
	\begin{tikzpicture}[very thick,scale=0.25,color=green!50!black, baseline=-0.07cm]
	\draw (0.5,-0.5) node[Odot] (unit) {}; 
	\fill (0,0.6) circle (7.0pt) node (meet) {};
	\draw (unit) .. controls +(0,0.5) and +(0.5,-0.5) .. (0,0.6);
	\draw (0,-1.5) -- (0,1.5); 
	\end{tikzpicture} 
	%%%%%%%%%%%%%%%%%%%%% 
}%
\,,\;\;
\tikzzbox{%
	%%%%%%%%%%%%%%%%%%%
	\begin{tikzpicture}[very thick,scale=0.33,color=green!50!black, baseline=-0.44cm, rotate=180]
	\draw[-dot-] (3,0) .. controls +(0,1) and +(0,1) .. (2,0);
	\draw[-dot-] (2.5,0.75) .. controls +(0,1) and +(0,1) .. (1.5,0.75);
	\draw (1.5,0.75) -- (1.5,0); 
	\draw (2,1.5) -- (2,2.25); 
	\end{tikzpicture} 
	%%%%%%%%%%%%%%%%%%% 
}%
=
\tikzzbox{%
	%%%%%%%%%%%%%%%%%%%
	\begin{tikzpicture}[very thick,scale=0.33,color=green!50!black, baseline=-0.44cm, rotate=180]
	\draw[-dot-] (3,0) .. controls +(0,1) and +(0,1) .. (2,0);
	\draw[-dot-] (2.5,0.75) .. controls +(0,1) and +(0,1) .. (3.5,0.75);
	\draw (3.5,0.75) -- (3.5,0); 
	\draw (3,1.5) -- (3,2.25); 
	\end{tikzpicture} 
	%%%%%%%%%%%%%%%%%%% 
}%
\,,\;\;
\tikzzbox{%
	%%%%%%%%%%%%%%%%%%%
	\begin{tikzpicture}[very thick,scale=0.25,color=green!50!black, baseline=-0.07cm, rotate=180]
	\draw (0.5,-0.5) node[Odot] (unit) {}; 
	\fill (0,0.6) circle (7.0pt) node (meet) {};
	\draw (unit) .. controls +(0,0.5) and +(0.5,-0.5) .. (0,0.6);
	\draw (0,-1.5) -- (0,1.5); 
	\end{tikzpicture} 
	%%%%%%%%%%%%%%%%%%% 
}%
=
\tikzzbox{%
	%%%%%%%%%%%%%%%%%%%
	\begin{tikzpicture}[very thick,scale=0.25,color=green!50!black, baseline=-0.07cm, rotate=180]
	\draw (0,-1.5) -- (0,1.5); 
	\end{tikzpicture} 
	%%%%%%%%%%%%%%%%%%% 
}%
=
\tikzzbox{%
	%%%%%%%%%%%%%%%%%%%
	\begin{tikzpicture}[very thick,scale=0.25,color=green!50!black, baseline=-0.07cm, rotate=180]
	\draw (-0.5,-0.5) node[Odot] (unit) {}; 
	\fill (0,0.6) circle (7.0pt) node (meet) {};
	\draw (unit) .. controls +(0,0.5) and +(-0.5,-0.5) .. (0,0.6);
	\draw (0,-1.5) -- (0,1.5); 
	\end{tikzpicture} 
	%%%%%%%%%%%%%%%%%%% 
}%
\,,\;\;
\tikzzbox{%
	%%%%%%%%%%%%%%%%%%%%%%
	\begin{tikzpicture}[very thick,scale=0.25,color=green!50!black, baseline=-0.07cm]
	\draw[-dot-] (0,0) .. controls +(0,-1) and +(0,-1) .. (-1,0);
	\draw[-dot-] (1,0) .. controls +(0,1) and +(0,1) .. (0,0);
	\draw (-1,0) -- (-1,1.5); 
	\draw (1,0) -- (1,-1.5); 
	\draw (0.5,0.8) -- (0.5,1.5); 
	\draw (-0.5,-0.8) -- (-0.5,-1.5); 
	\end{tikzpicture}
	%%%%%%%%%%%%%%%%%%%%%%
}%
=
%\tikzzbox{%
%	%%%%%%%%%%%%%%%%%%%%%%
%	\begin{tikzpicture}[very thick,scale=0.25,color=green!50!black, baseline=-0.07cm]
%	\draw[-dot-] (0,1.5) .. controls +(0,-1) and +(0,-1) .. (1,1.5);
%	\draw[-dot-] (0,-1.5) .. controls +(0,1) and +(0,1) .. (1,-1.5);
%	\draw (0.5,-0.8) -- (0.5,0.8); 
%	\end{tikzpicture}
%	%%%%%%%%%%%%%%%%%%%%%%
%}%
%=
\tikzzbox{%
	%%%%%%%%%%%%%%%%%%%%%%
	\begin{tikzpicture}[very thick,scale=0.25,color=green!50!black, baseline=-0.07cm]
	\draw[-dot-] (0,0) .. controls +(0,1) and +(0,1) .. (-1,0);
	\draw[-dot-] (1,0) .. controls +(0,-1) and +(0,-1) .. (0,0);
	\draw (-1,0) -- (-1,-1.5); 
	\draw (1,0) -- (1,1.5); 
	\draw (0.5,-0.8) -- (0.5,-1.5); 
	\draw (-0.5,0.8) -- (-0.5,1.5); 
	\end{tikzpicture}
	%%%%%%%%%%%%%%%%%%%%%%
}%
\,,\;\;
%%%%%%%%%%%%%%%%%%%%%%
\begin{tikzpicture}[very thick,scale=0.25,color=green!50!black, baseline=-0.1cm]
\draw[-dot-] (0,0) .. controls +(0,-1) and +(0,-1) .. (1,0);
\draw[-dot-] (0,0) .. controls +(0,1) and +(0,1) .. (1,0);
\draw (0.5,-0.8) -- (0.5,-1.5); 
\draw (0.5,0.8) -- (0.5,1.5); 
\end{tikzpicture}
%%%%%%%%%%%%%%%%%%%%%%
\, = \, 
%%%%%%%%%%%%%%%%%%%%%%
\begin{tikzpicture}[very thick,scale=0.25,color=green!50!black, baseline=-0.1cm]
\draw (0.5,-1.5) -- (0.5,1.5); 
\end{tikzpicture}
%%%%%%%%%%%%%%%%%%%%%%
\,,\;\;
%%%%%%%%%%%%%%%%%%%%%%
\begin{tikzpicture}[very thick,scale=0.25,color=green!50!black, baseline=-0.1cm, >=stealth]
\draw[-dot-] (0,0) .. controls +(0,1) and +(0,1) .. (-1,0);
\draw[color=green!50!black] (1,0) .. controls +(0,-1) and +(0,-1) .. (0,0);
\draw (-1,0) -- (-1,-1.5); 
\draw[postaction={decorate}, decoration={markings,mark=at position 0.4 with {\arrow[draw=green!50!black]{<}}}] (1,0) -- (1,1.5); 
\draw (-0.5,1.5) node[Odot] (end) {}; 
\draw (-0.5,0.8) -- (end); 
\end{tikzpicture}
%%%%%%%%%%%%%%%%%%%%%%
= 
%%%%%%%%%%%%%%%%%%%%%%
\begin{tikzpicture}[very thick,scale=0.25,color=green!50!black, baseline=-0.1cm, >=stealth]
\draw[green!50!black] (0,0) .. controls +(0,-1) and +(0,-1) .. (-1,0);
\draw[-dot-] (1,0) .. controls +(0,1) and +(0,1) .. (0,0);
\draw[postaction={decorate}, decoration={markings,mark=at position 0.4 with {\arrow[draw=green!50!black]{<}}}] (-1,0) -- (-1,1.5); 
\draw (1,0) -- (1,-1.5); 
\draw (0.5,1.5) node[Odot] (end) {}; 
\draw (0.5,0.8) -- (end); 
\end{tikzpicture}
%%%%%%%%%%%%%%%%%%%%%%
\,.
\ee
An algebra is \textsl{separable} if its multiplication has a right inverse as a bimodule map; a Frobenius algebra is $\Delta$-separable if this is given by the comultiplication.  

\begin{example}
	\begin{enumerate}[leftmargin=*, label={(\arabic*)}]
		\item 
		A dualisable object in $\Vectk$ or $\sVectk$ is precisely a finite-dimensional (super) $\Bbbk$-vector space. 
		\item 
		\label{item:2dClosed}
		The centre of a separable symmetric Frobenius algebra in~$\mathcal C$ is a commutative Frobenius algebra, describing a \textsl{state sum model} in dimension $n=2$, cf.\ \cite{lp0602047} and \cite[Sect.\,3.2]{Mule1}. 
		The origin of this term is reviewed in Section~\ref{subsec:OrbifoldConstruction}. 
		
		Non-semisimple examples in $\sVect_\C$ include \textsl{B-twisted sigma models}, where the commutative Frobenius algebras are constructed from Dolbeault cohomology of Calabi--Yau manifolds. 
		Non-semisimple examples in $\Vectk$ are \textsl{Landau--Ginzburg models}~$\zz_W$ with underlying commutative Frobenius algebra $\Jac_W = \Bbbk[x_i]/(\partial_{x_i} W)$ for some polynomial~$W$ such that $\Jac_W$ is finite-dimensional, see e.g.\ \cite[Ch.\,16]{mirrorbook}. 
		\item 
		\textsl{State sum models} in $n=3$ dimensions are constructed from spherical fusion categories \cite{TVmodel, bwTV1, TVire1}. 
		They turn out to be special cases of $\Vectk$-valued \textsl{Reshetikhin--Turaev models} \cite{retu2, tur, BalsamTQFT2, TVireBook} built from modular fusion categories, which are however often ``anomalous'' in the sense that the bordism category needs refinement, see \cite{tur}. 
		
		Conjecturally, \textsl{Rozansky--Witten models} \cite{RW1996} built from compact holomorphic symplectic manifolds are also 3-dimensional TQFTs valued in $\sVect_\C$. 
		\item 
		\textsl{State sum models} in $n=4$ dimensions are constructed from spherical fusion 2-categories~$\mathfrak S$ \cite{DouglasReutter2018}, see also Example~\ref{exa:SSMTQFT}\ref{item:4dSSM} below. 
		\textsl{Crane--Yetter models} are the special case when 
		%arXiv_v2: 
			%$\mathfrak S$ is the delooping of a modular fusion category. 
			 $\mathfrak S = \Bar\mathcal M$ is the delooping of a modular fusion category~$\mathcal M$, i.e.\ the 2-category with a single object~$*$ and $\End_{\Bar\mathcal M}(*)=\mathcal M$. 
		
		Conjecturally, topological twists of $\mathcal N=4$ supersymmetric Yang--Mills theory are also 4-dimensional TQFTs \cite{kw0604151}. 
		\item[($n$)]
		\label{item:ClosedEuler}
		For any~$n$, \textsl{Dijkgraaf--Witten models} \cite{DijkgraafWitten1990, FreedQuinn1993} are $n$-dimensional TQFTs built from finite (gauge) groups. 
		They are special cases of state sum models. 
		
		A simpler 
		%arXiv_v2: 
			%non-trivial 
		class of examples are \textsl{Euler TQFTs} $\zz_\psi^{\textrm{eu}} \colon \Bordor_{n,n-1} \lra \Vectk$ for some 
		%arXiv_v2: 
			%scalar $\psi\in\Bbbk^\times$: they 
			 $\psi\in\Bbbk^\times$. They 
		assign~$\Bbbk$ to every object, and on bordisms~$M$ we have 
		\be 
		\zz_\psi^{\textrm{eu}}(M) := \psi^{\chi(M) - \frac{1}{2} \chi(\partial M)}
		\ee 
		where~$\chi$ is the Euler characteristic, see e.g.\ \cite{Quinnlectures}. 
		The \textsl{trivial closed TQFT} is the special case with $\psi=1$. 
	\end{enumerate}
	\label{ex:ClosedTQFTs}
\end{example}

\section{Defect TQFTs}
\label{sec:DefectTQFTs}

The kind of ``defect'' which motivates ``defect TQFT'' is rooted in physics, where it describes a localised substance or physical system which is typically radically different from its immediate surroundings, and it often separates other regions, or mediates between them. 
Domain walls in ferromagnets are a standard (non-topological) example. 
We formalise this in terms of bordism categories $\Borddef$ whose morphisms are represented by bordisms which are stratified into submanifolds that in turn are labelled by prescribed ``defect data''. 
The main idea is captured by the following example for $n=2$: 
\be
\label{eq:DefectBordism}
\tikzzbox{
	%%%%%%%%%%%%%%%%%%%%%% 
	\begin{tikzpicture}[thick,scale=2.83, color=black, >=stealth, baseline=1.14cm]
	\coordinate (p1) at (-0.55,0);
	\coordinate (p2) at (-0.2,0);
	\coordinate (p3) at (0.2,0);
	\coordinate (p4) at (0.55,0);
	\coordinate (p5) at (0.175,0.8);
	\coordinate (p6) at (-0.175,0.8);
	\coordinate (s) at (0.175,0.07);
	\coordinate (d) at (0,0.5);
	%
	%
	% Back surface colouring
	\fill [BackSurface] 
	(p1) .. controls +(0,0.1) and +(0,0.1) ..  (p2)
	-- (p2) .. controls +(0,0.35) and +(0,0.35) ..  (p3)
	-- (p3) .. controls +(0,0.1) and +(0,0.1) ..  (p4)
	-- (p4) .. controls +(0,0.5) and +(0,-0.5) ..  (p5)
	-- (p5) .. controls +(0,0.1) and +(0,0.1) ..  (p6)
	-- (p6) .. controls +(0,-0.5) and +(0,0.5) ..  (p1)
	;
	%
	%: Front surface colouring: 
	\fill [Surface] 
	(p1) .. controls +(0,-0.1) and +(0,-0.1) ..  (p2)
	-- (p2) .. controls +(0,0.35) and +(0,0.35) ..  (p3)
	-- (p3) .. controls +(0,-0.1) and +(0,-0.1) ..  (p4)
	-- (p4) .. controls +(0,0.5) and +(0,-0.5) ..  (p5)
	-- (p5) .. controls +(0,-0.1) and +(0,-0.1) ..  (p6)
	-- (p6) .. controls +(0,-0.5) and +(0,0.5) ..  (p1)
	;
	%
	%%%%%%%%%%%%%%%%%%%%%%%%%5
	% defects: 
	\draw[black!40!red, very thick, postaction={decorate}, decoration={markings,mark=at position .5 with {\arrow[draw=black!40!red]{>}}}] (-0.52,0.2).. controls +(0.01,0.08) and +(-0.01,0.06) .. (-0.15,0.18);
	\draw[black!40!red, very thick, opacity=0.3] (-0.52,0.2).. controls +(0.01,-0.07) and +(-0.01,-0.05) .. (-0.15,0.18);
	\fill[black!40!red] (-0.32,0.28) circle (0pt) node {\scalebox{.4}{$X_1$}};
	\draw[blue!50!red, very thick, postaction={decorate}, decoration={markings,mark=at position .7 with {\arrow[draw=blue!50!red]{>}}}] (0.35,0.06) .. controls +(0,0.3) and +(0,-0.2) .. (0,0.45); 
	\fill[blue!50!red] (0.21,0.32) circle (0pt) node {\scalebox{.4}{$X_2$}};
	\fill[brown!50!blue] (0.35,0.073) circle (0.55pt) node {};
	\draw[blue!70!red, very thick, postaction={decorate}, decoration={markings,mark=at position .6 with {\arrow[draw=blue!70!red]{>}}}] (0,0.45) .. controls +(0,0) and +(0,0) .. (0,0.73);
	\fill[blue!70!red] (0.05,0.6) circle (0pt) node {\scalebox{.4}{$X_3$}};
	\fill[black] (0,0.45) circle (0.55pt) node[above] {};
	\fill[black] (0.02,0.45) circle (0pt) node[left] {{\tiny$\theta$}};
	%
	%%%%%%%%%%%%%%%%%%%%%%%%%5
	%
	\draw[thick] (p2) .. controls +(0,0.35) and +(0,0.35) ..  (p3); 
	\draw[thick] (p4) .. controls +(0,0.5) and +(0,-0.5) ..  (p5); 
	\draw[thick] (p6) .. controls +(0,-0.5) and +(0,0.5) ..  (p1);
	\draw[very thick, red!80!black] (p1) .. controls +(0,-0.1) and +(0,-0.1) ..  (p2); 
	\draw[very thick, red!80!black] (p3) .. controls +(0,-0.1) and +(0,-0.1) ..  (p4); 
	\draw[very thick, red!80!black] (p5) .. controls +(0,0.1) and +(0,0.1) ..  (p6); 
	\draw[very thick, red!80!black] (p5) .. controls +(0,-0.1) and +(0,-0.1) ..  (p6); 
	\draw[very thick, red!80!black] (p1) .. controls +(0,0.1) and +(0,0.1) ..  (p2); 
	\draw[very thick, red!80!black] (p3) .. controls +(0,0.1) and +(0,0.1) ..  (p4); 
	\fill[red!80!black] (0,0.07) circle (0pt) node[below] {{\tiny$E'$}};
	\fill[red!80!black] (0.26,0.88) circle (0pt) node[below] {{\tiny$E''$}};
	\fill[gray] (0.35,0.56) circle (0pt) node[below] {{\tiny$N$}};
	\fill[brown!50!blue] (0.35,0.073) circle (0.55pt) node {};
	\fill[blue!70!red] (0,0.723) circle (0.55pt) node {};
	\fill[red!80!black] (-0.47,0.12) circle (0pt) node {\scalebox{.4}{$u_1$}};
	\fill[red!80!black] (-0.2,0.4) circle (0pt) node {\scalebox{.4}{$u_2$}};
	\end{tikzpicture}
	%%%%%%%%%%%%%%%%%%%%%%
}
\circ 
\tikzzbox{
	%%%%%%%%%%%%%%%%%%%%%% 
	\begin{tikzpicture}[thick,scale=2.83,color=black, >=stealth, baseline=-1.14cm]
	\coordinate (p1) at (-0.55,0);
	\coordinate (p2) at (-0.2,0);
	\coordinate (p3) at (0.2,0);
	\coordinate (p4) at (0.55,0);
	\coordinate (p5) at (0.175,-0.8);
	\coordinate (p6) at (-0.175,-0.8);
	\coordinate (s) at (0.175,0.07);
	\coordinate (d) at (0,-0.5);
	%
	%
	% Back surface colouring
	\fill [BackSurface] 
	(p1) .. controls +(0,0.1) and +(0,0.1) ..  (p2)
	-- (p2) .. controls +(0,-0.35) and +(0,-0.35) ..  (p3)
	-- (p3) .. controls +(0,0.1) and +(0,0.1) ..  (p4)
	-- (p4) .. controls +(0,-0.5) and +(0,0.5) ..  (p5)
	-- (p5) .. controls +(0,0.1) and +(0,0.1) ..  (p6)
	-- (p6) .. controls +(0,0.5) and +(0,-0.5) ..  (p1)
	;
	%
	% Front surface colouring: 
	\fill [Surface] 
	(p1) .. controls +(0,-0.1) and +(0,-0.1) ..  (p2)
	-- (p2) .. controls +(0,-0.35) and +(0,-0.35) ..  (p3)
	-- (p3) .. controls +(0,-0.1) and +(0,-0.1) ..  (p4)
	-- (p4) .. controls +(0,-0.5) and +(0,0.5) ..  (p5)
	-- (p5) .. controls +(0,-0.1) and +(0,-0.1) ..  (p6)
	-- (p6) .. controls +(0,0.5) and +(0,-0.5) ..  (p1)
	;
	%
	%%%%%%%%%%%%%%%%%%%%%%%%%5
	% defects: 
	\draw[brown!50!red, very thick, postaction={decorate}, decoration={markings,mark=at position .5 with {\arrow[draw=brown!50!red]{>}}}] (-0.52,-0.2).. controls +(0.01,-0.03) and +(-0.01,-0.06) .. (-0.15,-0.18);
	\draw[brown!50!red, very thick, opacity=0.3] (-0.52,-0.2).. controls +(0.01,0.07) and +(-0.01,0.04) .. (-0.155,-0.18);
	\fill[brown!50!red] (-0.32,-0.27) circle (0pt) node {\scalebox{.4}{$X_6$}};
	\draw[blue!50!red, very thick, postaction={decorate}, decoration={markings,mark=at position .6 with {\arrow[draw=blue!50!red]{>}}}] (0,-0.45) .. controls +(0,0.2) and +(0,-0.3) .. (0.35,-0.073); 
	\fill[blue!50!red] (0.26,-0.32) circle (0pt) node {\scalebox{.4}{$X_2$}};
	\draw[brown!50!black, very thick, postaction={decorate}, decoration={markings,mark=at position .5 with {\arrow[draw=brown!50!black]{>}}}] (0.13,-0.75) .. controls +(0,0.2) and +(0.1,0) .. (0,-0.45);
	\fill[brown!50!black] (0.16,-0.55) circle (0pt) node {\scalebox{.4}{$X_4$}};
	\draw[brown!50!blue, very thick, postaction={decorate}, decoration={markings,mark=at position .5 with {\arrow[draw=brown!50!blue]{>}}}] (-0.13,-0.75) .. controls +(0,0.2) and +(-0.1,0) .. (0,-0.45);
	\fill[brown!50!blue] (-0.16,-0.55) circle (0pt) node {\scalebox{.4}{$X_5$}};
	\fill[black] (0,-0.45) circle (0.55pt) node[above] {};
	\fill[black] (0,-0.44) circle (0pt) node[below] {{\tiny$\varphi$}};
	%
	%%%%%%%%%%%%%%%%%%%%%%%%%5
	%
	\draw[thick] (p2) .. controls +(0,-0.35) and +(0,-0.35) ..  (p3); 
	\draw[thick] (p4) .. controls +(0,-0.5) and +(0,0.5) ..  (p5); 
	\draw[thick] (p6) .. controls +(0,0.5) and +(0,-0.5) ..  (p1); 
	\draw[very thick, red!80!black] (p1) .. controls +(0,-0.1) and +(0,-0.1) ..  (p2); 
	\draw[very thick, red!80!black] (p3) .. controls +(0,-0.1) and +(0,-0.1) ..  (p4) ; 
	\draw[very thick, red!80!black] (p5) .. controls +(0,-0.1) and +(0,-0.1) ..  (p6); 
	\draw[very thick, red!80!black] (p1) .. controls +(0,0.1) and +(0,0.1) ..  (p2); 
	\draw[very thick, red!80!black] (p3) .. controls +(0,0.1) and +(0,0.1) ..  (p4); 
	\draw[very thick, red!80!black] (p5) .. controls +(0,0.1) and +(0,0.1) ..  (p6); 
	\fill[red!80!black] (0,0.07) circle (0pt) node[below] {{\tiny$E'$}};
	\fill[red!80!black] (0.23,-0.73) circle (0pt) node[below] {{\tiny$E$}};
	\fill[gray] (0.33,-0.45) circle (0pt) node[below] {{\tiny$M$}};
	\fill[red!80!black] (-0.47,-0.12) circle (0pt) node {\scalebox{.4}{$u_1$}};
	\fill[red!80!black] (-0.2,-0.4) circle (0pt) node {\scalebox{.4}{$u_2$}};
	\fill[red!80!black] (0,-0.6) circle (0pt) node {\scalebox{.4}{$u_3$}};
	\fill[blue!50!red] (0.35,-0.076) circle (0.55pt) node {};
	\fill[brown!50!blue] (-0.13,-0.75) circle (0.55pt) node {};
	\fill[brown!50!black] (0.13,-0.75) circle (0.55pt) node {};
	\end{tikzpicture}
	%%%%%%%%%%%%%%%%%%%%%%
}
= 
\;
\tikzzbox{%
	%%%%%%%%%%%%%%%%%%%%%% 
	\begin{tikzpicture}[color=blue!50!black, scale = 0.4, >=stealth, baseline = 0cm]
	%% Coordinates
	\coordinate (l1) at (0,0);
	\coordinate (l2) at (2,0);
	\coordinate (r2) at (6,0);
	\coordinate (r1) at (8,0);
	\coordinate (ld1) at (1,0);
	\coordinate (ld2) at (7,0);
	\coordinate (t1) at (2.5,-1.25);
	\coordinate (t2) at (5.5,-1.25);
	\coordinate (c1) at (3,-3);
	\coordinate (c2) at (5,-3);
	\coordinate (2t1) at (2.5,1.25);
	\coordinate (2t2) at (5.5,1.25);
	\coordinate (2c1) at (3.25,3);
	\coordinate (2c2) at (4.75,3);
	\coordinate (d1) at (4,-2.63);
	\coordinate (d2) at (4.25,-3.36);
	\coordinate (d3) at (2.1,1.38);
	\coordinate (s1) at (3.5,-2.25);
	\coordinate (s2) at (4.5,-2.25);
	\coordinate (s3) at (3.5,-1.5);
	\coordinate (s4) at (4.5,-1);
	\coordinate (s5) at (1,0.5);
	\coordinate (s6) at (7.5,0);
	\coordinate (s7) at (2.3,1.8);
	\coordinate (s8) at (5,1.5);
	\coordinate (s9) at (3.5,-2.7);
	\coordinate (s10) at (4.5,-2.7);
	\coordinate (b1) at (3.2,-2.75);
	\coordinate (b2) at (3.8,-2.65);
	\coordinate (b3) at (4.67,-2.7);
	\coordinate (b4) at (3.5,2.7);
	\coordinate (b5) at (4.2,2.64);
	\coordinate (phi) at (4,-1.5);
	%
	% Filling (back)
	\fill[orange!30!white, opacity=0.8] (l1) .. controls +(0,3) and +(0,3) .. (r1) .. controls +(0,-3) and +(0,-3) .. (l1) ;
	%
	% Filling front
	\fill[orange!30!white, opacity=0.8] (l1) .. controls +(0,3) and +(0,3) .. (r1) .. controls +(0,-3) and +(0,-3) .. (l1) ;
	%
	%%%%
	% Outer curve
	\draw[black, thick] (l1) .. controls +(0,3) and +(0,3) .. (r1) .. controls +(0,-3) and +(0,-3) .. (l1) ;
	% Inner curve
	\fill [white] (l2) .. controls +(1,1) and +(-1,1) .. (r2) .. controls +(-1,-1) and +(1,-1) .. (l2) ;
	\draw[black, thick] (l2) .. controls +(1,1) and +(-1,1) .. (r2) .. controls +(-1,-1) and +(1,-1) .. (l2) ;
	\draw[black, thick] (l2) -- ($(l2) + (-0.25, 0.25)$) ;
	\draw[black, thick] (r2) -- ($(r2) + (0.25, 0.25)$) ;
	%
	%%%%%%%%%%%%%%%%%%%%%%%%%%%%%%%%%
	% "trunk1":
	\fill[white] (c1) -- (c2) .. controls +(0,1) and +(0,0) .. (t2) -- (t1) .. controls +(0,0) and +(0,1) .. (c1) ;
	\fill[orange!30!white, opacity=1] (c1) .. controls +(0,-0.5) and +(0,-0.5) .. (c2) .. controls +(0,1) and +(0,0) .. (t2) -- (t1) .. controls +(0,0) and +(0,1) .. (c1) ;
	\fill[orange!15!white, opacity=1] (c1) .. controls +(0,0.5) and +(0,0.5) ..  (c2) -- (c2) .. controls +(0,-0.5) and +(0,-0.5) .. (c1);
	\draw[black, thick] (c1) .. controls +(0,1) and +(0,0) .. (t1);
	\draw[black, thick] (c2) .. controls +(0,1) and +(0,0) .. (t2);
	%%%%%%%%%%%%%%%%%%%%%%%%%%%%%%%%%
	%
	%%%%%%%%%%%%%%%%%%%%%%%%%%%%%%%%%
	% "trunk2":
	\fill[white] (2c1) -- (2c2) .. controls +(0,-1) and +(0,0) .. (2t2) -- (2t1) .. controls +(0,0) and +(0,-1) .. (2c1) ;
	\fill[orange!30!white, opacity=1] (2c1) .. controls +(0,0.5) and +(0,0.5) .. (2c2) .. controls +(0,-1) and +(0,0) .. (2t2) -- (2t1) .. controls +(0,0) and +(0,-1) .. (2c1) ;
	\fill[orange!15!white, opacity=1] (2c1) .. controls +(0,-0.5) and +(0,-0.5) ..  (2c2) -- (2c2) .. controls +(0,0.5) and +(0,0.5) .. (2c1);
	\draw[black, thick] (2c1) .. controls +(0,-1) and +(0,0) .. (2t1);
	\draw[black, thick] (2c2) .. controls +(0,-1) and +(0,0) .. (2t2);
	%%%%%%%%%%%%%%%%%%%%%%%%%%%%%%%%%
	%
	% wrapping cycles: 
	\draw[brown!50!red, very thick, postaction={decorate}, decoration={markings,mark=at position .7 with {\arrow[draw=brown!50!red]{>}}}] (1,-1.53).. controls +(0.2,0.5) and +(-0.6,-0.3) .. (2.3,-0.27);
	\draw[brown!50!red, very thick, opacity=0.3] (1,-1.53).. controls +(0.5,0.2) and +(-0.3,-0.6) .. (2.3,-0.27);
	\fill[brown!50!red] (1.1,-0.7) circle (0pt) node {\scalebox{.4}{$X_6$}};
	\draw[black!40!red, very thick, opacity=0.3] (1.5,1.8).. controls +(0.5,-0.1) and +(-0.1,0.5) .. (2.8,0.5);
	\draw[black!40!red, very thick, postaction={decorate}, decoration={markings,mark=at position .5 with {\arrow[draw=black!40!red]{>}}}] (1.5,1.8).. controls +(0.1,-0.4) and +(-0.5,0.1) .. (2.8,0.5);
	\fill[black!40!red] (1.8,0.85) circle (0pt) node {\scalebox{.4}{$X_1$}};
	%
	%
	% boundary circles: 
	\draw[very thick, red!80!black] (c1) .. controls +(0,0.5) and +(0,0.5) ..  (c2); 
	\draw[very thick, red!80!black] (c1) .. controls +(0,-0.5) and +(0,-0.5) ..  (c2);
	\draw[very thick, red!80!black] (2c1) .. controls +(0,0.5) and +(0,0.5) ..  (2c2); 
	\draw[very thick, red!80!black] (2c1) .. controls +(0,-0.5) and +(0,-0.5) ..  (2c2); 
	%
	% line and point defects: 
	\draw[brown!50!black, very thick, postaction={decorate}, decoration={markings,mark=at position .4 with {\arrow[draw=brown!50!black]{>}}}] (b3) .. controls +(0,1.6) and +(-0.5,0) .. (phi);
	\fill[brown!50!black] ($(b3)+(0.2,0.75)$) circle (0pt) node {\scalebox{.4}{$X_4$}};
	\draw[brown!50!blue, very thick, postaction={decorate}, decoration={markings,mark=at position .4 with {\arrow[draw=brown!50!blue]{<}}}] (b1) .. controls +(0,1.5) and +(0.4,0) .. (phi);
	\fill[brown!50!blue] ($(b1)+(0.4,0.4)$) circle (0pt) node {\scalebox{.4}{$X_5$}};
	\draw[blue!50!red, very thick, postaction={decorate}, decoration={markings,mark=at position .7 with {\arrow[draw=blue!50!red]{>}}}] (phi) .. controls +(0.2,1.4) and +(0,-1.6) .. (7,0); 
	\draw[blue!50!red, very thick] (7,0) .. controls +(0,0.8) and +(0,-0.5) .. ($(b5)+(0.3,-1.3)$);
	\draw[blue!70!red, very thick, postaction={decorate}, decoration={markings,mark=at position .7 with {\arrow[draw=blue!70!red]{>}}}] ($(b5)+(0.3,-1.3)$) .. controls +(0,0.8) and +(0,-0.5) .. (b5);
	\fill[blue!70!red] ($(b5)+(-0.2,-0.7)$) circle (0pt) node {\scalebox{.4}{$X_3$}};
	\fill[blue!50!red] (6.3,-1.1) circle (0pt) node {\scalebox{.4}{$X_2$}};
	\fill[brown!50!blue] (b1) circle (4pt) node {};
	\fill[brown!50!black] (b3) circle (4pt) node {};
	\fill[blue!70!red] (b5) circle (4pt) node {};
	\fill[black] (phi) circle (4pt) node[above] {};
	\fill[black] ($(phi)+(0,0.1)$) circle (0pt) node[below] {{\tiny$\varphi$}};
	\fill[black] ($(b5)+(0.3,-1.3)$) circle (4pt) node[above] {};
	\fill[black] ($(b5)+(0.43,-1.3)$) circle (0pt) node[left] {{\tiny$\theta$}};
	\fill[red!80!black] (1,0.3) circle (0pt) node {\scalebox{.4}{$u_1$}};
	\fill[red!80!black] (3.3,1.5) circle (0pt) node {\scalebox{.4}{$u_2$}};
	\fill[red!80!black] (4.15,-2.3) circle (0pt) node {\scalebox{.4}{$u_3$}};
	\fill[red!80!black] (5.4,-2.55) circle (0pt) node[below] {{\tiny$E$}};
	\fill[red!80!black] (5.4,+3.6) circle (0pt) node[below] {{\tiny$E''$}};
	\fill[gray] (7.3,-1.8) circle (0pt) node[below] {{\tiny$N\!\sqcup_{E'}\!M$}};
	\end{tikzpicture}
	%%%%%
}
\ee 
Note that all top-dimensional strata have an orientation induced from the underlying manifold, while lower-dimensional strata come with their own orientation. 
Every stratum is also labelled: 
the prescribed \textsl{set of defect data}~$\mathds{D}$ consists of sets $D_0, D_1, \dots, D_n$, and every $j$-stratum is labelled by an element in~$D_j$. 
Moreover, $\mathds{D}$ also comes with adjacency rules about how strata may meet; for example in~\eqref{eq:DefectBordism} the ``source'' and ``target'' of $X_1\in D_1$ are $u_1\in D_2$ and $u_2\in D_2$, respectively, and the 1-strata adjacent to the 0-stratum labelled by $\varphi\in D_0$ must carry labels $X_2,X_4,X_5\in D_1$ with the given cyclic order and orientations. 
The set~$D_n$ is interpreted as a set of closed (or ``bulk'') TQFTs, while elements of~$D_{n-k}$ label ``defects'' of codimension~$k$; in particular, $D_{n-1}$ is comprised of
%arXiv_v2: 
	% ``boundary conditions'' and more general ``domain walls''. 
	  ``domain walls'' (including ``boundary conditions'', which are domain walls that have the trivial bulk TQFT on one side). 

The construction of the symmetric monoidal category $\Borddef$ is made precise in \cite[Sect.\,2.1--2.2]{CRS1} (see \cite{RunkelSuszekDefects, dkr1107.0495} for important earlier work on $n=2$, and \cite{2dDefectTQFTLectureNotes} for a review), here we point out only a few more basic aspects. 
Objects are $(n-1)$-dimensional closed manifolds~$E$ with $\mathds{D}$-labelled stratification such that $j$-strata of~$E$ are boundary components of $(j+1)$-strata of bordisms~$M$ with (co)domain~$E$ that end transversally on $\partial M$. 
By definition, all strata~$\sigma$ of a bordism~$M$ are open away from the boundary of~$M$, i.e.\ $\partial\sigma = \sigma\cap \partial M$, and all strata of objects are open. 

We only admit stratifications which locally are either cylinders over (lower-dimensional) defect balls, or cones over defect spheres. 
For example, every 2-dimensional defect bordism locally looks like one of these neighbourhoods: 
\be
%%%%%%%%%%%%%%%%%%%%%% 
\tikzzbox{\begin{tikzpicture}[very thick,scale=2.4,color=black, baseline=1.05cm, >=stealth]
	\clip (0.5,0.5) ellipse (0.5cm and 0.5cm);
	% 2-strata:
	\fill [orange!40!white, opacity=0.7] (0,0) -- (0,1) -- (1,1) -- (1,0);
	%
	% 2-strata labels:
	\fill[red!80!black] (0.5,0.5) circle (0pt) node {{\scriptsize $u_1$}};
	\end{tikzpicture}}%%popende
%%%%%%%%%%%%%%%%%%%%%% 
\qquad 
%%%%%%%%%%%%%%%%%%%%%% 
\tikzzbox{\begin{tikzpicture}[very thick,scale=2.4,color=black, baseline=1.05cm, >=stealth]
	\clip (0.5,0.5) ellipse (0.5cm and 0.5cm);
	% 2-strata:
	\fill [orange!40!white, opacity=0.7] (0,0) -- (0,1) -- (1,1) -- (1,0);
	\fill [orange!40!white, opacity=0.7] (0,0) -- (0,1) -- (0.5,1) -- (0.5,0);
	%
	% 2-strata labels:
	\fill[blue!50!black] (0.48,0.5) circle (0pt) node[right] {{\footnotesize $X$}};
	\fill[red!80!black] (0.75,0.25) circle (0pt) node {{\scriptsize $u_1$}};
	\fill[red!80!black] (0.25,0.25) circle (0pt) node {{\scriptsize $u_2$}};
	%
	% 1-strata_
	\draw[
	color=blue!50!black, 
	very thick,
	>=stealth, 
	postaction={decorate}, decoration={markings,mark=at position .5 with {\arrow[draw]{>}}}
	] 
	(0.5,0) --  (0.5,1);
	\end{tikzpicture}}%%popende
%%%%%%%%%%%%%%%%%%%%%% 
\qquad 
%%%%%%%%%%%%%%%%%%%%%%
\tikzzbox{\begin{tikzpicture}[very thick,scale=0.5,color=blue!50!black, baseline=-0.15cm]
	%	\fill[orange!40!white, opacity=0.7] (0,0) circle (2.5);
	\fill[orange!25!white, opacity=0.7] (0,0) -- (0:2.5) arc(0:120:2.5) -- (0,0);
	\fill[orange!40!white, opacity=0.7] (0,0) -- (120:2.5) arc(120:240:2.5) -- (0,0);
	\fill[orange!50!white, opacity=0.7] (0,0) -- (240:2.5) arc(240:360:2.5) -- (0,0);
	%%1-strata:  
	\draw[
	color=blue!50!black, 
	>=stealth,
	decoration={markings, mark=at position 0.5 with {\arrow{>}},
	}, postaction={decorate}
	] 
	(0,0) -- (0:2.5);
	\draw[
	color=blue!50!black, 
	>=stealth,
	decoration={markings, mark=at position 0.5 with {\arrow{<}},
	}, postaction={decorate} 
	] 
	(0,0) -- (120:2.5);
	\draw[
	color=blue!50!black, 
	>=stealth,
	decoration={markings, mark=at position 0.5 with {\arrow{<}},
	}, postaction={decorate}
	] 
	(0,0) -- (240:2.5); 
	%
	%0-stratum: 
	\fill[color=green!50!black] (0,0) circle (5pt) node[above] {{\footnotesize $\varphi$}};
	%
	%1-strata: 
	\fill[blue!50!black] (0.65,0.4) circle (0pt) node[right] {{\footnotesize $X_m$}};
	\fill[blue!50!black] (-0.85,-1.35) circle (0pt) node[right] {{\footnotesize $X_2$}};
	\fill[blue!50!black] (-2,1) circle (0pt) node[right] {{\footnotesize $X_1$}};
	%
	%2-strata: 
	\fill[red!80!black] (0.9,1.3) circle (0pt) node {{\scriptsize $u_2$}};
	\fill[blue!50!black] (0.35,-0.5) circle (0pt) node {\rotatebox{40}{{\scriptsize $\dots$}}};
	\fill[red!80!black] (-1.35,0) circle (0pt) node {\rotatebox{0}{{\scriptsize $u_1$}}};
	\end{tikzpicture}}%%popende%
%%%%%%%%%%%%%%%%%%%%%%  
\ee
where $u_i\in D_2$, $X_j\in D_1$, $\varphi\in D_0$, and $m\in\Z_{\geqslant 0}$. 
Again, which labels are allowed in such neighbourhoods is encoded in the adjacency rules of~$\mathds{D}$. 
We refer to \cite{CRS1} for the general case, and to \cite{dkr1107.0495, CMS} for details on the cases $n=2$ and $n=3$; see also Remark~\ref{rem:DefectDataFromCategories} for a higher categorical source of defect data. 

For a fixed choice of symmetric monoidal category~$\mathcal C$ and defect data~$\mathds{D}$ we have
\begin{definition}
	An \textsl{$n$-dimensional defect TQFT} is a symmetric monoidal functor
	\be
	\label{eq:Zdef}
	\zz \colon \Borddef \lra \mathcal C \, . 
	\ee 
\end{definition}

If $D_n = \{*\}$ is a one-element set and $D_j=\varnothing$ for $j\leqslant n-1$, then a defect TQFT~\eqref{eq:Zdef} reduces to a closed TQFT~\eqref{eq:Zclosed} (after forgetting the label~$*$). 
Slightly more generally, for arbitrary~$\mathds{D}$ we have a non-full embedding $\Bordor_{n,n-1} \longhookrightarrow \Borddef$ for every $u\in D_n$, which views any bordism as trivially stratified and $u$-labelled. 
Pre-composition produces closed TQFTs from defect TQFTs. 

\begin{example}
	For $n=1$ we may choose~$D_1$ to consist of finite-dimensional $\Bbbk$-vector spaces, $D_0$ of their linear maps $f\colon V\lra W$, while the adjacency rules simply read off the (co)domain~$V$ or~$W$ of~$f$. 
	Then immersing defect bordisms into~$\R^2$ and interpreting them as string diagrams in $\Vectk$, e.g.\
	\be 
	%%%%%%%%%%%%%%%%%%%%%% 
	\tikzzbox{\begin{tikzpicture}[thick,scale=1,color=gray!60!blue, baseline=-0cm, >=stealth]
		%
		% circle: 
		\draw[red!50, very thick] (0,0) circle (0.8);
		\fill[red!50] (90:0.6) circle (0pt) node {{\tiny$X$}};
		\fill[red!50] (180:0.6) circle (0pt) node {{\tiny$Y$}};
		\draw[->, very thick, red!50] (-0.100,0.8) -- (0.101,0.8) node[below] {}; 
		\draw[->, very thick, red!50] (-0.8,-0.100) -- (-0.8,0.101) node[below] {}; 
		\fill[black] (-150:0.8) circle (2.25pt) node[left] {{\tiny$q$}};
		\fill[black] (+150:0.8) circle (2.25pt) node[left] {{\tiny$p$}};
		%
		% strand: 
		\draw[string=red!50, very thick] (0.9,-1) .. controls +(0,0.25) and +(0,-0.25) .. (1.1,-0.3);
		\draw[string=red!50, very thick] (1.1,-0.3) .. controls +(0,0.15) and +(0,-0.15) .. (1.1,0.3);
		\draw[string=red!50, very thick] (1.1,0.3) .. controls +(0,0.15) and +(0,-0.15) .. (1.2,1);
		\fill[red!50] (1.33,0.6) circle (0pt) node {{\tiny$U$}};
		\fill[red!50] (1.28,-0.05) circle (0pt) node {{\tiny$W$}};
		\fill[red!50] (1.13,-0.7) circle (0pt) node {{\tiny$V$}};
		\fill[red!30] (0.9,-1) circle (2pt) node[left] {};
		\fill[red!30] (1.2,1) circle (2pt) node[left] {};
		\fill[black] (1.1,-0.3) circle (2.25pt) node[right] {{\tiny$f$}};
		\fill[black] (1.1,+0.3) circle (2.25pt) node[right] {{\tiny$g$}};
		\end{tikzpicture}}%%popende
	%%%%%%%%%%%%%%%%%%%%%% 
	\lmt 
	\Big( \operatorname{tr}(p\circ q) \cdot g\circ f \colon V\lra U \Big),
	\ee 
	gives a defect TQFT $\Bord_{1,0}^{\textrm{def}}(\mathds{D})\lra \Vectk$. 
\end{example}

\begin{example}
	For $n=2$, fix a rigid symmetric monoidal category~$\mathcal C$. 
	\begin{enumerate}[leftmargin=*, label={(\arabic*)}]
		\item 
		For $D^{\textrm{{\tiny triv}}_2}_2 = \{*\}$, $D^{\textrm{{\tiny triv}}_2}_1 = \operatorname{Ob}(\mathcal C)$ and $D^{\textrm{{\tiny triv}}_2}_0 = \operatorname{Mor}(\mathcal C)$, the \textsl{trivial defect TQFT} $\zz^{\textrm{{\tiny triv}}}_2$ simply disregards all 2-strata and interprets the remaining defects as a string diagram in~$\mathcal C$, e.g.\
		\be 
		\zz^{\textrm{{\tiny triv}}}_2 \colon 
		\tikzzbox{%
			%%%%%%%%%%%%%%%%%%%%%% 
			\begin{tikzpicture}[color=blue!50!black, scale = 0.4, >=stealth, baseline = 0cm]
			%% Coordinates
			\coordinate (l1) at (0,0);
			\coordinate (l2) at (2,0);
			\coordinate (r2) at (6,0);
			\coordinate (r1) at (8,0);
			\coordinate (ld1) at (1,0);
			\coordinate (ld2) at (7,0);
			\coordinate (t1) at (2.5,-1.25);
			\coordinate (t2) at (5.5,-1.25);
			\coordinate (c1) at (3,-3);
			\coordinate (c2) at (5,-3);
			\coordinate (2t1) at (2.5,1.25);
			\coordinate (2t2) at (5.5,1.25);
			\coordinate (2c1) at (3.25,3);
			\coordinate (2c2) at (4.75,3);
			\coordinate (d1) at (4,-2.63);
			\coordinate (d2) at (4.25,-3.36);
			\coordinate (d3) at (2.1,1.38);
			\coordinate (s1) at (3.5,-2.25);
			\coordinate (s2) at (4.5,-2.25);
			\coordinate (s3) at (3.5,-1.5);
			\coordinate (s4) at (4.5,-1);
			\coordinate (s5) at (1,0.5);
			\coordinate (s6) at (7.5,0);
			\coordinate (s7) at (2.3,1.8);
			\coordinate (s8) at (5,1.5);
			\coordinate (s9) at (3.5,-2.7);
			\coordinate (s10) at (4.5,-2.7);
			\coordinate (b1) at (3.2,-2.75);
			\coordinate (b2) at (3.8,-2.65);
			\coordinate (b3) at (4.67,-2.7);
			\coordinate (b4) at (3.5,2.7);
			\coordinate (b5) at (4.2,2.64);
			\coordinate (phi) at (3.5,-1.3);
			%
			% Filling (back)
			\fill[orange!30!white, opacity=0.8] (l1) .. controls +(0,3) and +(0,3) .. (r1) .. controls +(0,-3) and +(0,-3) .. (l1) ;
			%
			% Filling front
			\fill[orange!30!white, opacity=0.8] (l1) .. controls +(0,3) and +(0,3) .. (r1) .. controls +(0,-3) and +(0,-3) .. (l1) ;
			%
			%%%%
			% Outer curve
			\draw[black, thick] (l1) .. controls +(0,3) and +(0,3) .. (r1) .. controls +(0,-3) and +(0,-3) .. (l1) ;
			% Inner curve
			\fill [white] (l2) .. controls +(1,1) and +(-1,1) .. (r2) .. controls +(-1,-1) and +(1,-1) .. (l2) ;
			\draw[black, thick] (l2) .. controls +(1,1) and +(-1,1) .. (r2) .. controls +(-1,-1) and +(1,-1) .. (l2) ;
			\draw[black, thick] (l2) -- ($(l2) + (-0.25, 0.25)$) ;
			\draw[black, thick] (r2) -- ($(r2) + (0.25, 0.25)$) ;
			%
			%%%%%%%%%%%%%%%%%%%%%%%%%%%%%%%%%
			% "trunk1":
			\fill[white] (c1) -- (c2) .. controls +(0,1) and +(0,0) .. (t2) -- (t1) .. controls +(0,0) and +(0,1) .. (c1) ;
			\fill[orange!30!white, opacity=1] (c1) .. controls +(0,-0.5) and +(0,-0.5) .. (c2) .. controls +(0,1) and +(0,0) .. (t2) -- (t1) .. controls +(0,0) and +(0,1) .. (c1) ;
			\fill[orange!15!white, opacity=1] (c1) .. controls +(0,0.5) and +(0,0.5) ..  (c2) -- (c2) .. controls +(0,-0.5) and +(0,-0.5) .. (c1);
			\draw[black, thick] (c1) .. controls +(0,1) and +(0,0) .. (t1);
			\draw[black, thick] (c2) .. controls +(0,1) and +(0,0) .. (t2);
			%%%%%%%%%%%%%%%%%%%%%%%%%%%%%%%%%
			%
			%%%%%%%%%%%%%%%%%%%%%%%%%%%%%%%%%
			% "trunk2":
			\fill[white] (2c1) -- (2c2) .. controls +(0,-1) and +(0,0) .. (2t2) -- (2t1) .. controls +(0,0) and +(0,-1) .. (2c1) ;
			\fill[orange!30!white, opacity=1] (2c1) .. controls +(0,0.5) and +(0,0.5) .. (2c2) .. controls +(0,-1) and +(0,0) .. (2t2) -- (2t1) .. controls +(0,0) and +(0,-1) .. (2c1) ;
			\fill[orange!15!white, opacity=1] (2c1) .. controls +(0,-0.5) and +(0,-0.5) ..  (2c2) -- (2c2) .. controls +(0,0.5) and +(0,0.5) .. (2c1);
			\draw[black, thick] (2c1) .. controls +(0,-1) and +(0,0) .. (2t1);
			\draw[black, thick] (2c2) .. controls +(0,-1) and +(0,0) .. (2t2);
			%%%%%%%%%%%%%%%%%%%%%%%%%%%%%%%%%
			%
			% boundary circles: 
			\draw[very thick, red!80!black] (c1) .. controls +(0,0.5) and +(0,0.5) ..  (c2); 
			\draw[very thick, red!80!black] (c1) .. controls +(0,-0.5) and +(0,-0.5) ..  (c2);
			\draw[very thick, red!80!black] (2c1) .. controls +(0,0.5) and +(0,0.5) ..  (2c2); 
			\draw[very thick, red!80!black] (2c1) .. controls +(0,-0.5) and +(0,-0.5) ..  (2c2); 
			%
			% line and point defects: 
			\draw[brown!50!black, very thick, postaction={decorate}, decoration={markings,mark=at position .27 with {\arrow[draw=brown!50!black]{>}}}] (b3) .. controls +(0,1.5) and +(0,1.5) .. (b2);
			\fill[brown!50!black] ($(b3)+(0.2,0.75)$) circle (0pt) node {{\tiny$V$}};
			\draw[brown!50!blue, very thick, postaction={decorate}, decoration={markings,mark=at position .7 with {\arrow[draw=brown!50!blue]{>}}}] (b1) .. controls +(0,0.1) and +(0,-0.1) .. (phi);
			\fill[brown!50!blue] ($(b1)+(-0.15,0.95)$) circle (0pt) node {{\tiny$X$}};
			\draw[brown!50!orange, very thick, postaction={decorate}, decoration={markings,mark=at position .7 with {\arrow[draw=brown!50!orange]{>}}}] (phi) .. controls +(-1,0.1) and +(0,-0.6) .. (1,0); 
			\draw[brown!50!orange, very thick] (1,0) .. controls +(0,1) and +(0,-2.5) .. (b4);
			\fill[brown!50!orange] (1.4,-1) circle (0pt) node {{\tiny$Y$}};
			\draw[blue!50!red, very thick, postaction={decorate}, decoration={markings,mark=at position .7 with {\arrow[draw=blue!50!red]{<}}}] (phi) .. controls +(1,0.1) and +(0,-0.6) .. (7,0); 
			\draw[blue!50!red, very thick] (7,0) .. controls +(0,1) and +(0,-2.5) .. (b5);
			\fill[blue!50!red] (6,-1.1) circle (0pt) node {{\tiny$Z$}};
			\fill[brown!50!blue] (b1) circle (4pt) node {};
			\fill[brown!50!black] (b2) circle (4pt) node {};
			\fill[brown!50!black] (b3) circle (4pt) node {};
			\fill[brown!50!orange] (b4) circle (4pt) node {};
			\fill[blue!50!red] (b5) circle (4pt) node {};
			\fill[black] (phi) circle (4pt) node[above] {};
			\fill[black] ($(phi)+(0,-0.2)$) circle (0pt) node[above] {{\tiny$\varphi$}};
			\end{tikzpicture}
			%%%%%
		}
		\lmt 
		\Big( \varphi \otimes \ev_V \colon X \otimes V^\dagger \otimes V \lra Y \otimes Z^\dagger \Big) \,. 
		\ee 
		Here in general both the symmetric monoidal structure and the rigidity of~$\mathcal C$ are needed. 
		Note that $\zz^{\textrm{{\tiny triv}}}_2$ is ``trivial'' only on 2-strata. 
		\item 
		\label{item:2dDefectSSM}
		We define~$D^{\textrm{ss}_2}_2$ to be the set of separable symmetric Frobenius algebras in~$\mathcal C$, while~$D^{\textrm{ss}_2}_1$ and~$D^{\textrm{ss}_2}_0$ are their finite-dimensional bimodules and bimodule maps, respectively. 
		By combining closed state sum models with~$\zztriv_2$ above, one obtains the \textsl{defect state sum model} $\zz^{\textrm{ss}}_2 \colon \Bord_{2,1}^{\textrm{def}}(\mathds{D}^{\textrm{ss}_2})\lra \mathcal C$, see e.g.\ \cite{dkr1107.0495}. 
		Below in Section~\ref{subsec:OrbifoldCompletion} we will be in a better position to give details on~$\zz^{\textrm{ss}}_2$, as a special case of the orbifold construction. 
		In particular, by restricting $\zz^{\textrm{ss}}_2$ to defect bordisms all of whose 2-strata are labelled by the trivial Frobenius algebra $\one\in\mathcal C$, we essentially recover $\zz^{\textrm{{\tiny triv}}}_2$. 
	\end{enumerate}
	\label{exa:2dDefectTQFT}
\end{example}

\begin{example}
	For $n=2$ and $\mathcal C=\Vectk$, \textsl{Landau--Ginzburg models} give rise to a defect TQFT~$\zz^{\textrm{LG}}$ whose bulk theory label set~$D^{\textrm{LG}}_2$ consists of polynomials as in Example~\ref{ex:ClosedTQFTs}\ref{item:2dClosed}, but depending only on an even number of variables, cf.\ \cite{cm1208.1481, 2dDefectTQFTLectureNotes, CMM}. 
	The sets~$D^{\textrm{LG}}_1$ and~$D^{\textrm{LG}}_0$ are made up of matrix factorisations and their maps up to homotopy, respectively, as reviewed e.g.\ in \cite{cr1210.6363}. 
\end{example}

\begin{example}
	\begin{enumerate}[leftmargin=*, label={(\arabic*)}]
		\item 
		\label{item:RT}
		For $n=3$, $\mathcal C = \Vectk$ and an algebraically closed field~$\Bbbk$, \textsl{Reshetikhin--Turaev models} associated to a modular fusion category~$\mathcal M$ can be lifted to a defect TQFT~$\zz^{\textrm{RT}}_{\mathcal M}$ \cite{ks1012.0911, fsv1203.4568, CRS2, KMRS, CM3dOrbifoldCompletion}. 
		$D^{\textrm{RT}}_3$ consists of pairs $(\mathcal M,A)$ where~$A$ is a commutative $\Delta$-separable Frobenius algebra in~$\mathcal M$; the case $A=\one$ on trivially stratified bordisms recovers closed Reshetikhin--Turaev models. 
		The surface defect label set~$D^{\textrm{RT}}_2$ is made up of $\Delta$-separable symmetric Frobenius algebras~$F$ which have simultaneous bimodule structures, $D^{\textrm{RT}}_1$ consists of multimodules over such~$F$, and~$D^{\textrm{RT}}_0$ of multimodule maps. 
		One is naturally led to this somewhat intricate structure by carrying out the orbifold construction, see Section~\ref{subsec:OrbifoldCompletion}. 
		\item 
		A special case is the \textsl{trivial defect TQFT}~$\zz^{\textrm{{\tiny triv}}}_3$, which may be identified with (the Euler completion of, cf.\ Example~\ref{exa:EulerTQFT}) $\zz^{\textrm{RT}}_{\vectk}$ restricted to defect bordisms all of whose 3-strata are labelled with the trivial modular category $\vectk$ and trivial algebra~$\Bbbk$, see \cite{CRS3}. 
		Again, $\zz^{\textrm{{\tiny triv}}}_3$ is quite non-trivial away from top-dimensional strata, where it is essentially given by~$\zz^{\textrm{ss}}_2$. 
		The \textsl{3-dimensional defect state sum model}~$\zz^{\textrm{ss}}_3$ is situated ``between'' $\zz^{\textrm{{\tiny triv}}}_3$ and~$\zz^{\textrm{RT}}_{\mathcal M}$. 
		This is explained in Section~\ref{subsec:OrbifoldCompletion} as another special case of the orbifold construction. 
	\end{enumerate}
	\label{ex:3dDefectTQFT}
\end{example}

\begin{example}
	\label{exa:EulerTQFT}
	For arbitrary dimension~$n$, the Euler TQFT~$\zz^{\textrm{eu}}_\psi$ of Example~\ref{ex:ClosedTQFTs}$(n)$ can be generalised as follows \cite{CRS1}. 
	Fix $\Psi := (\psi_1,\dots,\psi_n) \in (\Bbbk^\times)^n$. 
	Let $\Bord_{n,n-1}^{\textrm{def}}$ denote the category of stratified bordisms \textsl{without} any labels for strata. 
	The \textsl{Euler defect TQFT} $\zz^{\textrm{eu}}_\Psi \colon \Bord_{n,n-1}^{\textrm{def}} \lra \Vectk$ assigns~$\Bbbk$ to every object, while for a defect bordism~$M$ we set
	\be 
	\label{eq:DefectEuiler}
	\zz^{\textrm{eu}}_\Psi(M) 
		= 
		\prod_{j=1}^n \; \prod_{\textrm{$j$-strata\,$\sigma_j\!\subset\! M$}} 
			\psi_j^{\chi(\sigma_j) - \frac{1}{2}\chi(\partial \sigma_j)} \, . 
	\ee 
	
	For an arbitrary defect TQFT~$\zz$, its \textsl{Euler completion}~$\zz^\odot$ is constructed by allowing for extra insertions of invertible point defects~$\psi_j$ on all $j$-strata, in complete analogy to~\eqref{eq:DefectEuiler}. 
	This is made precise in \cite[Sect.\,2.5]{CRS1}, where it is also shown that $(\zz^\odot)^\odot \cong \zz^\odot$ and $\zz^\odot \otimes \zz^{\textrm{eu}}_\Psi \cong \zz^\odot$ for every defect TQFT~$\zz$. 
\end{example}

All of the above examples have in common that their underlying defect data can be extracted from a corresponding higher category. 
In general, it is expected that a defect TQFT~$\zz$ as in~\eqref{eq:Zdef} gives rise to a $\mathcal C$-enriched $n$-category~$\mathcal D_\zz$ with coherent adjoints for all morphisms. 
These higher categories in practice often admit a symmetric monoidal structure, making them natural codomains of \textsl{fully extended} TQFTs, cf.\ \cite{l0905.0465, KapustinICM2010} and \cite[Sect.\,1]{CMM}. 

Objects of~$\mathcal D_\zz$ are bulk theories, i.e.\ elements of~$D_n$, $k$-morphisms with $1\leqslant k \leqslant n-1$ are defects of codimension~$k$, i.e.\ elements of $D_{n-k}$ and more generally $(n-k)$-fold cylinders over defect $k$-balls, 
while $n$-morphisms (contain~$D_0$ and) are obtained by evaluating~$\zz$ on defect $(n-1)$-spheres. 
Adjunctions come from orientation reversal and ``folding''. 
There is also a notion of Euler completion $\mathcal D_\zz^\odot$ on the level of higher categories, equivalent to $\mathcal D_{\zz^\odot}$. 

For $n=2$ and $\mathcal C = \Vectk$ it is a rigorous result that~$\mathcal D_\zz$ is a \textsl{(planar) pivotal 2-category}, i.e.\ all 1-morphisms $X\colon a\lra b$ have coherently isomorphic left and right adjoints $\dX\cong \Xd\colon b\lra a$ (see e.g.\ \cite{2dDefectTQFTLectureNotes} for a review): 

\begin{theorem}[\cite{dkr1107.0495}]
	\label{thm:DKR}
	Every defect TQFT $\zz\colon \Bord^{\textrm{def}}_{2,1}(\mathds{D}) \lra \Vectk$ gives rise to a $\Bbbk$-linear pivotal 2-category~$\mathcal D_\zz$. 
\end{theorem}

\begin{example}
	\begin{enumerate}[leftmargin=*, label={(\arabic*)}]
		\item 
		\label{item:BC}
		The pivotal 2-category associated to the $\mathcal C$-valued trivial defect TQFT~$\zz^{\textrm{{\tiny triv}}}_2$ is the delooping of the full subcategory~$\mathcal C^{\textrm{d}}$ of all dualisable objects in~$\mathcal C$, $\mathcal D_{\zz^{\textrm{{\tiny triv}}}_2} = \Bar\mathcal C^{\textrm{d}}$, whose single object~$*$ is the single element of~$D^{\textrm{{\tiny triv}}}_2$. 
		\item 
		\label{item:ssFrob}
		For state sum models, $\mathcal D_{\zzss_2}$ is the 2-category $\ssFrob(\mathcal C)$ of separable symmetric Frobenius algebras, finite-dimensional bimodules and bimodule maps in~$\mathcal C$, cf.\ \cite{dkr1107.0495}. 
		\item 
		\label{item:LG}
		The 2-category $\LG_\Bbbk$ associated to Landau--Ginzburg models admits a natural pivotal structure on its full subcategory of polynomials in an \textsl{even} number of variables \cite{cm1208.1481}. 
		(All of $\LG_\Bbbk$ is only ``graded pivotal'', cf.\ \cite[Def.\,7.1]{cm1208.1481}, this relates to the fact that Landau--Ginzburg models are \textsl{spin} TQFTs \cite{RSpinLorantNils}.) 
	\end{enumerate}
	\label{exa:Pivotal2categories}
\end{example}

For $n=3$ and $\mathcal C = \Vectk$ the higher category associated to a defect TQFT has also been rigorously constructed. 
Recall \cite{GPS, Gurskibook} that every 3-category is equivalent to a \textsl{Gray category}~$\mathcal T$, i.e.\ a category enriched in 2-categories and strict 2-functors with the Gray tensor product. 
Hence only the interchange law is not necessarily strict in~$\mathcal T$. 
A \textsl{Gray category with duals} \cite{BMS} then has pivotal Hom 2-categories and compatible adjoints for all 1-morphisms. 

\begin{theorem}[\cite{CMS}]
	\label{thm:CMS}
	Every defect TQFT $\zz\colon \Bord^{\textrm{def}}_{3,2}(\mathds{D}) \lra \Vectk$ gives rise to a $\Bbbk$-linear Gray category with duals~$\mathcal D_\zz$. 
\end{theorem}

\begin{example}
	\begin{enumerate}[leftmargin=*, label={(\arabic*)}]
		\item 
		\label{item:BssFrob}
		The 3-category associated to the trivial defect TQFT~$\zztriv_3$ valued in~$\mathcal C$ is the delooping of~$\mathcal D_{\zzss_2}$, i.e.\ $\mathcal D_{\zztriv_3} = \Bar\ssFrob(\mathcal C)$. 
		\item 
		The 3-category associated to the defect state sum model~$\zzss_3$ is described in Section~\ref{subsec:OrbifoldCompletion} below. 
		The 3-category $\textrm{sFus}_\Bbbk$ of spherical fusion categories, bimodule categories with trace, bimodule functors and natural transformations introduced in \cite{Bimodtrace} is a sub-3-category of~$\mathcal D_{\zzss_3}$. 
		\item 
		The 3-category associated to defect Reshetikhin--Turaev theory~$\zz^{\textrm{RT}}_{\mathcal M}$ was constructed in \cite{fsv1203.4568, KMRS, CM3dOrbifoldCompletion}. 
		Its objects and $k$-morphisms are given by the sets~$D^{\textrm{RT}}_3$ and~$D^{\textrm{RT}}_{3-k}$ of Example~\ref{ex:3dDefectTQFT}\ref{item:RT}, respectively, see Section~\ref{subsec:OrbifoldCompletion} below. 
		\item 
		Conjecturally, the 3-category associated to Rozansky--Witten models is the one described in \cite{KRS,KR0909.3643}. 
		The pivotal symmetric monoidal structure of the homotopy sub-2-category for affine target spaces $T^*\C^n$ is worked out rigorously in \cite{BCR, BCFR}. 
	\end{enumerate}
	\label{exa:GrayCatDual}
\end{example}

\begin{remark}
	\label{rem:DefectDataFromCategories}
	The defect data~$\mathds{D}$ of a TQFT $\zz \colon \Borddef \lra \mathcal C$ can essentially be reconstructed from the associated higher category~$\mathcal D_\zz$. 
	Conversely, from any $n$-category with duals~$\mathcal D$ one naturally extracts defect data~$\mathds{D}^{\mathcal D}$ whose label sets $D^{\mathcal D}_j$ consist of $(n-j)$-cells in~$\mathcal D$, and the adjacency rules are supplied by source and target maps as well as other composition rules in~$\mathcal D$. 
\end{remark}

\section{Orbifolds of defect TQFTs}
\label{sec:Orbifolds}

The orbifold construction takes as input a defect TQFT $\zz \colon \Borddef \lra \mathcal C$ for arbitrary $n\geqslant 1$, as well as a set of defect labels $\A_j \in D_j$ for $j\in\{1,\dots,n\}$ and $\A^+_0, \A^-_0 \in D_0$ that are subject to constraints described below. 
As output it produces a closed TQFT $\zz_\A \colon \Bordor_{n,n-1} \lra \mathcal C$ which on any given bordism~$M$ is constructed in three main steps: 
\begin{enumerate}[leftmargin=*, label={(\arabic*)}]
	\item 
	\label{item:Orb1}
	choose a nice stratification~$M^t$ of~$M$, 
	\item 
	\label{item:Orb2}
	label the strata of~$M^t$ with~$\A$, obtaining a morphism $M^{t,\A}$ in $\Borddef$, 
	\item 
	\label{item:Orb3}
	define $\zz_\A(M)$ by taking the colimit of $\zz(M^{t,\A})$ over all stratifications~$M^t$. 
\end{enumerate}
This is made precise below. 
In Section~\ref{subsec:OrbifoldData} we take ``nice stratifications'' to be Poincar\'{e} dual to oriented triangulations, and explain how independence of choice of triangulation imposes (algebraic) defining conditions on ``orbifold data''~$\A$, thus covering steps~\ref{item:Orb1} and~\ref{item:Orb2} from above. 
In Section~\ref{subsec:OrbifoldConstruction}, we describe step~\ref{item:Orb3} in detail, and we provide several examples of orbifold TQFTs~$\zz_\A$, including state sum models and gaugings of (higher) symmetry groups as special cases. 
Finally, in Section~\ref{subsec:OrbifoldCompletion}, we explain how the above orbifold construction can be generalised, at least for $n\leqslant 3$, to produce a single orbifold \textsl{defect} TQFT that subsumes all the orbifold closed TQFTs $\{\zz_\A\}_\A$ as well as all their defects. 
Algebraically, this is captured by the $n$-category $(\mathcal D_\zz)_{\textrm{orb}}$ of \textsl{representations} of all orbifold data~$\A$.

\subsection{Orbifold data}
\label{subsec:OrbifoldData}

There are several contenders for ``nice stratifications'' of bordisms in relation to orbifold TQFTs. 
For example, the ``admissible skeleta'' of \cite{TVireBook, CMRSS1} are a good choice in practice for low dimensions. 
Here we will exclusively consider the more traditional choice of stratifications that are  Poincar\'{e} dual to triangulations, which are particularly useful for the general development of the theory. 

%arXiv_v2: 
	%Recall, e.g.\ from the textbook \cite{Munkres} or the summary in \cite[Sect.\,3.1]{CRS1}, that for $n\in\Z_{+}$ the \textsl{standard $n$-simplex} is $\Delta^n := \{ \sum_{i=1}^{n+1} t_i e_i \,|\, t_i\in\R_{\geqslant 0}, \; \sum_i t_i = 1 \} \longhookrightarrow \R^n$, 
	 Recall that $\Delta^n := \{ \sum_{i=1}^{n+1} t_i e_i \,|\, t_i\in\R_{\geqslant 0}, \; \sum_i t_i = 1 \} \longhookrightarrow \R^n$ is the \textsl{standard $n$-simplex}, 
where $\{e_i\}_i$ is the standard basis of~$\R^{n+1}$. 
%arXiv_v2: 
	See e.g.\ \cite[Sect.\,3.1]{CRS1} for more details on the (affine) simplicial notions that we use in the following. 
By an \textsl{oriented $n$-simplex} we mean an $n$-simplex with a total order on the set of its vertices up to equivalence (given by even permutations of vertices). 
For example,  
\be 
\label{eq:OrientedSimplices}
\Delta^1_+
= 
%%%%%%%%%%%%%%%%%%%%%% 
\tikzzbox{\begin{tikzpicture}[thick,scale=2.021,color=gray!60!blue, baseline=-0.1cm, >=stealth]
	%				style={x={(-0.6cm,-0.4cm)},y={(1cm,-0.2cm)},z={(0cm,0.9cm)}}]
	\coordinate (v1) at (0,0);
	\coordinate (v2) at (1,0);
	\fill[color=gray!60!blue] (v1) circle (0.9pt) node[above] (0up) {};
	\fill[color=gray!60!blue] (v2) circle (0.9pt) node[above] (0up) {};
	\draw[string=gray!60!blue, very thick] (v1) -- (v2);
	\fill[color=gray!60!blue] (v1) circle (0.9pt) node[above] (0up) {{\scriptsize$1$}};
	\fill[color=gray!60!blue] (v2) circle (0.9pt) node[above] (0up) {{\scriptsize$2$}};
	\end{tikzpicture}}%%popende
%%%%%%%%%%%%%%%%%%%%%% 
\, , \quad 
\Delta^2_+
= 
%%%%%%%%%%%%%%%%%%%%%% 
\tikzzbox{\begin{tikzpicture}[thick,scale=2.021,color=gray!60!blue, baseline=0.75cm, >=stealth]
	%				style={x={(-0.6cm,-0.4cm)},y={(1cm,-0.2cm)},z={(0cm,0.9cm)}}]
	\coordinate (v1) at (0,0);
	\coordinate (v2) at (1,0);
	\coordinate (v3) at (0,1);
	\fill [blue!30,opacity=0.545] (v1) -- (v2) -- (v3);
	\fill[color=gray!60!blue] (v1) circle (0.9pt) node[above] (0up) {};
	\fill[color=gray!60!blue] (v2) circle (0.9pt) node[above] (0up) {};
	\fill[color=gray!60!blue] (v3) circle (0.9pt) node[above] (0up) {};
	\draw[string=gray!60!blue, very thick] (v1) -- (v2);
	\draw[string=gray!60!blue, very thick] (v1) -- (v3);
	\draw[string=gray!60!blue, very thick] (v2) -- (v3);
	\fill[color=gray!60!blue] (0.35,0.35) circle (0pt) node (0up) {$\circlearrowleft$};
	\fill[color=gray!60!blue] (v1) circle (0.9pt) node[left] (0up) {{\scriptsize$1$}};
	\fill[color=gray!60!blue] (v2) circle (0.9pt) node[above] (0up) {{\scriptsize$2$}};
	\fill[color=gray!60!blue] (v3) circle (0.9pt) node[left] (0up) {{\scriptsize$3$}};
	\end{tikzpicture}}%%popende
%%%%%%%%%%%%%%%%%%%%%% 
\, , \quad 
\Delta^3_+
= 
%%%%%%%%%%%%%%%%%%%%%% 
\tikzzbox{\begin{tikzpicture}[thick,scale=2.321,color=gray!60!blue, baseline=-0.3cm, >=stealth, 
	style={x={(-0.6cm,-0.4cm)},y={(1cm,-0.2cm)},z={(0cm,0.9cm)}}]
	\coordinate (v1) at (1,0,0);
	\coordinate (v2) at (1,1,0);
	\coordinate (v3) at (0,0,0);
	\coordinate (v4) at (0.25,0.1,0.75);
	\fill [blue!20,opacity=0.545] (v1) -- (v2) -- (v3);
	\fill [blue!20,opacity=0.545] (v4) -- (v2) -- (v3);
	\fill [blue!20,opacity=0.545] (v1) -- (v4) -- (v3);
	\fill[color=gray!60!blue] (v3) circle (0.9pt) node[left] (0up) {{\scriptsize$3$}};
	\fill[color=gray!60!blue] (v1) circle (0.9pt) node[above] (0up) {{\scriptsize$1$}};
	\fill[color=gray!60!blue] (v2) circle (0.9pt) node[right] (0up) {{\scriptsize$2$}};
	\fill[color=gray!60!blue] (v4) circle (0.9pt) node[right] (0up) {{\scriptsize$4$}};
	\draw[string=gray!60!blue, very thick] (v1) -- (v3);
	\draw[string=gray!60!blue, very thick] (v2) -- (v3);
	\draw[string=gray!60!blue, very thick] (v3) -- (v4);
	\fill [blue!20,opacity=0.545] (v1) -- (v2) -- (v4);
	\draw[string=gray!60!blue, very thick] (v2) -- (v4);
	\draw[string=gray!60!blue, very thick] (v1) -- (v4);
	\draw[string=gray!60!blue, very thick] (v1) -- (v2);
	\end{tikzpicture}}%%popende
%%%%%%%%%%%%%%%%%%%%%% 
\ee 
represent oriented $n$-simplices for $n\in\{1,2,3\}$, and similarly for the oppositely oriented simplices~$\Delta^n_-$ that are represented, say, by swapping (only)~$1$ and~$2$. 
Note that before taking equivalence classes, a total order on vertices induces orientations on all subsimplices. 
To avoid clutter, we usually denote oriented $n$-simplices simply by~$\Delta^n$, and we note that the boundary~$\partial \Delta^n$ of~$\Delta^n$ consists of precisely $n+1$ $(n-1)$-simplices. 

A \textsl{simplicial complex}~$C$ is a finite collection of simplices which is closed with respect to taking faces, and such that for all $\delta,\delta'$ in~$C$, $\delta\cap\delta'$ is either empty or a face of~$\delta$ and~$\delta'$. 
Since~$\Delta^n$ can naturally be viewed as a topological space~$|\Delta^n|$ (with topology induced by~$\R^n$), for any simplicial complex~$C$ we get a topological space~$|C|$, called its geometric realisation. 

By definition, an \textsl{(oriented) triangulation}~$t$ of an $n$-manifold~$M$ is a simplicial complex~$C$ with a total ordering on its vertices, together with a homeomorphism $|C| \lra M$; we also ask that all simplices~$\delta$ satisfy either $\delta\subset\partial M$ or $\delta^\circ \cap \partial M = \varnothing$. 
It is customary to depict~$t$ as the image of~$|C|$ in~$M$, thus ``approximating''~$M$ by a mesh of $n$-simplices glued along their faces. 
For example, 
\be 
\label{eq:TriaTorus}
\tikzzbox{%
	%%%%%%%%%%%%%%%%%%%%%% 
	\begin{tikzpicture}[color=blue!50!black, scale = 0.4, baseline = 0cm]
	%% Coordinates
	\coordinate (l1) at (-4,0);
	\coordinate (l2) at (-2,0);
	\coordinate (r2) at (+2,0);
	\coordinate (r1) at (+4,0);
	\coordinate (ld1) at (-3,0);
	\coordinate (ld2) at (+3,0);
	\coordinate (t1) at (-1.5,-1.75);
	\coordinate (t2) at (+1.5,-1.75);
	\coordinate (c1) at (-0.75,-3);
	\coordinate (c2) at (+0.75,-3);
	\coordinate (s1) at (-3.4,-1.25);
	\coordinate (s2) at (-2.7,-1.25);
	\coordinate (s3) at (-0.55,-1.25);
	\coordinate (s4) at (0,-1.25);
	\coordinate (s5) at (+0.55,-1.25);
	\coordinate (s6) at (+2.7,-1.25);
	\coordinate (s7) at (+3.4,-1.25);
	\coordinate (s8) at (-2.3,-1.87);
	\coordinate (s9) at (-0.55,-2.74);
	\coordinate (s10) at (0,-2.63);
	\coordinate (s11) at (+0.55,-2.74);
	\coordinate (s12) at (-0.3,-3.35);
	\coordinate (s13) at (+0.3,-3.35);
	\coordinate (s14) at (-0.55,-0.73);
	\coordinate (s15) at (+0.55,-0.73);
	\coordinate (s16) at (-3.4,1.25);
	\coordinate (s17) at (-2.7,+1.25);
	\coordinate (s18) at (-0.55,+1.25);
	\coordinate (s19) at (+0.55,+1.25);
	\coordinate (s20) at (+2.7,+1.25);
	\coordinate (s21) at (+3.4,1.25);
	\coordinate (s22) at (-0.55,2.23);
	\coordinate (s23) at (+0.55,2.23);
	\coordinate (s25) at (+2.3,-1.87);
	\coordinate (s26) at (-1.6,0.35);
	\coordinate (s27) at (-0.55,0.73);
	\coordinate (s28) at (0,0.75);
	\coordinate (s29) at (+0.55,0.73);
	\coordinate (s30) at (+1.6,0.35);
	%
	% Filling (back)
	\fill[orange!30!white, opacity=0.8] (l1) .. controls +(0,3) and +(0,3) .. (r1) .. controls +(0,-3) and +(0,-3) .. (l1) ;
	%
	% Filling front
	\fill[orange!30!white, opacity=0.8] (l1) .. controls +(0,3) and +(0,3) .. (r1) .. controls +(0,-3) and +(0,-3) .. (l1) ;
	%
	% 0-simplices: 
	\fill[blue] (s2) circle (3pt) node {};
	\fill[blue] (s3) circle (3pt) node {};
	\fill[blue] (s4) circle (3pt) node {};
	\fill[blue] (s5) circle (3pt) node {};
	\fill[blue] (s6) circle (3pt) node {};
	\fill[blue] (s9) circle (3pt) node {};
	\fill[blue] (s17) circle (3pt) node {};
	\fill[blue] (s18) circle (3pt) node {};
	\fill[blue] (s19) circle (3pt) node {};
	\fill[blue] (s20) circle (3pt) node {};
	%
	% 1-simplices: 
	\draw[blue, thick] (s1) -- (s7);
	\draw[blue, thick] (s16) -- (s21);
	\draw[blue, thick] (s16) -- (s1);
	\draw[blue, thick] (s7) -- (s21);
	\draw[blue, thick] (s2) -- (s17);
	\draw[blue, thick] (s2) -- (s16);
	\draw[blue, thick] (s2) -- (s14);
	\draw[blue, thick] (s3) -- (s14);
	\draw[blue, thick] (s4) -- (s14);
	\draw[blue, thick] (s6) -- (s15);
	\draw[blue, thick] (s5) -- (s15);
	\draw[blue, thick] (s4) -- (s15);
	\draw[blue, thick] (s2) -- (s8);
	\draw[blue, thick] (s3) -- (s8);
	\draw[blue, thick] (s6) -- (s25);
	\draw[blue, thick] (s5) -- (s25);
	\draw[blue, thick] (s17) -- (s22);
	\draw[blue, thick] (s18) -- (s22);
	\draw[blue, thick] (s19) -- (s22);
	\draw[blue, thick] (s19) -- (s23);
	\draw[blue, thick] (s20) -- (s23);
	\draw[blue, thick] (s6) -- (s20);
	\draw[blue, thick] (s6) -- (s21);
	\draw[blue, thick] (s17) -- (s26);
	\draw[blue, thick] (s18) -- (s27);
	\draw[blue, thick] (s18) -- (s28);
	\draw[blue, thick] (s19) -- (s29);
	\draw[blue, thick] (s20) -- (s30);
	\draw[blue, very thin] (s19) -- (s23);
	%
	%%%%
	% Outer curve
	\draw[black, thick] (l1) .. controls +(0,3) and +(0,3) .. (r1) .. controls +(0,-3) and +(0,-3) .. (l1) ;
	% Inner curve
	\fill [white] (l2) .. controls +(1,1) and +(-1,1) .. (r2) .. controls +(-1,-1) and +(1,-1) .. (l2) ;
	\draw[black, thick] (l2) .. controls +(1,1) and +(-1,1) .. (r2) .. controls +(-1,-1) and +(1,-1) .. (l2) ;
	\draw[black, thick] (l2) -- ($(l2) + (-0.25, 0.25)$) ;
	\draw[black, thick] (r2) -- ($(r2) + (0.25, 0.25)$) ;
	%
	%%%%%%%%%%%%%%%%%%%%%%%%%%%%%%%%%
	% "trunk":
	\fill[white] (c1) -- (c2) .. controls +(0,1) and +(0,0) .. (t2) -- (t1) .. controls +(0,0) and +(0,1) .. (c1) ;
	\fill[orange!30!white, opacity=1] (c1) .. controls +(0,-0.5) and +(0,-0.5) .. (c2) .. controls +(0,1) and +(0,0) .. (t2) -- (t1) .. controls +(0,0) and +(0,1) .. (c1) ;
	\fill[orange!15!white, opacity=1] (c1) .. controls +(0,0.5) and +(0,0.5) ..  (c2) -- (c2) .. controls +(0,-0.5) and +(0,-0.5) .. (c1);
	\draw[black, thick] (c1) .. controls +(0,1) and +(0,0) .. (t1);
	\draw[black, thick] (c2) .. controls +(0,1) and +(0,0) .. (t2);
	%%%%%%%%%%%%%%%%%%%%%%%%%%%%%%%%%
	%
	% 1-simplices: 
	\draw[blue, thick] (s3) -- (s9);
	\draw[blue, thick] (s3) -- (s10);
	\draw[blue, thick] (s4) -- (s10);
	\draw[blue, thick] (s4) -- (s11);
	\draw[blue, thick] (s12) -- ($(s12)+(0,0.7)$);
	\draw[blue, thick] (s13) -- ($(s13)+(0,0.7)$);
	\draw[blue, thick] (s5) -- (s11);
	\draw[blue, thick] (s9) -- ($(s9)+(-0.2,0)$);
	\draw[blue, thick] (s11) -- ($(s11)+(+0.2,0)$);
	\draw[blue, thick] (s13) -- ($(s13)+(-0.15,0.68)$);
	%
	% boundary circle: 
	\draw[very thick, red!80!black] (c1) .. controls +(0,0.5) and +(0,0.5) ..  (c2); 
	\draw[very thick, red!80!black] (c1) .. controls +(0,-0.5) and +(0,-0.5) ..  (c2); 
	%
	% 0-simplices: 
	\fill[blue] (s9) circle (3pt) node {};
	\fill[blue] (s10) circle (3pt) node {};
	\fill[blue] (s11) circle (3pt) node {};
	\fill[blue] (s12) circle (3pt) node {};
	\fill[blue] (s13) circle (3pt) node {};
	\end{tikzpicture}
	%%%%%
}
\;\; 
\textrm{ is a triangulation of }
\;\; 
\tikzzbox{%
	%%%%%%%%%%%%%%%%%%%%%% 
	\begin{tikzpicture}[color=blue!50!black, scale = 0.4, baseline = 0cm]
	%% Coordinates
	\coordinate (l1) at (-4,0);
	\coordinate (l2) at (-2,0);
	\coordinate (r2) at (+2,0);
	\coordinate (r1) at (+4,0);
	\coordinate (ld1) at (-3,0);
	\coordinate (ld2) at (+3,0);
	\coordinate (t1) at (-1.5,-1.75);
	\coordinate (t2) at (+1.5,-1.75);
	\coordinate (c1) at (-0.75,-3);
	\coordinate (c2) at (+0.75,-3);
	%
	% Filling (back)
	\fill[orange!30!white, opacity=0.8] (l1) .. controls +(0,3) and +(0,3) .. (r1) .. controls +(0,-3) and +(0,-3) .. (l1) ;
	%
	% Filling front
	\fill[orange!30!white, opacity=0.8] (l1) .. controls +(0,3) and +(0,3) .. (r1) .. controls +(0,-3) and +(0,-3) .. (l1) ;
	%
	%%%%
	% Outer curve
	\draw[black, thick] (l1) .. controls +(0,3) and +(0,3) .. (r1) .. controls +(0,-3) and +(0,-3) .. (l1) ;
	% Inner curve
	\fill [white] (l2) .. controls +(1,1) and +(-1,1) .. (r2) .. controls +(-1,-1) and +(1,-1) .. (l2) ;
	\draw[black, thick] (l2) .. controls +(1,1) and +(-1,1) .. (r2) .. controls +(-1,-1) and +(1,-1) .. (l2) ;
	\draw[black, thick] (l2) -- ($(l2) + (-0.25, 0.25)$) ;
	\draw[black, thick] (r2) -- ($(r2) + (0.25, 0.25)$) ;
	%
	%%%%%%%%%%%%%%%%%%%%%%%%%%%%%%%%%
	% "trunk":
	\fill[white] (c1) -- (c2) .. controls +(0,1) and +(0,0) .. (t2) -- (t1) .. controls +(0,0) and +(0,1) .. (c1) ;
	\fill[orange!30!white, opacity=1] (c1) .. controls +(0,-0.5) and +(0,-0.5) .. (c2) .. controls +(0,1) and +(0,0) .. (t2) -- (t1) .. controls +(0,0) and +(0,1) .. (c1) ;
	\fill[orange!15!white, opacity=1] (c1) .. controls +(0,0.5) and +(0,0.5) ..  (c2) -- (c2) .. controls +(0,-0.5) and +(0,-0.5) .. (c1);
	\draw[black, thick] (c1) .. controls +(0,1) and +(0,0) .. (t1);
	\draw[black, thick] (c2) .. controls +(0,1) and +(0,0) .. (t2);
	%%%%%%%%%%%%%%%%%%%%%%%%%%%%%%%%%
	%
	% boundary circle: 
	\draw[very thick, red!80!black] (c1) .. controls +(0,0.5) and +(0,0.5) ..  (c2); 
	\draw[very thick, red!80!black] (c1) .. controls +(0,-0.5) and +(0,-0.5) ..  (c2); 
	\end{tikzpicture}
	%%%%%
}
= T^2 \setminus B^2 \, , 
\ee 
where we suppress orientations, and we do not show the triangulation on the rear side of the torus. 

An important result is that every two triangulations of a given piecewise linear (PL) manifold can be related by a finite sequence of ``moves'' that make only local changes. 
To make this precise on the level of an $n$-dimensional simplicial complex~$C$, let $K\subset C$ be an $n$-dimensional subcomplex together with an isomorphism $\varphi \colon K \lra F$ to an $n$-dimensional subcomplex $F\subset \partial \Delta^{n+1}$ with precisely~$k$ $n$-simplices. 
The associated \textsl{$k$-$(n+2-k)$ Pachner move} is the replacement 
\be 
C \lmt \big(C\setminus K\big) \cup_{\varphi} \big(\partial \Delta^{n+1} \setminus \overset{\circ}{F} \big)
\ee 
which exchanges~$K$ by the $n+2-k$ faces on ``the other side of~$\partial\Delta^{n+1}$''. 
Pachner's theorem \cite{Pachpaper} then states that if two triangulated PL manifolds are PL isomorphic, then there exists a finite sequence of Pachner moves between them. 
From this it is straightforward to obtain an analogous result for oriented triangulations, cf.\ \cite[Prop.\,3.3]{CRS1}. 
From now on we assume all triangulations to be oriented. 

\begin{example}
	\begin{enumerate}[leftmargin=*, label={(\arabic*)}]
		\item
		\label{item:1dPachnerMove}
		There is only one type of Pachner move (and its inverse) for $n=1$: 
		$\partial\Delta^2$ has precisely three faces, and the 2-1 Pachner move replaces two 1-simplices joined at one vertex with a single 1-simplex (opposite to that vertex in $\partial\Delta^2$), 
		\be 
		\label{eq:1dPachner}
		%%%%%%%%%%%%%%%%%%%%%% 
		\tikzzbox{\begin{tikzpicture}[thick,scale=2.021,color=gray!60!blue, baseline=-0.1cm, >=stealth]
			%				style={x={(-0.6cm,-0.4cm)},y={(1cm,-0.2cm)},z={(0cm,0.9cm)}}]
			\coordinate (v1) at (0,0);
			\coordinate (v2) at (1,0);
			\coordinate (vb) at (0.5,0);
			b
			\fill[color=gray!60!blue] (v1) circle (0.9pt) node[above] (0up) {};
			\fill[color=gray!60!blue] (v2) circle (0.9pt) node[above] (0up) {};
			\fill[color=gray!60!blue] (v2) circle (0.9pt) node[above] (0up) {};
			\draw[gray!60!blue, very thick] (v1) -- (v2);
			\fill[color=gray!60!blue] (v1) circle (0.9pt) node[above] (0up) {{\scriptsize$a$}};
			\fill[color=gray!60!blue] (vb) circle (0.9pt) node[above] (0up) {{\scriptsize$b$}};
			\fill[color=gray!60!blue] (v2) circle (0.9pt) node[above] (0up) {{\scriptsize$c$}};
			\end{tikzpicture}}%%popende
		%%%%%%%%%%%%%%%%%%%%%%  
		\stackrel{\text{2-1}}{\longleftrightarrow}
		%%%%%%%%%%%%%%%%%%%%%% 
		\tikzzbox{\begin{tikzpicture}[thick,scale=2.021,color=gray!60!blue, baseline=-0.1cm, >=stealth]
			%				style={x={(-0.6cm,-0.4cm)},y={(1cm,-0.2cm)},z={(0cm,0.9cm)}}]
			\coordinate (v1) at (0,0);
			\coordinate (v2) at (1,0);
			\fill[color=gray!60!blue] (v1) circle (0.9pt) node[above] (0up) {};
			\fill[color=gray!60!blue] (v2) circle (0.9pt) node[above] (0up) {};
			\draw[gray!60!blue, very thick] (v1) -- (v2);
			\fill[color=gray!60!blue] (v1) circle (0.9pt) node[above] (0up) {{\scriptsize$a$}};
			\fill[color=gray!60!blue] (v2) circle (0.9pt) node[above] (0up) {{\scriptsize$c$}};
			\end{tikzpicture}}%%popende
		%%%%%%%%%%%%%%%%%%%%%% 
		\ee 
		where $a,b,c$ are pairwise distinct real numbers that give a total order (and induced orientation). 
		Note that there is one oriented Pachner move for each of the three relative orders that~$b$ can have with respect to $a<c$, and there are three further moves for $a>c$. 
		\item 
		In dimension $n=2$, there are 2-2 and 1-3 Pachner moves, corresponding to partitions of the four faces of $\partial\Delta^3$: 
		\be 
		\label{eq:2dPachner}
		\!
		%%%%%%%%%%%%%%%%%%%%%% 
		\tikzzbox{\begin{tikzpicture}[thick,scale=1.5,color=gray!60!blue, baseline=0.5cm, >=stealth]
			\coordinate (v1) at (0,0);
			\coordinate (v2) at (1.5,0);
			\coordinate (v3) at (1.5,1);
			\coordinate (v4) at (0,1);
			\fill [blue!15,opacity=1] (v1) -- (v2) -- (v3) -- (v4);
			\draw[color=gray!60!blue, very thick] (v1) -- (v2);
			\draw[color=gray!60!blue, very thick] (v1) -- (v3);
			\draw[color=gray!60!blue, very thick] (v2) -- (v3);
			\draw[color=gray!60!blue, very thick] (v3) -- (v4);
			\draw[color=gray!60!blue, very thick] (v1) -- (v4);
			\fill[color=gray!60!blue] (v1) circle (1.3pt) node[left] (0up) {};
			\fill[color=gray!60!blue] (v2) circle (1.3pt) node[right] (0up) {};
			\fill[color=gray!60!blue] (v3) circle (1.3pt) node[left] (0up) {};
			\fill[color=gray!60!blue] (v4) circle (1.3pt) node[right] (0up) {};
			\fill[color=gray!60!blue] (v1) circle (0.9pt) node[left] (0up) {{\scriptsize$a$}};
			\fill[color=gray!60!blue] (v2) circle (0.9pt) node[right] (0up) {{\scriptsize$b$}};
			\fill[color=gray!60!blue] (v3) circle (0.9pt) node[right] (0up) {{\scriptsize$c$}};
			\fill[color=gray!60!blue] (v4) circle (0.9pt) node[left] (0up) {{\scriptsize$d$}};
			\end{tikzpicture}}%%popende
		%%%%%%%%%%%%%%%%%%%%%% 
		\!\stackrel{\text{2-2}}{\longleftrightarrow}\!
		%%%%%%%%%%%%%%%%%%%%%% 
		\tikzzbox{\begin{tikzpicture}[thick,scale=1.5,color=gray!60!blue, baseline=0.5cm, >=stealth]
			\coordinate (v1) at (0,0);
			\coordinate (v2) at (1.5,0);
			\coordinate (v3) at (1.5,1);
			\coordinate (v4) at (0,1);
			\fill [blue!15,opacity=1] (v1) -- (v2) -- (v3) -- (v4);
			\draw[color=gray!60!blue, very thick] (v1) -- (v2);
			\draw[color=gray!60!blue, very thick] (v2) -- (v4);
			\draw[color=gray!60!blue, very thick] (v2) -- (v3);
			\draw[color=gray!60!blue, very thick] (v3) -- (v4);
			\draw[color=gray!60!blue, very thick] (v1) -- (v4);
			\fill[color=gray!60!blue] (v1) circle (1.3pt) node[left] (0up) {};
			\fill[color=gray!60!blue] (v2) circle (1.3pt) node[right] (0up) {};
			\fill[color=gray!60!blue] (v3) circle (1.3pt) node[left] (0up) {};
			\fill[color=gray!60!blue] (v4) circle (1.3pt) node[right] (0up) {};
			\fill[color=gray!60!blue] (v1) circle (0.9pt) node[left] (0up) {{\scriptsize$a$}};
			\fill[color=gray!60!blue] (v2) circle (0.9pt) node[right] (0up) {{\scriptsize$b$}};
			\fill[color=gray!60!blue] (v3) circle (0.9pt) node[right] (0up) {{\scriptsize$c$}};
			\fill[color=gray!60!blue] (v4) circle (0.9pt) node[left] (0up) {{\scriptsize$d$}};
			\end{tikzpicture}}%%popende
		%%%%%%%%%%%%%%%%%%%%%% 
		\!\! , \quad
		%%%%%%%%%%%%%%%%%%%%%% 
		\tikzzbox{\begin{tikzpicture}[thick,scale=1.5,color=gray!60!blue, baseline=0.5cm, >=stealth]
			\coordinate (v1) at (0,0);
			\coordinate (v2) at (1.4,0);
			\coordinate (v3) at (0.7,1.1);
			\fill [blue!15,opacity=1] (v1) -- (v2) -- (v3);
			\draw[color=gray!60!blue, very thick] (v1) -- (v2);
			\draw[color=gray!60!blue, very thick] (v1) -- (v3);
			\draw[color=gray!60!blue, very thick] (v2) -- (v3);
			\fill[color=gray!60!blue] (v1) circle (1.3pt) node[left] (0up) {};
			\fill[color=gray!60!blue] (v2) circle (1.3pt) node[right] (0up) {};
			\fill[color=gray!60!blue] (v3) circle (1.3pt) node[left] (0up) {};
			\fill[color=gray!60!blue] (v1) circle (0.9pt) node[above] (0up) {{\scriptsize$a$}};
			\fill[color=gray!60!blue] (v2) circle (0.9pt) node[above] (0up) {{\scriptsize$b$}};
			\fill[color=gray!60!blue] (v3) circle (0.9pt) node[left] (0up) {{\scriptsize$c$}};
			\end{tikzpicture}}%%popende
		%%%%%%%%%%%%%%%%%%%%%% 
		\stackrel{\text{1-3}}{\longleftrightarrow}
		%%%%%%%%%%%%%%%%%%%%%% 
		\tikzzbox{\begin{tikzpicture}[thick,scale=1.5,color=gray!60!blue, baseline=0.5cm, >=stealth]
			\coordinate (v1) at (0,0);
			\coordinate (v2) at (1.4,0);
			\coordinate (v3) at (0.7,1.1);
			\coordinate (v4) at (0.7,0.42);
			\fill [blue!15,opacity=1] (v1) -- (v2) -- (v3);
			\draw[color=gray!60!blue, very thick] (v1) -- (v2);
			\draw[color=gray!60!blue, very thick] (v1) -- (v3);
			\draw[color=gray!60!blue, very thick] (v2) -- (v3);
			\draw[color=gray!60!blue, very thick] (v4) -- (v1);
			\draw[color=gray!60!blue, very thick] (v4) -- (v2);
			\draw[color=gray!60!blue, very thick] (v4) -- (v3);
			\fill[color=gray!60!blue] (v1) circle (1.3pt) node[left] (0up) {};
			\fill[color=gray!60!blue] (v2) circle (1.3pt) node[right] (0up) {};
			\fill[color=gray!60!blue] (v3) circle (1.3pt) node[left] (0up) {};
			\fill[color=gray!60!blue] (v4) circle (1.3pt) node[left] (0up) {};
			\fill[color=gray!60!blue] (v1) circle (0.9pt) node[above] (0up) {{\scriptsize$a$}};
			\fill[color=gray!60!blue] (v2) circle (0.9pt) node[above] (0up) {{\scriptsize$b$}};
			\fill[color=gray!60!blue] (v3) circle (0.9pt) node[left] (0up) {{\scriptsize$c$}};
			\fill[color=gray!60!blue] (v4) circle (0.9pt) node[below] (0up) {{\scriptsize$d$}};
			\end{tikzpicture}}%%popende
		%%%%%%%%%%%%%%%%%%%%%% 
		\ee 
		\item 
		In dimension $n=3$, there are 2-3 and 1-4 Pachner moves, corresponding to partitions of the five faces of $\partial\Delta^4$: 
		\be 
		\label{eq:3dPachner}
		\!\!\!
		%%%%%%%%%%%%%%%%%%%%%% 
		\tikzzbox{\begin{tikzpicture}[thick,scale=2.321,color=gray!60!blue, baseline=-0.3cm, >=stealth, 
			style={x={(-0.6cm,-0.4cm)},y={(1cm,-0.2cm)},z={(0cm,0.9cm)}}]
			\coordinate (v1) at (1,0,0);
			\coordinate (v2) at (1,1,0);
			\coordinate (v3) at (0,0,0);
			\coordinate (v4) at (0.25,0.1,0.75);
			\coordinate (v0) at (0.25,0.1,-0.75);
			\fill [blue!20,opacity=0.545] (v1) -- (v2) -- (v3);
			\fill [blue!20,opacity=0.545] (v4) -- (v2) -- (v3);
			\fill [blue!20,opacity=0.545] (v1) -- (v4) -- (v3);
			\fill[color=gray!60!blue] (v3) circle (0.9pt) node[right] (0up) {{\scriptsize$c$}};
			\fill[color=gray!60!blue] (v1) circle (0.9pt) node[below] (0up) {{\scriptsize$a$}};
			\fill[color=gray!60!blue] (v2) circle (0.9pt) node[below] (0up) {{\scriptsize$b$}};
			\fill[color=gray!60!blue] (v4) circle (0.9pt) node[right] (0up) {{\scriptsize$d$}};
			\fill[color=gray!60!blue] (v0) circle (0.9pt) node[left] (0up) {{\scriptsize$e$}};
			\fill[color=gray!60!blue] (v3) circle (0.9pt) node[left] (0up) {};
			\fill [blue!20,opacity=0.545] (v0) -- (v1) -- (v3);
			\fill [blue!20,opacity=0.545] (v0) -- (v2) -- (v3);
			\draw[color=gray!60!blue, very thick] (v1) -- (v3);
			\draw[color=gray!60!blue, very thick] (v2) -- (v3);
			\draw[color=gray!60!blue, very thick] (v3) -- (v4);
			%
			% lower tetrahedron: 
			%
			\draw[color=gray!60!blue, very thick] (v0) -- (v3);
			\fill [blue!20,opacity=0.545] (v0) -- (v1) -- (v2);
			\draw[color=gray!60!blue, very thick] (v0) -- (v1);
			\draw[color=gray!60!blue, very thick] (v0) -- (v2);
			\fill [blue!20,opacity=0.545] (v1) -- (v2) -- (v4);
			\draw[color=gray!60!blue, very thick] (v2) -- (v4);
			\draw[color=gray!60!blue, very thick] (v1) -- (v4);
			\draw[color=gray!60!blue, very thick] (v1) -- (v2);
			\fill[color=gray!60!blue] (v0) circle (0.9pt) node[below] (0up) {};
			\fill[color=gray!60!blue] (v1) circle (0.9pt) node[below] (0up) {};
			\fill[color=gray!60!blue] (v2) circle (0.9pt) node[below] (0up) {};
			\fill[color=gray!60!blue] (v4) circle (0.9pt) node[above] (0up) {};
			\end{tikzpicture}}%%popende
		%%%%%%%%%%%%%%%%%%%%%% 
		%
		\stackrel{\text{2-3}}{\longleftrightarrow}
		%
		%%%%%%%%%%%%%%%%%%%%%% 
		\tikzzbox{\begin{tikzpicture}[thick,scale=2.321,color=gray!60!blue, baseline=-0.3cm, >=stealth, 
			style={x={(-0.6cm,-0.4cm)},y={(1cm,-0.2cm)},z={(0cm,0.9cm)}}]
			\coordinate (v1) at (1,0,0);
			\coordinate (v2) at (1,1,0);
			\coordinate (v3) at (0,0,0);
			\coordinate (v4) at (0.25,0.1,0.75);
			\coordinate (v0) at (0.25,0.1,-0.75);
			\fill [blue!20,opacity=0.545] (v0) -- (v1) -- (v4);
			\fill [blue!20,opacity=0.545] (v0) -- (v2) -- (v4);
			\fill [blue!20,opacity=0.545] (v0) -- (v3) -- (v4);
			\fill[color=gray!60!blue] (v3) circle (0.9pt) node[right] (0up) {{\scriptsize$c$}};
			\fill[color=gray!60!blue] (v1) circle (0.9pt) node[below] (0up) {{\scriptsize$a$}};
			\fill[color=gray!60!blue] (v2) circle (0.9pt) node[below] (0up) {{\scriptsize$b$}};
			\fill[color=gray!60!blue] (v4) circle (0.9pt) node[right] (0up) {{\scriptsize$d$}};
			\fill[color=gray!60!blue] (v0) circle (0.9pt) node[left] (0up) {{\scriptsize$e$}};
			\fill [blue!20,opacity=0.545] (v1) -- (v2) -- (v3);
			\fill [blue!20,opacity=0.545] (v4) -- (v2) -- (v3);
			\fill [blue!20,opacity=0.545] (v1) -- (v4) -- (v3);
			\fill[color=gray!60!blue] (v3) circle (0.9pt) node[left] (0up) {};
			\fill [blue!20,opacity=0.545] (v0) -- (v1) -- (v3);
			\fill [blue!20,opacity=0.545] (v0) -- (v2) -- (v3);
			\draw[color=gray!60!blue, very thick] (v1) -- (v3);
			\draw[color=gray!60!blue, very thick] (v2) -- (v3);
			\draw[color=gray!60!blue, very thick] (v3) -- (v4);
			\draw[color=gray!60!blue, very thick] (v0) -- (v3);
			\draw[color=gray!60!blue, very thick] (v0) -- (v4);
			\fill [blue!20,opacity=0.545] (v0) -- (v1) -- (v2);
			\draw[color=gray!60!blue, very thick] (v0) -- (v1);
			\draw[color=gray!60!blue, very thick] (v0) -- (v2);
			\fill [blue!20,opacity=0.545] (v1) -- (v2) -- (v4);
			\draw[color=gray!60!blue, very thick] (v2) -- (v4);
			\draw[color=gray!60!blue, very thick] (v1) -- (v4);
			\draw[color=gray!60!blue, very thick] (v1) -- (v2);
			\fill[color=gray!60!blue] (v0) circle (0.9pt) node[below] (0up) {};
			\fill[color=gray!60!blue] (v1) circle (0.9pt) node[below] (0up) {};
			\fill[color=gray!60!blue] (v2) circle (0.9pt) node[below] (0up) {};
			\fill[color=gray!60!blue] (v4) circle (0.9pt) node[above] (0up) {};
			\end{tikzpicture}}%%popende
		%%%%%%%%%%%%%%%%%%%%%% 
		\!\! , \quad 
		%%%%%%%%%%%%%%%%%%%%%% 
		\tikzzbox{\begin{tikzpicture}[thick,scale=2.321,color=gray!60!blue, baseline=-0.3cm, >=stealth, 
			style={x={(-0.6cm,-0.4cm)},y={(1cm,-0.2cm)},z={(0cm,0.9cm)}}]
			\coordinate (v1) at (1,0,0);
			\coordinate (v2) at (1,1,0);
			\coordinate (v3) at (0,0,0);
			\coordinate (v4) at (0.25,0.1,0.75);
			%\coordinate (v0) at (0.25,0.1,-0.75);
			%
			\fill [blue!20,opacity=0.545] (v1) -- (v2) -- (v3);
			\fill [blue!20,opacity=0.545] (v4) -- (v2) -- (v3);
			\fill [blue!20,opacity=0.545] (v1) -- (v4) -- (v3);
			\fill[color=gray!60!blue] (v3) circle (0.9pt) node[right] (0up) {{\scriptsize$c$}};
			\fill[color=gray!60!blue] (v1) circle (0.9pt) node[below] (0up) {{\scriptsize$a$}};
			\fill[color=gray!60!blue] (v2) circle (0.9pt) node[below] (0up) {{\scriptsize$b$}};
			\fill[color=gray!60!blue] (v4) circle (0.9pt) node[right] (0up) {{\scriptsize$d$}};
			\fill[color=gray!60!blue] (v3) circle (0.9pt) node[left] (0up) {};
			\draw[color=gray!60!blue, very thick] (v1) -- (v3);
			\draw[color=gray!60!blue, very thick] (v2) -- (v3);
			\draw[color=gray!60!blue, very thick] (v3) -- (v4);
			\fill [blue!20,opacity=0.545] (v1) -- (v2) -- (v4);
			\draw[color=gray!60!blue, very thick] (v2) -- (v4);
			\draw[color=gray!60!blue, very thick] (v1) -- (v4);
			\draw[color=gray!60!blue, very thick] (v1) -- (v2);
			\fill[color=gray!60!blue] (v1) circle (0.9pt) node[below] (0up) {};
			\fill[color=gray!60!blue] (v2) circle (0.9pt) node[below] (0up) {};
			\fill[color=gray!60!blue] (v4) circle (0.9pt) node[above] (0up) {};
			\end{tikzpicture}}%%popende
		%%%%%%%%%%%%%%%%%%%%%% 
		%
		\stackrel{\text{1-4}}{\longleftrightarrow}
		%
		%%%%%%%%%%%%%%%%%%%%%% 
		\tikzzbox{\begin{tikzpicture}[thick,scale=2.321,color=gray!60!blue, baseline=-0.3cm, >=stealth, 
			style={x={(-0.6cm,-0.4cm)},y={(1cm,-0.2cm)},z={(0cm,0.9cm)}}]
			\coordinate (v1) at (1,0,0);
			\coordinate (v2) at (1,1,0);
			\coordinate (v3) at (0,0,0);
			\coordinate (v4) at (0.25,0.1,0.75);
			%\coordinate (v5) at (0.25,0.1,0.375);
			\coordinate (v5) at (0.7,0.3,0.175);
			\fill [blue!20,opacity=0.545] (v1) -- (v2) -- (v3);
			\fill [blue!20,opacity=0.545] (v4) -- (v2) -- (v3);
			\fill [blue!20,opacity=0.545] (v1) -- (v4) -- (v3);
			\fill[color=gray!60!blue] (v3) circle (0.9pt) node[left] (0up) {};
			\fill[color=gray!60!blue] (v3) circle (0.9pt) node[right] (0up) {{\scriptsize$c$}};
			\fill[color=gray!60!blue] (v1) circle (0.9pt) node[below] (0up) {{\scriptsize$a$}};
			\fill[color=gray!60!blue] (v2) circle (0.9pt) node[below] (0up) {{\scriptsize$b$}};
			\fill[color=gray!60!blue] (v4) circle (0.9pt) node[right] (0up) {{\scriptsize$d$}};
			\fill[color=gray!60!blue] (v5) circle (0.9pt) node[below] (0up) {{\scriptsize$e$}};
			\draw[color=gray!60!blue, very thick] (v1) -- (v5);
			\draw[color=gray!60!blue, very thick] (v2) -- (v5);
			\draw[color=gray!60!blue, very thick] (v3) -- (v5);
			\draw[color=gray!60!blue, very thick] (v4) -- (v5);
			\fill[color=gray!60!blue] (v5) circle (0.9pt) node[below] (0up) {};
			\draw[color=gray!60!blue, very thick] (v1) -- (v3);
			\draw[color=gray!60!blue, very thick] (v2) -- (v3);
			\draw[color=gray!60!blue, very thick] (v3) -- (v4);
			\fill [blue!20,opacity=0.545] (v1) -- (v2) -- (v4);
			\draw[color=gray!60!blue, very thick] (v2) -- (v4);
			\draw[color=gray!60!blue, very thick] (v1) -- (v4);
			\draw[color=gray!60!blue, very thick] (v1) -- (v2);
			\fill[color=gray!60!blue] (v1) circle (0.9pt) node[below] (0up) {};
			\fill[color=gray!60!blue] (v2) circle (0.9pt) node[below] (0up) {};
			\fill[color=gray!60!blue] (v4) circle (0.9pt) node[above] (0up) {};
			\end{tikzpicture}}%%popende
		%%%%%%%%%%%%%%%%%%%%%% 
		\ee 
	\end{enumerate}
	\label{exa:PachnerMoves}
\end{example}

\medskip 

We can now explain what we mean by ``nice stratification'' in step~\ref{item:Orb1} of the orbifold construction. 
Let~$M$ be an oriented $n$-bordism with an oriented triangulation~$t$. 
By definition, the $j$-strata of the \textsl{Poincar\'{e} dual stratification}~$M^t$ are transversal to the $(n-j)$-simplices of~$t$, and oriented such that for all strata~$\sigma$ in~$M^t$ the orientation of~$\sigma$ together with that of its dual simplex (in that order) gives the orientation of~$M$. 
Moreover, if a stratum~$\sigma$ intersects $\partial M$, it does so transversally. %, or $\sigma\subset\partial M$. 
Hence if~$M$ is $|\Delta_+^1|$, $|\Delta_+^2|$, or $|\Delta_+^3|$, then~$M^t$ is given by 
\be 
%%%%%%%%%%%%%%%%%%%%%% 
\tikzzbox{
}%%popende
%%%%%%%%%%%%%%%%%%%%%% 
\, , 
\ee 
respectively. 
Another example (suppressing orientations to avoid clutter) is:
\be 
(M,t) = 
\tikzzbox{%
	%%%%%%%%%%%%%%%%%%%%%% 
	%
	%%%%%
}
\ee 

\begin{example}
	The Poincar\'{e} duals of $n$-dimensional Pachner moves relate different stratifications of the $n$-ball with identical boundary stratification: 
	\begin{enumerate}[leftmargin=*, label={(\arabic*)}]
		\item 
		The duals of the three 1-dimensional Pachner moves for $a<c$ in~\eqref{eq:1dPachner} are
		\be 
		\!\!\!
		\label{eq:1dDualPacher}
		%%%%%%%%%%%%%%%%%%%%%% 
		\tikzzbox{%
}%%popende
		%%%%%%%%%%%%%%%%%%%%%% 
		\, ,  
		\ee 
		corresponding to $a<b<c$, $b<a<c$, $a<c<b$ in~\eqref{eq:1dPachner}, respectively. 
		\item 
		The duals of the 2-dimensional Pachner moves for $a<b<c<d$ in~\eqref{eq:2dPachner} are 
		\be 
		%%%%%%%%%%%%%%%%%%%%%% 
		\tikzzbox{%
}%%popende
		%%%%%%%%%%%%%%%%%%%%%% 
		\ee 
		and there are various other oriented Pachner moves for the other total orders of the vertices in~\eqref{eq:2dPachner}. 
		\item 
		The dual of one the oriented 2-3 Pachner moves in~\eqref{eq:3dPachner} is
		\be 
		\label{eq:3dDualPachner}
		%%%%%%%%%%%%%%%%%%%%%% 
		\tikzzbox{%
}%%popende
		%%%%%%%%%%%%%%%%%%%%%% 
		\ee  
		and analogously for the other dual 2-3 and 1-4 moves, which are harder to picture. 
	\end{enumerate}
\end{example}

We are now ready to give the first definition of ``orbifold datum'', namely as a collection of defects that encode invariance under choice of triangulation (see \cite[Def.\,3.5]{CRS1} for details): 

\begin{definition}
	An \textsl{orbifold datum~$\A$ for a defect TQFT} $\zz \colon \Borddef \lra \mathcal C$ consists of elements $\A_j \in D_j$ for all $j\in\{1,\dots,n\}$ and $\A_0^+, \A_0^- \in D_0$ subject to: 
	\begin{itemize}[leftmargin=3mm, itemsep=3pt]
		\item[] 
		\textsl{Compatibility}: $\A_j$ can label $j$-strata in the dual of an (oriented) triangulation, such that all adjacent $k$-strata can be labelled by~$\A_k$  for $j\in\{1,\dots,n\}$, 
		and analogously for $\A_0^\pm$ labelling the 0-strata dual to the two oriented $n$-simplices. 
		\item[]
		\textsl{Invariance}: Let~$B$ and~$B'$ be two stratified $n$-balls of a dual Pachner move. 
		Viewing them as bordisms with domain~$\varnothing$, and labelling their strata with~$\A$ gives two morphisms $B_\A, B'_\A$ in $\Borddef$, and we demand 
		\be 
		\label{eq:Invariance}
		\zz(B_\A) = \zz(B'_\A) \, . 
		\ee 
	\end{itemize} 
\end{definition} 

\begin{remark}
	\label{rem:E1algebras1}
	Since there are only finitely many Pachner moves, there are only finitely many invariance conditions on orbifold data. 
	Moreover, the compatibility condition implies that $\A_{n-1}$-labelled $(n-1)$-strata separate two $\A_n$-labelled $n$-strata (because the dual simplex~$\Delta^1$ has two faces), while $\A_{n-2}$-labelled $(n-2)$-strata have three adjacent $\A_{n-1}$-strata (because~$\Delta^2$ has three faces). 
	More generally, $\A_j$-labelled $j$-strata have precisely $n-j+1$ adjacent $(j+1)$-strata, because $\Delta^{n-j}$ has $n-j+1$ faces. 
\end{remark}

\begin{example} 
	\begin{enumerate}[leftmargin=*, label={(\arabic*)}]
		\item 
		\label{item:OrbData1}
		For an orbifold datum $\A = (\A_1,\A_0^+,\A_0^-)$ for a 1-dimensional TQFT~$\zz$, the invariance condition from the first move in~\eqref{eq:1dDualPacher} reads 
		\be 
		\label{eq:1dInv}
		\zz\Big( 
		%%%%%%%%%%%%%%%%%%%%%% 
		\tikzzbox{\begin{tikzpicture}[thick,scale=1.2,color=gray!60!blue, baseline=-0.13cm, >=stealth]
			\coordinate (d1) at (0.5,0);
			\coordinate (d2) at (1,0);
			\draw[red!50, opacity=0.6, very thick] (0,0) -- (1.5,0);
			\fill[color=green!60!black] (d1) circle (1.6pt) node[above] (0up) {\tiny{$+$}};
			\fill[color=green!60!black] (d2) circle (1.6pt) node[above] (0up) {\tiny{$+$}};
			\fill[color=green!60!black] (d1) circle (1.6pt) node[below] (0up) {\tiny{$\A_0^+$}};
			\fill[color=green!60!black] (d2) circle (1.6pt) node[below] (0up) {\tiny{$\A_0^+$}};
			\end{tikzpicture}}%%popende
		%%%%%%%%%%%%%%%%%%%%%% 
		\Big) 
		=
		\zz\Big( 
		%%%%%%%%%%%%%%%%%%%%%% 
		\tikzzbox{\begin{tikzpicture}[thick,scale=1.2,color=gray!60!blue, baseline=-0.13cm, >=stealth]
			\coordinate (d1) at (0.5,0);
			\draw[red!50, opacity=0.6, very thick] (0,0) -- (1,0);
			\fill[color=green!60!black] (d1) circle (1.6pt) node[above] (0up) {\tiny{$+$}};
			\fill[color=green!60!black] (d1) circle (1.6pt) node[below] (0up) {\tiny{$\A_0^+$}};
			\end{tikzpicture}}%%popende
		%%%%%%%%%%%%%%%%%%%%%%
		\Big) 
		,
		\ee 
		where we suppress the labels~$\A_1$ for 1-strata. 
		Hence 
		\be 
		\zz(\A_0^+) := \zz(
		%%%%%%%%%%%%%%%%%%%%%% 
		\tikzzbox{\begin{tikzpicture}[thick,scale=1.2,color=gray!60!blue, baseline=-0.13cm, >=stealth]
			\coordinate (d1) at (0.25,0);
			\draw[red!50, opacity=0.6, very thick] (0,0) -- (0.5,0);
			\fill[color=green!60!black] (d1) circle (1.6pt) node[above] (0up) {\tiny{$+$}};
			\fill[color=green!60!black] (d1) circle (1.6pt) node[below] (0up) {\tiny{$\A_0^+$}};
			\end{tikzpicture}}%%popende
		%%%%%%%%%%%%%%%%%%%%%%
		) 
		\colon \zz(\A_1) \lra \zz(\A_1) := \zz 
		(
		%%%%%%%%%%%%%%%%%%%%%% 
		\tikzzbox{\begin{tikzpicture}[thick,scale=1.2,color=gray!60!blue, baseline=-0.13cm, >=stealth]
			\coordinate (d1) at (0.25,0);
			\fill[color=red!50, very thick] (d1) circle (1.6pt) node[above] (0up) {\tiny{$+$}};
			\fill[color=red!50, very thick] (d1) circle (1.6pt) node[below] (0up) {\tiny{$\A_1$}};
			\end{tikzpicture}}%%popende
		%%%%%%%%%%%%%%%%%%%%%%
		)
		\in \mathcal C
		\ee 
		is an \textsl{idempotent} in the codomain~$\mathcal C$ of~$\zz$. 
		The other conditions from Example~\ref{exa:PachnerMoves}\ref{item:1dPachnerMove} imply that $\zz(\A_0^-) = \zz(\A_0^+)$. 
		\item 
		\label{item:OrbData2}
		An orbifold datum $\A = (\A_2,\A_1,\A_0^+,\A_0^-)$ for a 2-dimensional TQFT~$\zz$ satisfies the invariance conditions 
		\be 
		\label{eq:2dInv}
		\!\!\!
		\zz\Bigg(
		%%%%%%%%%%%%%%%%%%%%%% 
		\tikzzbox{\begin{tikzpicture}[thick,scale=1,color=gray!60!blue, baseline=0.38cm, >=stealth]
			\coordinate (a1) at (0.5, 0.75);
			\coordinate (a2) at (1, 0.25);
			\coordinate (e1) at (-0.25,0.5);
			\coordinate (e2) at (0.5,1.25);
			\coordinate (e3) at (1,-0.25);
			\coordinate (e4) at (1.75,0.5);
			\fill [orange!12] (-0.25,-0.25) -- (1.75,-0.25) -- (1.75, 1.25) -- (-0.25,1.25);
			\draw[string=green!60!black, very thick] (a2) -- (a1);
			\draw[string=green!60!black=green!60!black, very thick] (e1) -- (a1);
			\draw[string=green!60!black, very thick] (a1) -- (e2);
			\draw[string=green!60!black, very thick] (e3) -- (a2);
			\draw[string=green!60!black, very thick] (e4) -- (a2);
			\fill[color=green!60!black] (a1) circle (2pt) node[below] (0up) {};
			\fill[color=green!60!black] (a2) circle (2pt) node[right] (0up) {};
						\fill[color=green!60!black] ($(a1)+(-0.1,0)$) circle (0pt) node[below] (0up) {{\tiny$\A_0^+$}};
						\fill[color=green!60!black] ($(a2)+(0,-0.1)$) circle (0pt) node[right] (0up) {{\tiny$\A_0^+$}};
			\end{tikzpicture}}%%popende
		%%%%%%%%%%%%%%%%%%%%%% 
		\Bigg) 
		= 
		\zz\Bigg(
		%%%%%%%%%%%%%%%%%%%%%% 
		\tikzzbox{\begin{tikzpicture}[thick,scale=1,color=gray!60!blue, baseline=0.38cm, >=stealth]
			\coordinate (a1) at (0.5, 0.25);
			\coordinate (a2) at (1, 0.75);
			\coordinate (e1) at (-0.25,0.5);
			\coordinate (e2) at (0.5,-0.25);
			\coordinate (e3) at (1,1.25);
			\coordinate (e4) at (1.75,0.5);

			\fill [orange!12] (-0.25,-0.25) -- (1.75,-0.25) -- (1.75, 1.25) -- (-0.25,1.25);
			\draw[string=green!60!black, very thick] (a1) -- (a2);
			\draw[string=green!60!black, very thick] (e1) -- (a1);
			\draw[string=green!60!black, very thick] (e2) -- (a1);
			\draw[string=green!60!black, very thick] (a2) -- (e3);
			\draw[string=green!60!black, very thick] (e4) -- (a2);
			\fill[color=green!60!black] (a1) circle (2pt) node[left] (0up) {};
			\fill[color=green!60!black] (a2) circle (2pt) node[left] (0up) {};
			\fill[color=green!60!black] ($(a1)+(0.04,-0.12)$) circle (0pt) node[left] (0up) {{\tiny$\mathcal A_0^+$}};
			\fill[color=green!60!black] ($(a2)+(0,0.05)$) circle (0pt) node[below] (0up) {{\tiny$\mathcal A_0^+$}};
			\end{tikzpicture}}%%popende
		%%%%%%%%%%%%%%%%%%%%%% 
		\Bigg) 
		, \quad 
		\zz\Bigg(
		%%%%%%%%%%%%%%%%%%%%%% 
		\tikzzbox{\begin{tikzpicture}[thick,scale=1,color=gray!60!blue, baseline=0.38cm, >=stealth]
			\coordinate (a1) at (0.7, 0.5);
			\coordinate (e1) at (0.7,-0.25);
			\coordinate (e2) at (-0.05,1.25);
			\coordinate (e3) at (1.45,1.25);
			\fill [orange!12] (-0.25,-0.25) -- (1.65,-0.25) -- (1.65, 1.25) -- (-0.25,1.25);
			\draw[string=green!60!black, very thick] (e1) -- (a1);
			\draw[string=green!60!black, very thick] (a1) -- (e2);
			\draw[string=green!60!black, very thick] (e3) -- (a1);
			\fill[color=green!60!black] (a1) circle (2pt) node[left] (0up) {};
						\fill[color=green!60!black] ($(a1)+(0,0)$) circle (0pt) node[right] (0up) {{\tiny$\A_0^+$}};
			\end{tikzpicture}}%%popende
		%%%%%%%%%%%%%%%%%%%%%% 
		\Bigg) 
		= 
		\zz\Bigg(
		%%%%%%%%%%%%%%%%%%%%%% 
		\tikzzbox{\begin{tikzpicture}[thick,scale=1,color=gray!60!blue, baseline=0.38cm, >=stealth]
			\coordinate (a1) at (0.7, 0.3);
			\coordinate (a2) at (0.5, 0.7);
			\coordinate (a3) at (0.9, 0.7 );
			\coordinate (e1) at (0.7,-0.25);
			\coordinate (e2) at (-0.05,1.25);
			\coordinate (e3) at (1.45,1.25);
			\fill [orange!12] (-0.25,-0.25) -- (1.65,-0.25) -- (1.65, 1.25) -- (-0.25,1.25);
			\draw[string=green!60!black, very thick] (a1) -- (a2);
			\draw[string=green!60!black, very thick] (a2) -- (a3);
			\draw[string=green!60!black, very thick] (a3) -- (a1);
			\draw[string=green!60!black, very thick] (e1) -- (a1);
			\draw[string=green!60!black, very thick] (a2) -- (e2);
			\draw[string=green!60!black, very thick] (e3) -- (a3);
			\fill[color=green!60!black] (a1) circle (2pt) node[left] (0up) {};
			\fill[color=green!60!black] (a2) circle (2pt) node[left] (0up) {};
			\fill[color=green!60!black] (a3) circle (2pt) node[left] (0up) {};
						\fill[color=green!60!black] ($(a1)+(0,0)$) circle (0pt) node[right] (0up) {{\tiny$\A_0^+$}};
						\fill[color=green!60!black] ($(a2)+(0,0)$) circle (0pt) node[left] (0up) {{\tiny$\A_0^-$}};
						\fill[color=green!60!black] ($(a3)+(0,0)$) circle (0pt) node[right] (0up) {{\tiny$\A_0^+$}};
			\end{tikzpicture}}%%popende
		%%%%%%%%%%%%%%%%%%%%%% 
		\Bigg) 
		\ee 
		(where we suppress the labels $\A_1,\A_2$ for all 1- and 2-strata) as well as those coming from all the other total orders on vertices in~\eqref{eq:2dPachner}. 
		It follows from the first identity in~\eqref{eq:2dInv} that~$\zz(\A_0^+)$ is an associative multiplication on~$\A_1$ viewed as a 1-morphism $\A_1\colon \A_2\lra\A_2$ in~$\mathcal D_\zz$ (recall the discussion around Theorem~\ref{thm:DKR}). 
		Here we identified~$\A_0^+$ with the defect bordism 
		\be 
		%%%%%%%%%%%%%%%%%%%%%%
		\begin{tikzpicture}[very thick, scale=0.45,color=green!50!black, baseline, >=stealth]
		\fill [color=orange!12, draw=white]  (0,0) circle (2);
		\draw[very thick, color=red!80!black] (0,0) circle (2);
		\fill (-45:2) circle (4.0pt) node[left] {};
		\fill (225:2) circle (4.0pt) node[left] {};
		\fill (90:2) circle (4.0pt) node[left] {};
		\draw[
		color=green!50!black, 
		>=stealth, 
		postaction={decorate}, decoration={markings,mark=at position .5 with {\fill circle (2pt);}}, 
		decoration={markings,mark=at position .21 with {\arrow[draw]{>}}},
		decoration={markings,mark=at position .85 with {\arrow[draw]{<}}}
		] 
		(-45:2) .. controls +(0,2) and +(0,2) .. (225:2);
		\draw[
		color=green!50!black, 
		>=stealth, 
		postaction={decorate}, 
		decoration={markings,mark=at position .7 with {\arrow[draw]{>}}}
		] 
		(0,0) -- (90:2); 
		\draw[line width=1] (0,-0.45) node[line width=0pt] (sch) {{\scriptsize $\mathcal A_0^+$}};
		\end{tikzpicture}
		%%%%%%%%%%%%%%%%%%%%%%
		\colon 
		\varnothing
		\lra 
		%%%%%%%%%%%%%%%%%%%%%%
		\begin{tikzpicture}[very thick, scale=0.45,color=green!50!black, baseline, >=stealth]
		\draw[very thick, color=red!80!black] (0,0) circle (2);
		\fill (-45:2) circle (4.0pt) node[right] {{\tiny $(\mathcal A_1,-)$}};
		\fill (225:2) circle (4.0pt) node[left] {{\tiny $(\mathcal A_1,-)$}};
		\fill (90:2) circle (4.0pt) node[below] {{\tiny $(\mathcal A_1,+)$}};
		\end{tikzpicture}
		%%%%%%%%%%%%%%%%%%%%%%
		\, . 
		\ee 
		The other invariance conditions similarly impose that~$\zz(\A_0^-)$ is a coassociative comultiplication on~$\A_1$, and more generally all the invariance conditions~\eqref{eq:Invariance} for $n=2$ are equivalent to $(\A_1, \zz(\A_0^+), \zz(\A_0^-))$ being a \textsl{$\Delta$-separable symmetric Frobenius algebra} in~$\mathcal D_\zz$, see \cite[Prop.\,3.4]{cr1210.6363} for details. 
		\item 
		\label{item:OrbData3}
		An orbifold datum $\A = (\A_3,\A_2,\A_1,\A_0^+,\A_0^-)$ for a 3-dimensional TQFT~$\zz$ satisfies the invariance condition 
		\be 
		\label{eq:3dInv}
		\!\!\!
		\zz\left( 
		%%%%%%%%%%%%%%%%%%%%%% 
		\tikzzbox{\begin{tikzpicture}[thick,scale=1.77,color=blue!50!black, baseline=0.8cm, >=stealth,
			style={x={(-0.55cm,-0.4cm)},y={(1cm,-0.2cm)},z={(0cm,0.9cm)}}]
			%: where to put leftmost T-line: 
			\pgfmathsetmacro{\yy}{0.2}
			% bottom vertices: 
			\coordinate (P) at (0.5, \yy, 0);
			\coordinate (R) at (0.625, 0.5 + \yy/2, 0);
			\coordinate (L) at (0.5, 0, 0);
			\coordinate (R1) at (0.25, 1, 0);
			\coordinate (R2) at (0.5, 1, 0);
			\coordinate (R3) at (0.75, 1, 0);
			\coordinate (N1) at ($(R1) + (R2) - (R)$);
			\coordinate (N2) at ($(R) + 2*(R2) - 2*(R)$);
			\coordinate (N3) at ($(R3) + (R2) - (R)$);
			\coordinate (N4) at ($2*(R3) - (R)$);
			% top vertices: 
			\coordinate (Lo) at (0.5, 0, 2);
			\coordinate (Po) at (0.5, \yy, 2);
			\coordinate (O1) at ($(N1) + (0,0,2)$);
			\coordinate (O2) at ($(N2) + (0,0,2)$);
			\coordinate (O3) at ($(N3) + (0,0,2)$);
			\coordinate (O4) at ($(N4) + (0,0,2)$);
			\coordinate (A2) at ($(O2) - (R3) + (R)$);
			\coordinate (A3) at ($(O3) - 2*(R3) + 2*(R)$);
			\coordinate (A4) at ($(O4) - 3*(R3) + 3*(R)$);
			% middle vertices: 
			\coordinate (Pt) at (0.5, \yy, 1);
			\coordinate (Rt) at ($(A2) + (0,0,-1)$);
			\coordinate (Lt) at (0.5, 0, 1);
			\coordinate (M1) at ($(N1) + (0,0,1)$);
			\coordinate (M2) at ($(N2) + (0,0,1)$);
			\coordinate (M3) at ($(N3) + (0,0,1)$);
			\coordinate (M4) at ($(N4) + (0,0,1)$);
			%
			%alpha: 
			\coordinate (alphabottom) at (0.5, 0.5, 0.5);
			\coordinate (alphatop) at (0.5, 0.5, 1.5);
			%
			% A-planes:
			\fill [green!50,opacity=0.545] (P) -- (alphabottom) -- (R);
			\fill [green!50,opacity=0.545] (R) -- (alphabottom) -- ($(R3) + (0,0,1)$) -- (R3);
			\fill [green!50,opacity=0.545] (P) -- (alphabottom) -- (Rt) -- (A2) -- (O1) -- (N1);
			\fill [green!50,opacity=0.545] (alphabottom) -- (Rt) -- (A2) -- (A3) -- (alphatop) -- (Pt);
			\fill [green!50,opacity=0.545] (alphatop) -- (A4) -- (A3);
			\fill [green!50,opacity=0.545] (R) -- (alphabottom) -- (Rt) -- (A2) -- (O2) -- (N2);
			\draw[color=green!60!black, ultra thick, rounded corners=0.5mm, postaction={decorate}, decoration={markings,mark=at position .17 with {\arrow[draw=green!60!black]{>}}}] (alphabottom) -- (Rt) -- (A2);
			\fill [green!50,opacity=0.545] (R) -- (alphabottom) -- (Pt) -- (alphatop) -- ($(R3) + (0,0,1)$)-- (R3);
			\fill [green!50,opacity=0.545] (R3) -- ($(R3) + (0,0,1)$) -- (alphatop) -- (A3) -- (O3) -- (N3);
			\draw[string=green!60!black, ultra thick] (alphatop) -- (A3); 	%T-line
			\fill [green!50,opacity=0.545] (R3) -- ($(R3) + (0,0,1)$) -- (alphatop) -- (A4) -- (O4) -- (N4);
			\fill [green!50,opacity=0.545] (L) -- (P) -- (alphabottom) -- (Pt) -- (alphatop) -- (Po) -- (Lo);
			%
			%T-lines: 
			%\draw[string=green!60!black, ultra thick] (alphabottom) -- (Pt);
			\draw[string=green!60!black, ultra thick] (P) -- (alphabottom);
			\draw[string=green!60!black, ultra thick] (R) -- (alphabottom);
			\draw[string=green!60!black, ultra thick] (R3) -- ($(R3) + (0,0,1)$);
			\draw[string=green!60!black, ultra thick] ($(R3) + (0,0,1)$) -- (alphatop);
			\draw[string=green!60!black, ultra thick] (alphatop) -- (A4);
			\draw[color=green!60!black, ultra thick, rounded corners=0.5mm, postaction={decorate}, decoration={markings,mark=at position .33 with {\arrow[draw=green!60!black]{>}}}] (alphabottom) -- (Pt) -- (alphatop);
			\draw[color=green!60!black, ultra thick, rounded corners=0.5mm] (R3) --  ($(R3) + (0,0,1)$) -- (alphatop) ;
			%
			%: labels: 
			\fill[color=green!60!black] (alphabottom) circle (1.2pt) node[left] (0up) { {\scriptsize$\A_0^+$} };
			\fill[color=green!60!black] (alphatop) circle (1.2pt) node[left] (0up) { {\scriptsize$\A_0^+$} };
			%
			% black auxiliary boundaries: 
			\draw [black,opacity=0.4141, densely dotted, semithick] (Lt) -- (Pt) -- (0.75, 1, 1) -- (M4);
			\draw [black,opacity=0.4141, densely dotted, semithick] (0.75, 1, 1) -- (M3);
			\draw [black,opacity=0.4141, densely dotted, semithick] (Pt) -- (Rt) -- (M2);
			\draw [black,opacity=0.4141, densely dotted, semithick] (Rt) -- (M1);
			%
			% black boundaries\draw 
			\draw [black!80!white,opacity=1, thin] (R) -- (P);
			\draw [black,opacity=1, very thin] (A2) -- (A3);
			\draw [black,opacity=1, very thin] (A4) -- (A3);
			\draw [black,opacity=1, very thin] (A2) -- (O2) -- (N2) -- (R);
			\draw [black,opacity=1, very thin] (A3) -- (O3) -- (N3) -- (R3);
			\draw [black,opacity=1, very thin] (A4) -- (O4) -- (N4) -- (R3);
			\draw [black,opacity=1, very thin] (R3) -- (R);
			\draw [black,opacity=1, very thin] (L) -- (P);
			\draw [black,opacity=1, very thin] (Po) -- (Lo) -- (L);
			\draw [black,opacity=1, very thin] (R3) -- (R);
			\draw [black,opacity=1, very thin] (A2) -- (O1) -- (N1) -- (P);
			\end{tikzpicture}}%%popende
		%%%%%%%%%%%%%%%%%%%%%% 
		\right)
		=
		\zz\left( 
		%%%%%%%%%%%%%%%%%%%%%% 
		\tikzzbox{\begin{tikzpicture}[thick,scale=1.77,color=green!60!black, baseline=1.7cm, >=stealth, 
			style={x={(-0.6cm,-0.4cm)},y={(1cm,-0.2cm)},z={(0cm,0.9cm)}}]
			%: where to put leftmost T-line: 
			\pgfmathsetmacro{\yy}{0.2}
			% bottom vertices: 
			\coordinate (P) at (0.5, \yy, 0);
			\coordinate (R) at (0.625, 0.5 + \yy/2, 0);
			\coordinate (L) at (0.5, 0, 0);
			\coordinate (R1) at (0.25, 1, 0);
			\coordinate (R2) at (0.5, 1, 0);
			\coordinate (R3) at (0.75, 1, 0);
			\coordinate (N1) at ($(R1) + (R2) - (R)$);
			\coordinate (N2) at ($(R) + 2*(R2) - 2*(R)$);
			\coordinate (N3) at ($(R3) + (R2) - (R)$);
			\coordinate (N4) at ($2*(R3) - (R)$);
			% 1-st floor vertices: 
			\coordinate (1P) at ($(P) + (0,0,1)$);
			\coordinate (1L) at (0.5, 0, 1);
			\coordinate (1N1) at ($(N1) + (0,0,1)$);
			\coordinate (1N2) at ($(N2) + (0,0,1)$);
			\coordinate (1N3) at ($(N3) + (0,0,1)$);
			\coordinate (1N4) at ($(N4) + (0,0,1)$);
			\coordinate (1R) at ($(R) + (0,0,1)$);
			\coordinate (1RR) at ($(R3) + (-0.25,0,1)$);
			% 2-nd floor vertices: 
			\coordinate (2P) at ($(P) + (0,0,2)$);
			\coordinate (2L) at (0.5, 0, 2);
			\coordinate (2N1) at ($(N1) + (0,0,2)$);
			\coordinate (2N2) at ($(N2) + (0,0,2)$);
			\coordinate (2N3) at ($(N3) + (0,0,2)$);
			\coordinate (2N4) at ($(N4) + (0,0,2)$);
			\coordinate (2RR) at ($(1RR) + (0,0,1)$);
			\coordinate (2R) at ($(2N3) - 2*(2N3) + 2*(2RR)$);
			% 3-rd floor vertices: 
			\coordinate (3P) at ($(P) + (0,0,3)$);
			\coordinate (3L) at (0.5, 0, 3);
			\coordinate (3N1) at ($(N1) + (0,0,3)$);
			\coordinate (3N2) at ($(N2) + (0,0,3)$);
			\coordinate (3N3) at ($(N3) + (0,0,3)$);
			\coordinate (3N4) at ($(N4) + (0,0,3)$);
			\coordinate (3RR) at ($(2RR) + (-0.25,0,1)$);
			\coordinate (3R) at ($(2R) + (0,0,1)$);
			%
			%alpha: 
			\coordinate (alpha1) at (0.5, 0.75, 0.5);
			\coordinate (alpha2) at (0.5, 0.5, 1.5);
			\coordinate (alpha3) at (0.25, 0.75, 2.5);
			%
			%psi
			\coordinate (psi) at (0.5, 0.75, 1.5);
			%
			% A-planes:
			\fill [green!50,opacity=0.545] (P) -- (1P) -- (alpha2) -- (2P) -- (3P) -- (3L) -- (L);
			\fill [green!50,opacity=0.545] (P) -- (1P) -- (alpha2) -- (2R) -- (alpha3) -- (3RR) -- (3N1) -- (N1);
			\fill [green!50,opacity=0.545] (alpha3) -- (3RR) -- (3R);
			\fill [green!50,opacity=0.545] (R) -- (alpha1) -- (1RR) -- (2RR) -- (alpha3) -- (3RR) -- (3N2) -- (N2);
			\draw[string=green!60!black, ultra thick] (alpha3) -- (3RR); 	%T-line
			\fill [green!50,opacity=0.545] (R3) -- (alpha1) -- (1RR) -- (2RR) -- (alpha3) -- (3R) -- (3N3) -- (N3);
			\fill [green!50,opacity=0.545] (R3) -- (alpha1) -- (R);
			\fill [green!50,opacity=0.545] (alpha1) -- (1R) -- (alpha2) -- (2R) -- (alpha3) -- (2RR) -- (1RR);
			\fill[color=green!60!black] (alpha3) circle (1.2pt) node[left] (0up) {{\tiny$\A_0^+$}}; 	%alpha3
			\draw[color=green!60!black, ultra thick, rounded corners=0.5mm, postaction={decorate}, decoration={markings,mark=at position .34 with {\arrow[draw=green!60!black]{>}}}] (R3) -- (alpha1) -- (1RR) -- (2RR) -- (alpha3);
			\fill [green!50,opacity=0.545] (alpha2) -- (2P) -- (3P) -- (3R) -- (alpha3) -- (2R);
			\draw[string=green!60!black, ultra thick] (alpha3) -- (3R); 		%T-line
			\draw[color=green!60!black, ultra thick, rounded corners=0.5mm, postaction={decorate}, decoration={markings,mark=at position .3 with {\arrow[draw=green!60!black]{>}}}] (alpha2) -- (2R) -- (alpha3);
			\fill [green!50,opacity=0.545] (R3) -- (alpha1) -- (1R) -- (alpha2) -- (2P) -- (3P) -- (3N4) -- (N4);
			\fill [green!50,opacity=0.545] (P) -- (1P) -- (alpha2) -- (1R) -- (alpha1) -- (R);
			\draw[color=green!60!black, ultra thick, rounded corners=0.5mm, postaction={decorate}, decoration={markings,mark=at position .3 with {\arrow[draw=green!60!black]{>}}}] (P) -- (1P) -- (alpha2);
			\draw[color=green!60!black, ultra thick, rounded corners=0.5mm, postaction={decorate}, decoration={markings,mark=at position .3 with {\arrow[draw=green!60!black]{>}}}] (alpha1) -- (1R) -- (alpha2);
			\draw[color=green!60!black, ultra thick, rounded corners=0.5mm, postaction={decorate}, decoration={markings,mark=at position .3 with {\arrow[draw=green!60!black]{>}}}] (alpha2) -- (2P) -- (3P);
			\draw[string=green!60!black, ultra thick] (R) -- (alpha1); 		%T-line
			\draw[string=green!60!black, ultra thick] (R3) -- (alpha1); 		%T-line
			%
			%% labels
			\fill[color=green!60!black] (alpha1) circle (1.2pt) node[left] (0up) { {\scriptsize$\A_0^+$} };
			\fill[color=green!60!black] (alpha2) circle (1.2pt) node[left] (0up) { {\scriptsize$\A_0^+$} };
			%
			% black auxiliary boundaries: 
			\draw [black,opacity=0.4141, densely dotted, semithick] (2L) -- (2P) -- (2N4);
			\draw [black,opacity=0.4141, densely dotted, semithick] (2P) -- (2N1);
			\draw [black,opacity=0.4141, densely dotted, semithick] (2R) -- (2RR) -- (2N2);
			\draw [black,opacity=0.4141, densely dotted, semithick] (2RR) -- (2N3);
			\draw [black,opacity=0.4141, densely dotted, semithick] (1L) -- (1P) -- (1N4);
			\draw [black,opacity=0.4141, densely dotted, semithick] (1R) -- (1RR) -- (1N3);
			\draw [black,opacity=0.4141, densely dotted, semithick] (1RR) -- (1N2);
			\draw [black,opacity=0.4141, densely dotted, semithick] (1P) -- (1N1);
			%
			% black boundaries: 
			\draw [black,opacity=1, very thin] (3P) -- (3L) -- (L) -- (P); 
			\draw [black,opacity=1, very thin] (N1) -- (P); 
			\draw [black,opacity=1, very thin] (3RR) -- (3N1) -- (N1) -- (3N1); 
			\draw [black,opacity=1, very thin] (3RR) -- (3R); 
			\draw [black,opacity=1, very thin] (3RR) -- (3N2) -- (N2) -- (R); 
			\draw [black,opacity=1, very thin] (3R) -- (3N3) -- (N3) -- (R3); 
			\draw [black,opacity=1, very thin] (R3) -- (R); 
			\draw [black,opacity=1, very thin] (3P) -- (3RR);
			\draw [black,opacity=1, very thin] (3P) -- (3N4) -- (N4) -- (R); 
			\draw [black,opacity=1, very thin] (P) -- (R); 
			\end{tikzpicture}}%%popende
		%%%%%%%%%%%%%%%%%%%%%% 
		\right)
		\ee 
		(where we suppress the labels $\A_1,\A_2,\A_3$ for all 1-, 2, and 3-strata) as well as those coming from the other dual 2-3 moves in~\eqref{eq:3dDualPachner} and those dual to the 1-4 moves in~\eqref{eq:3dPachner}. 
		
		Expressed internally to the 3-category~$\mathcal D_\zz$ (recall Theorem~\ref{thm:CMS}), 
		%arXiv_v2: 
			%it follows that the 1-morphism $\A_2\colon \A_3\lra \A_3$ has the structure of a (not necessarily unital) $E_1$-algebra with multiplication $\A_1\colon \A_2\circ \A_2\lra \A_2$ and associator $\zz(\A_0^+)$. 
			%Indeed, the condition~\eqref{eq:3dInv} precisely states that the pentagon axiom for $\zz(\A_0^+)$ holds. 
			 it follows that the 1-morphism $\A_2\colon \A_3\lra \A_3$ comes with the structure of a (not necessarily unital) $E_1$-algebra. 
			 Indeed, the multiplication $\A_1\colon \A_2\circ \A_2\lra \A_2$ is associative up to the associator $\zz(\A_0^+)$, and the condition~\eqref{eq:3dInv} precisely states that the pentagon axiom for $\zz(\A_0^+)$ holds. 
		More generally, all the invariance conditions~\eqref{eq:Invariance} for $n=3$ are equivalent to $(\A_2, \A_1, \zz(\A_0^+), \zz(\A_0^-))$ being a categorification of $\Delta$-separable symmetric Frobenius algebras in~$\mathcal D_\zz$, where the uncategorified defining conditions only hold up to coherent 3-isomorphisms built from $\zz(\A_0^\pm)$ and the adjunction data in~$\mathcal D_\zz$, see \cite{CRS1, CM3dOrbifoldCompletion} for details. 
	\end{enumerate}
	\label{exa:OrbDataGeneral}
\end{example}

Remark~\ref{rem:E1algebras1} and Example~\ref{exa:OrbDataGeneral} illustrate that orbifold data for~$\zz$ naturally give rise to $E_1$-algebras 
%arXiv_v2: 
	%(with extra structure) 
in the associated higher 
%arXiv_v2: 
	%category~$\mathcal D_\zz$. 
	 category~$\mathcal D_\zz$, i.e.\ algebras that are associative up to higher coherences which are part of their structure. 
Hence we have the following expected equivalent characterisation, which is rigorous for $n=2$ and $n=3$, see \cite[Sect.\,3.3]{cr1210.6363} and \cite[Sect.\,4.2]{CRS1}, respectively: 

\begin{definition}
	An orbifold datum for an $n$-dimensional defect TQFT~$\zz$ consists of an object $\A_n\in\mathcal D_\zz$ together with $(n-j)$-morphisms~$\A_j$, $j\in\{1,\dots,n-1\}$, and $n$-morphisms $\A_0^+,\A_0^-$ which can label the dual stratifications of simplices in the graphical calculus of~$\mathcal D_\zz$, such that dual Pachner moves become identities of $n$-morphisms in that calculus. 
\end{definition}

In short, orbifold data~$\A$ for~$\zz$ are $E_1$-algebras in the monoidal $(n-1)$-category $\mathcal D_\zz(\A_{n},\A_{n})$, subject to further constraints from Pachner moves. 

\begin{example}[Orbifold data for state sum models]
	We consider the trivial $n$-dimensional defect TQFT $\zz^{\textrm{{\tiny triv}}}_n$ valued in $\mathcal C = \Vectk$. 
	The following is preparation for the construction of state sum models in Example~\ref{exa:SSMTQFT} below. 
	\begin{enumerate}[leftmargin=*, label={(\arabic*)}]
		\item 
		\label{item:SSMOrb1}
		For $n=1$, an orbifold datum for $\zz^{\textrm{{\tiny triv}}}_1$ is equivalent to $1\in\Bbbk$, the only linear idempotent on $\zz^{\textrm{{\tiny triv}}}_1(\textrm{pt}) = \Bbbk$, cf.\ Example~\ref{exa:OrbDataGeneral}\ref{item:OrbData1}. 
		\item 
		For $n=2$, an orbifold datum in $\mathcal D_{\zz^{\textrm{{\tiny triv}}}_2} = \Bar\vectk$ is a $\Delta$-separable symmetric Frobenius $\Bbbk$-algebra, cf.\ Examples~\ref{exa:Pivotal2categories}\ref{item:BC} and~\ref{exa:OrbDataGeneral}\ref{item:OrbData2}. 
		More generally, all separable symmetric Frobenius $\Bbbk$-algebras are obtained by taking the Euler completion, cf.\ Example~\ref{exa:EulerTQFT} and \cite{Mule1, CM3dOrbifoldCompletion}. 
		\item 
		\label{item:SSMOrb3}
		For $n=3$, orbifold data in the Euler completion of $\mathcal D_{\zz^{\textrm{{\tiny triv}}}_3} = \Bar\ssFrob(\Vectk)$ can be obtained from spherical fusion categories, cf.\ Examples~\ref{exa:GrayCatDual}\ref{item:BssFrob}, \ref{exa:OrbDataGeneral}\ref{item:OrbData3} and \cite[Sect.\,4]{CRS1}. 
		\item 
		For $n=4$, orbifold data in the Euler completion of $\mathcal D_{\zz^{\textrm{{\tiny triv}}}_4}$ can be obtained from spherical fusion 2-categories, as explained in \cite{DouglasReutter2018, LukasNilsVincentas}. 
	\end{enumerate}  
	\label{exa:OrbForSSM}
\end{example}

\begin{example}[Orbifold data from (higher) group actions]
	\label{exa:OrbifoldDataFromGroupActions}
	Let~$\zz$ be an $n$-dimensional defect TQFT, and let~$G$ be a finite group which we view as a discrete strict monoidal $(n-1)$-category (whose only morphisms are identities). 
	A \textsl{$G$-action} on~$\zz$ is an $n$-functor $\rho\colon \Bar G \lra \mathcal D_\zz$. 
	Assuming~$\mathcal D_\zz$ to have appropriate direct sums we set $\A^\rho_n = \rho(*)$, $\A^\rho_{n-1} = \bigoplus_{g\in G}\rho(g)$ and define $\A^\rho_j$ for $j\leqslant n-2$ in terms of the coherence data of~$\rho$. 
	In particular, $\A^\rho_{n-2}$ is obtained from sums over the 2-isomorphisms $\rho(g)\circ\rho(h) \lra \rho(gh)$. 
	This gives rise to an $E_1$-algebra~$\A^\rho$ in~$\mathcal D_\zz$, which may or may not be an orbifold datum. 
	
	For $n=2$, orbifold data of type~$\A^\rho$ are studied in \cite{BCP2}, including twists of the (co)multiplication $(\A^\rho_0)^\pm$ by elements in group 
	%arXiv_v2: 
		%cohomolgy 
		 cohomology 
	$H^2(G;\Bbbk^\times)$. 
	In the context of B-twisted sigma models and Landau--Ginzburg models, (the underlying algebras of) $\A^\rho$ had been considered earlier in \cite[Sect.\,2.2]{PolishchukKernelAlgebras} and \cite[Sect.\,7.1]{cr1210.6363} (in the 2-categories of \cite{cw1007.2679} and \cite{cm1208.1481}), respectively. 
	
	For $n=3$, orbifold data of type~$\A^\rho$ are constructed from ribbon crossed $G$-categories, cf.\ \cite[Sect.\,5]{CRS3}. 
	A related analysis in the context of once-extended TQFTs is carried out in \cite{SchweigertWoike}. 
	
	More generally, actions $\Bar\mathcal G \lra \mathcal D_\zz$ of $n$-groups~$\mathcal G$ give candidates of orbifold data; 
	%arXiv_v2: 
		for $p\in\Z_{\geqslant 1}$, a \textsl{$p$-form symmetry} is the special case when $\mathcal G = \Bar^p H$ is the $p$-fold delooping of an abelian group~$H$. 
	Just as in the group case, there may be obstructions; see \cite[Rem.\,7.5(ii)]{cr1210.6363} for an example where~$\A^\rho$ is a $\Delta$-separable Frobenius algebra which is however not symmetric. 
\end{example}

\begin{example}[Orbifold data from invertible spheres]
	\label{exa:OrbFromInvertibleSpheres}
	Let~$\zz$ be an $n$-dimensional defect TQFT, and let $X\in\mathcal D_\zz(a,b)$ be a 1-morphism such that $X$-labelled $(n-1)$-spheres are invertible $n$-morphisms in the graphical calculus of~$\mathcal D_\zz$, cf.\ \cite[Rem.\,3.19]{CRS1}. 
	%arXiv_v2: 
		%Then $\A_n^X = a$, $\A_{n-1}^X = X^\dagger \circ X$, and~$\A_j^X$ for $j\leqslant n-2$ defined in terms of (higher) adjunction data of~$X$ give rise to an orbifold datum for~$\zz$. 
		 Then defining $\A_n^X = a$, $\A_{n-1}^X = X^\dagger \circ X$, and~$\A_j^X$ for $j\leqslant n-2$ in terms of (higher) adjunction data of~$X$ gives rise to an orbifold datum for~$\zz$. 
	Here~$\A_j^X$ can be thought of as the way $n-j+1$ $j$-spheres can touch at a $j$-stratum. 
	
	For $n=2$, invertibility of the two oriented $X$-labelled 1-spheres means that the left and right quantum dimensions of~$X$ are invertible. 
	Writing $\boldsymbol{\star} = \dim_{\textrm{r}}(X)^{-1}$, it is checked in \cite[Sect.\,4]{cr1210.6363} that 
	\be 
	(\A^X_0)^+ = 
	%%%%%%%%%%%%%%%%%%%%%%
	\begin{tikzpicture}[very thick,scale=0.9,color=blue!50!black, baseline=.9cm]
	\fill [orange!12] (1.25,0.34) -- (3.75,0.34) -- (3.75,1.655) -- (1.25,1.665);
	\draw[line width=0pt] 
	(3,0) node[line width=0pt] (D) {{\small$X^\dagger$}}
	(2,0) node[line width=0pt] (s) {{\small$X\vphantom{\X^\dagger}$}}; 
	\draw[redirected] (D) .. controls +(0,1) and +(0,1) .. (s);
	\draw[line width=0pt] (2.5,0.4) node[line width=0pt] (D) {{\tiny$a$}};
	\draw[line width=0pt] (2.5,1.4) node[line width=0pt] (D) {{\tiny$b$}};
	\draw[line width=0pt] (3.2,1.2) node[line width=0pt] (D) {{\tiny$a$}};
	\draw[line width=0pt] (1.8,1.2) node[line width=0pt] (D) {{\tiny$a$}};
	\draw[line width=0pt] 
	(3.45,0) node[line width=0pt] (re) {{\small$X\vphantom{\X^\dagger}$}}
	(1.55,0) node[line width=0pt] (li) {{\small$X^\dagger$}}; 
	\draw[line width=0pt] 
	(2.7,2) node[line width=0pt] (ore) {{\small$X\vphantom{\X^\dagger}$}}
	(2.3,2) node[line width=0pt] (oli) {{\small$X^\dagger$}}; 
	\draw (li) .. controls +(0,0.75) and +(0,-0.25) .. (2.3,1.25);
	\draw (2.3,1.25) -- (oli);
	\draw (re) .. controls +(0,0.75) and +(0,-0.25) .. (2.7,1.25);
	\draw (2.7,1.25) -- (ore);
	\end{tikzpicture}
	%%%%%%%%%%%%%%%%%%%%%%
	, \quad 
	(\A^X_0)^- = 
	%%%%%%%%%%%%%%%%%%%%%%
	\begin{tikzpicture}[very thick,scale=0.9,color=blue!50!black, baseline=-0.9cm, rotate=180]
	\fill [orange!12] (1.25,0.34) -- (3.75,0.34) -- (3.75,1.655) -- (1.25,1.665);
	\draw[line width=0pt] 
	(3,0) node[line width=0pt] (D) {{\small$X\vphantom{X^\dagger}$}}
	(2,0) node[line width=0pt] (s) {{\small${X^\dagger}$}}; 
	\draw[redirected] (D) .. controls +(0,1) and +(0,1) .. (s);
	\draw (2.5,1.13) node (D) {{\small$\boldsymbol{\star}$}}; 
	\draw[line width=0pt] (2.5,0.45) node[line width=0pt] (D) {{\tiny$a$}};
	\draw[line width=0pt] (2.5,1.45) node[line width=0pt] (D) {{\tiny$b$}};
	\draw[line width=0pt] (3.2,1.2) node[line width=0pt] (D) {{\tiny$a$}};
	\draw[line width=0pt] (1.8,1.2) node[line width=0pt] (D) {{\tiny$a$}};
	\draw[line width=0pt] 
	(3.45,0) node[line width=0pt] (re) {{\small${X^\dagger}$}}
	(1.55,0) node[line width=0pt] (li) {{\small$X\vphantom{X^\dagger}$}}; 
	\draw[line width=0pt] 
	(2.7,2) node[line width=0pt] (ore) {{\small${X^\dagger}$}}
	(2.3,2) node[line width=0pt] (oli) {{\small$X\vphantom{X^\dagger}$}}; 
	\draw (li) .. controls +(0,0.75) and +(0,-0.25) .. (2.3,1.25);
	\draw (2.3,1.25) -- (oli);
	\draw (re) .. controls +(0,0.75) and +(0,-0.25) .. (2.7,1.25);
	\draw (2.7,1.25) -- (ore);
	\end{tikzpicture}
	%%%%%%%%%%%%%%%%%%%%%%
	\ee 
	make $\A_1^X = X^\dagger\circ X$ into a $\Delta$-separable symmetric Frobenius algebra. 
	As shown in \cite{CRCR, OEReck}, there are such orbifold data which are neither related to state sum models nor to group actions, cf.\ Example~\ref{exa:OrbFromInvertibleQuantumDimension} below. 
\end{example}

\subsection{Orbifold construction}
\label{subsec:OrbifoldConstruction}

Let $\zz \colon \Borddef \lra \mathcal C$ be a 
%arXiv_v2: 
	%defect TQFT, 
	 defect TQFT where idempotents in~$\mathcal C$ split, 
and let~$\A$ be an orbifold datum for~$\zz$. 
From this we construct a closed TQFT $\zz_\A \colon \Bordor_{n,n-1} \lra \mathcal C$ as follows. 
For $\Sigma\in\Bordor_{n,n-1}$, choose a triangulation~$t$ of the cylinder $\Sigma \times [0,1] \colon \Sigma \lra \Sigma$, and denote the induced triangulations of its incoming and outgoing boundary components by~$\tau$ and~$\tau'$, respectively. 
By labelling the Poincar\'{e} dual stratification with~$\A$, we obtain a defect bordism $C_{\Sigma,\tau',\tau}^{t,\A} \colon \Sigma^{\tau,\A} \lra \Sigma^{\tau',\A}$ in $\Borddef$. 
For example, if $n=2$ and $\Sigma = S^1$, we can choose 
\be 
\label{eq:CylinderProjectorExample}
C_{\Sigma,\tau',\tau}^{t,\A} = 
\tikzzbox{%
	%%%%%%%%%%%%%%%%%%%%%% 
	\begin{tikzpicture}[thick,scale=0.5,color=green!50!black, baseline=1.15cm]
	\coordinate (p1) at (1.5,-0.5);
	\coordinate (p2) at (0,0);
	\coordinate (p3) at (1.5,0.5);
	\coordinate (p-1) at (2.5,-0.5);
	\coordinate (p-2) at (4,0);
	\coordinate (p-3) at (2.5,0.5);
	\coordinate (h) at (0,5);
	\coordinate (h2) at (0,2.5);
	\coordinate (c) at (3,-2.5);
	\coordinate (d1) at ($(3,0)+0.5*(h)$);
	\coordinate (d2) at ($(5,0)+0.5*(h)$);
	\coordinate (mu1) at (1.4,1.6);
	\coordinate (mu2) at (3.4,1.6);
	\coordinate (mu3) at (1.6,0.87);
	\coordinate (Delta1) at (0.7,0.1);
	\coordinate (Delta2) at (2.8,0.1);
	\coordinate (foldleft) at (0,0.5);
	\coordinate (foldright) at (4,1.3);
	%
	%
	% Filling back
	\fill [orange!20!white, opacity=0.8]
	(p3) .. controls +(-1,0) and +(0,0.5) .. ($(p2)$) -- ($(p2)+(h)$) .. controls +(0,0.5) and +(-1,0) .. ($(p3)+(h)$) -- ($(p-3)+(h)$) .. controls +(1,0) and +(0,0.5) .. ($(p-2)+(h)$) -- ($(p-2)$) .. controls +(0,0.5) and +(1,0) .. (p-3) -- (p3);
	\draw[very thick, red!80!black] 
	(p1) .. controls +(-1,0) and +(0,-0.5) .. (p2) 
	-- (p2) .. controls +(0,0.5) and +(-1,0) .. (p3) 
	--(p3) -- (p-3) .. controls +(1,0) and +(0,0.5) .. (p-2)
	-- (p-2) .. controls +(0,-0.5) and +(1,0) .. (p-1) -- (p1);
	%
	% Triangulation t BACK: 
	\draw[very thin, blue] ($(p2)+(h2)$) .. controls +(0,0.5) and +(-1,0) .. ($(p3)+(h2)$) 
	--($(p3)+(h2)$) -- ($(p-3)+(h2)$) .. controls +(1,0) and +(0,0.5) .. ($(p-2)+(h2)$); 
	\draw[very thin, blue] ($(p2)+(h2)$) .. controls +(0,-1) and +(0,0.5) .. (p-2);
	\draw[very thin, blue] ($(p2)+(h)$) .. controls +(0,-1) and +(0,0.5) .. ($(p-2)+(h2)$);
	%
	%%%%%%%%%%%%%%%%%%%%%%%%%%%%%%
	% dual stratification BACK
	%
	% 0-strata: 
	\fill (mu3) circle (3pt) node[left] (0up) { };
	\fill ($(mu3)+(0,-0.37)$) circle (3pt) node[left] (0up) { };
	\fill (3.7,1.11) circle (3pt) node[left] (0up) { };
	\fill ($(3.7,1.11)+(h2)$) circle (3pt) node[left] (0up) { };
	\fill (0.6,3.26) circle (3pt) node[left] (0up) { };
	% 1-strata: 
	\draw[very thick] (foldleft) .. controls +(0,0.5) and +(0,-0.5) .. (foldright);
	\draw[very thick] (mu3) -- ($(mu3)+(0,-0.4)$);
	\draw[very thick] ($(foldleft)+(h2)$) .. controls +(0,0.5) and +(0,-0.5) .. ($(foldright)+(h2)$);
	\draw[very thick] ($(3.7,1.11)+(h2)$) -- ($(3.7,1.11)+(0,0)$);
	\draw[very thick] (0.6,3.26) -- (0.6,5.47);
	%%%%%%%%%%%%%%%%%%%%%%%%%%%%%%
	%
	%%%%%%%%%%%%%%
	%Filling front
	\fill [orange!30!white, opacity=0.8]
	(p1) .. controls +(-1,0) and +(0,-0.5) .. (p2) -- ($(p2)+(h)$) .. controls +(0,-0.5) and +(-1,0) .. ($(p1)+(h)$) -- ($(p-1)+(h)$) .. controls +(1,0) and +(0,-0.5) .. ($(p-2)+(h)$) -- ($(p-2)$) .. controls +(0,-0.5) and +(1,0) .. (p-1) -- (p1);
	\draw[very thin, blue] (p2) -- ($(p2)+(h)$);
	\draw[very thin, blue] (p-2) -- ($(p-2)+(h)$);
	\draw[very thick, red!80!black] (p1) .. controls +(-1,0) and +(0,-0.5) .. (p2);
	%		-- (p2) .. controls +(0,0.5) and +(-1,0) .. (p3);
	%	--(p3) -- (p-3) .. controls +(1,0) and +(0,0.5) .. (p-2)
	\draw[very thick, red!80!black] (p-2) .. controls +(0,-0.5) and +(1,0) .. (p-1) -- (p1);
	\draw[very thick, red!80!black] ($(p1)+(h)$) .. controls +(-1,0) and +(0,-0.5) .. ($(p2)+(h)$) 
	-- ($(p2)+(h)$) .. controls +(0,0.5) and +(-1,0) .. ($(p3)+(h)$) 
	--($(p3)+(h)$) -- ($(p-3)+(h)$) .. controls +(1,0) and +(0,0.5) .. ($(p-2)+(h)$)
	-- ($(p-2)+(h)$) .. controls +(0,-0.5) and +(1,0) .. ($(p-1)+(h)$) -- ($(p1)+(h)$);
	%
	%%%%%%%%%%%%%%%%%%%%%%%%%%%%%%
	% Triangulation t FRONT: 
	%
	% middle circle: 
	\draw[very thin, blue] ($(p1)+(h2)$) .. controls +(-1,0) and +(0,-0.5) .. ($(p2)+(h2)$);
	\draw[very thin, blue] ($(p-2)+(h2)$) .. controls +(0,-0.5) and +(1,0) .. ($(p-1)+(h2)$) -- ($(p1)+(h2)$);
	% lower part: 
	\draw[very thin, blue] ($(p2)+(h2)$) .. controls +(0,-0.75) and +(0,0.2) .. (2,-0.5) -- ($(2,-0.5)+(h2)$) -- (p-2);
	\fill[blue] (p2) circle (1.5pt) node[left] (0up) { };
	\fill[blue] (p-2) circle (1.5pt) node[left] (0up) { };
	\fill[blue] (2,-0.5) circle (1.5pt) node[left] (0up) { };
	% upper part: 
	\draw[very thin, blue] ($(p2)+(h)$) .. controls +(0,-0.75) and +(0,0.2) .. ($(2,-0.5)+(h2)$) -- ($(2,-0.5)+(h)$) -- ($(p-2)+(h2)$);
	\fill[blue] ($(p2)+(h)$) circle (1.5pt) node[left] (0up) { };
	\fill[blue] ($(p-2)+(h)$) circle (1.5pt) node[left] (0up) { };
	\fill[blue] ($(2,-0.5)+(h)$) circle (1.5pt) node[left] (0up) { };
	\fill[blue] ($(p2)+(h2)$) circle (1.5pt) node[left] (0up) { };
	\fill[blue] ($(p-2)+(h2)$) circle (1.5pt) node[left] (0up) { };
	\fill[blue] ($(2,-0.5)+(h2)$) circle (1.5pt) node[left] (0up) { };
	%%%%%%%%%%%%%%%%%%%%%%%%%%%%%%
	%
	%%%%%%%%%%%%%%%%%%%%%%%%%%%%%%
	% dual stratification FRONT
	%
	% 0-strata: 
	\fill (Delta1) circle (3pt) node[left] (0up) { };
	\fill (mu1) circle (3pt) node[left] (0up) { };
	\fill (Delta2) circle (3pt) node[left] (0up) { };
	\fill (mu2) circle (3pt) node[left] (0up) { };
	\fill ($(Delta1)+(0,-0.6)$) circle (3pt) node[left] (0up) { };
	\fill ($(Delta2)+(0,-0.6)$) circle (3pt) node[left] (0up) { };
	\fill ($(Delta1)+(h2)$) circle (3pt) node[left] (0up) { };
	\fill ($(mu1)+(h2)$) circle (3pt) node[left] (0up) { };
	\fill ($(Delta2)+(h2)$) circle (3pt) node[left] (0up) { };
	\fill ($(mu2)+(h2)$) circle (3pt) node[left] (0up) { };
	\fill (0.6,5.47) circle (3pt) node[left] (0up) { };
	\fill ($(mu1)+(h2)+(0,0.4)$) circle (3pt) node[left] (0up) { };
	\fill ($(mu2)+(h2)+(0,0.4)$) circle (3pt) node[left] (0up) { };
	%
	% 1-strata
	\draw[very thick] (Delta1) .. controls +(0,0) and +(0,-0.15) .. (foldleft);
	\draw[very thick] (foldright) .. controls +(0,0.25) and +(0,0) .. (mu2);
	\draw[very thick] (Delta1) -- ($(Delta1)+(0,-0.63)$);
	\draw[very thick] (Delta2) -- ($(Delta2)+(0,-0.63)$);
	\draw[very thick] (Delta1) -- (mu1) -- (Delta2) -- (mu2);
	\draw[very thick] (mu1) -- ($(Delta1)+(h2)$);
	\draw[very thick] (mu2) -- ($(Delta2)+(h2)$);
	\draw[very thick] ($(Delta1)+(h2)$) .. controls +(0,0) and +(0,-0.15) .. ($(foldleft)+(h2)$);
	\draw[very thick] ($(foldright)+(h2)$) .. controls +(0,0.25) and +(0,0) .. ($(mu2)+(h2)$);
	\draw[very thick] ($(Delta1)+(h2)$) -- ($(mu1)+(h2)$) -- ($(Delta2)+(h2)$) -- ($(mu2)+(h2)$);
	\draw[very thick] ($(mu2)+(h2)+(0,0.4)$) -- ($(mu2)+(h2)+(0,0)$);
	\draw[very thick] ($(mu1)+(h2)+(0,0.4)$) -- ($(mu1)+(h2)+(0,0)$);
	%
	%%%%%%%%%%%%%%%%%%%%%%%%%%%
	%
	\end{tikzpicture}
	%%%%%%%%%%%%%%%%%%%%%% 
}
\colon 
%%%%%%%%%%%%%%%%%%%%%%
\begin{tikzpicture}[very thick, scale=0.45,color=green!50!black, baseline, >=stealth]
\draw[very thick, color=red!80!black] (0,0) circle (2);
\fill (-45:2) circle (4.0pt) node[right] {{\tiny $(\mathcal A_1,+)$}};
\fill (225:2) circle (4.0pt) node[left] {{\tiny $(\mathcal A_1,+)$}};
\fill (90:2) circle (4.0pt) node[below] {{\tiny $(\mathcal A_1,+)$}};
\end{tikzpicture}
%%%%%%%%%%%%%%%%%%%%%%
\lra 
%%%%%%%%%%%%%%%%%%%%%%
\begin{tikzpicture}[very thick, scale=0.45,color=green!50!black, baseline, >=stealth]
\draw[very thick, color=red!80!black] (0,0) circle (2);
\fill (-45:2) circle (4.0pt) node[right] {{\tiny $(\mathcal A_1,+)$}};
\fill (225:2) circle (4.0pt) node[left] {{\tiny $(\mathcal A_1,+)$}};
\fill (90:2) circle (4.0pt) node[below] {{\tiny $(\mathcal A_1,+)$}};
\end{tikzpicture}
%%%%%%%%%%%%%%%%%%%%%%
\ee  
where on the left we suppress $\A$-labels for the stratification, as well as orientations. 
The map 
$
\Phi_{\Sigma,\A}^{\tau',\tau} := \zz(C_{\Sigma,\tau',\tau}^{t,\A})\colon \zz(\Sigma^{\tau,\A}) \lra \zz(\Sigma^{\tau',\A})
$ 
does not depend on the triangulation~$t$ away from the boundary thanks to the invariance condition on~$\A$, and we have 
$
\Phi_{\Sigma,\A}^{\tau'',\tau} = \Phi_{\Sigma,\A}^{\tau'',\tau'} \circ \Phi_{\Sigma,\A}^{\tau',\tau}
$ 
for all triangulations $\tau,\tau',\tau''$ of~$\Sigma$. 
Then we define 
\be 
\label{eq:ZAobj}
\zz_\A(\Sigma) = \colim_{\tau,\tau'} \big(\Phi_{\Sigma,\A}^{\tau',\tau}\big) \,.
\ee 
Concretely, $\zz_\A$ can be computed (up to isomorphism) as the image of the idempotent $\Phi_{\Sigma,\A}^{\tau,\tau}$ for any triangulation~$\tau$ of~$\Sigma$, 
%arXiv_v2: 
	%if the colimit~\eqref{eq:ZAobj} exists (which it does e.g.\ if $\mathcal C = \Vectk$ or $\mathcal C = \sVectk$). 
	 which exists by assumption on~$\mathcal C$ (which in turn holds e.g.\ if $\mathcal C = \Vectk$ or $\mathcal C = \sVectk$). 

Similarly, for a morphism $M\colon \Sigma\lra \Sigma'$ in $\Bordor_{n,n-1}$, we may choose an arbitrary triangulation~$t$ which induces triangulations~$\tau$ and~$\tau'$ on~$\Sigma$ and~$\Sigma'$, respectively. 
Then by definition 
\be 
\label{eq:ZAmor} 
\zz_\A(M) = 
\Big(\!\!
\begin{tikzcd}[column sep=2em]
\zz_\A(\Sigma) \arrow[hook]{r} 
& \zz \big( \Sigma^{\tau,\A} \big) \arrow{rr}{\zz(M^{t,\A})}
& 
& \zz\big( \Sigma'^{\tau',\A} \big) \arrow[two heads]{r}
& \zz_\A(\Sigma') 
\end{tikzcd}  
\!\!\Big) 
\ee 
where the last map is part of the data of the colimit $\zz_\A(\Sigma'$), and the first map is obtained from the universal property of the colimit $\zz_\A(\Sigma)$. 
This means that if e.g.\ $\mathcal C=\Vectk$, $\zz_\A(M)$ is given by pre- and post-composing $\zz(M^{t,\A})$ with the inclusion and surjection maps which split the idempotents $\Phi_{\Sigma,\A}^{\tau,\tau}$ and $\Phi_{\Sigma',\A}^{\tau',\tau'}$, respectively.
%, for any choice of triangulation~$t$ of~$M$. 
As explained in more detail in \cite[Sect.\,3.2]{CRS1}, the thus defined functor~$\zz_\A$ inherits a symmetric monoidal structure from~$\zz$, and we have: 

%arXiv_v2: 
	%\begin{definition}
	%	Let~$\A$ be an orbifold datum for $\zz \colon \Borddef \lra \mathcal C$ such that the colimits~\eqref{eq:ZAobj} exist in~$\mathcal C$. 
	%	Then~\eqref{eq:ZAobj} and~\eqref{eq:ZAmor} assemble into the \textsl{orbifold (TQFT)} $\zz_\A\colon \Bordor_{n,n-1} \lra \mathcal C$. 
	%\end{definition}
	 \begin{definitiontheorem}
	 Let~$\A$ be an orbifold datum for a defect TQFT $\zz \colon \Borddef \lra \mathcal C$ such that the colimits~\eqref{eq:ZAobj} exist in~$\mathcal C$. 
	 Then~\eqref{eq:ZAobj} and~\eqref{eq:ZAmor} assemble into the \textsl{orbifold (TQFT)} $\zz_\A\colon \Bordor_{n,n-1} \lra \mathcal C$. 
	\end{definitiontheorem}

\begin{example}
	State sum models are (Euler completed) orbifolds of the trivial defect TQFT valued in $\mathcal C = \Vectk$: 
	\begin{enumerate}[leftmargin=*, label={(\arabic*)}]
		\item 
		There are no non-trivial state sum models in dimension $n=1$. 
		This is consistent with the fact that $\zz^{\textrm{{\tiny triv}}}_1(\Sigma) = \Bbbk$ for all objects (points) $\Sigma\in\Borddefblank_{1,0}(\mathds{D}^{\textrm{{\tiny triv}}_1})$, and the only idempotent on~$\Bbbk$ is~1, cf.\ Example~\ref{exa:OrbForSSM}\ref{item:SSMOrb1}.   
		\item 
		\label{item:2dSSM}
		In dimension $n=2$, orbifold data~$\A$ in $\mathcal D_{\zz^{\textrm{{\tiny triv}}}_2} = \Bar\vectk$ are $\Delta$-separable symmetric Frobenius $\Bbbk$-algebras, for which the orbifold construction~$\zz_\A$ coincides with the state sum model construction of \cite{bp9205031, FHK, lp0602047}. 
		Hence~$\zz_\A$ is the closed TQFT equivalently described by the commutative Frobenius $\Bbbk$-algebra which is the centre of~$\A$. 
		
		The construction of \cite{bp9205031, FHK, lp0602047} refined by \cite[Sect.\,3.2]{Mule1} in fact takes arbitrary separable symmetric Frobenius algebras as input, not necessarily $\Delta$-separable ones. 
		As explained in \cite{Mule1, CM3dOrbifoldCompletion}, such algebras  
		correspond to the Euler completion of $(\zz^{\textrm{{\tiny triv}}}_2)_\A$ (recall Example~\ref{exa:EulerTQFT}, see also Example~\ref{exa:2dDefectSSMs}). 
		This appearance of Euler completion continues in higher dimensions. 
		
		To explain the name ``state sum model'', let us choose a basis $\{a_i\}$ of the vector space~$\A_1$. 
		Hence there are scalars $\mu_{ij}^k, \Delta_i^{jk}$ such that $\A_0^+(a_i\otimes a_j) = \sum_{k} \mu_{ij}^k \cdot a_k$ and $\A_0^-(a_i) = \sum_{j,k} \Delta_i^{jk} \cdot a_j \otimes a_k$. 
		This in turn means that the main ingredient $(\zztriv_2)(M^{t,\A})$ in $(\zztriv_2)_\A(M)$ for any bordism~$M$, which is basically a string diagram between tensor powers of~$\A_1$ whose only vertices are~$\A_0^\pm$, is a \textsl{sum} (one for each $\A_1$-labelled strand) over the \textsl{states} $a_i, a_j$, etc.  
		\item 
		\label{item:3dSSM}
		In dimension $n=3$, orbifold data~$\A^{\mathcal S}$ in the Euler completion of $\mathcal D_{\zztriv_3} = \Bar\ssFrob(\Vectk)$ can be extracted from spherical fusion categories~$\mathcal S$. 
 		Indeed, if~$I$ is a set of representatives of isomorphism classes of simple objects in~$\mathcal S$, we have $\A^{\mathcal S}_3 = *$, $\A^{\mathcal S}_2 = \bigoplus_{i\in I} \Bbbk$ is a direct sum of trivial Frobenius algebras, $\A^{\mathcal S}_1 = \bigoplus_{i,j,k\in I} \mathcal S(i\otimes j,k)$ as an $\A^{\mathcal S}_2$-$(\A^{\mathcal S}_2\otimes_\Bbbk\A^{\mathcal S}_2)$-bimodule, and $(\A^{\mathcal S}_0)^\pm$ is basically given by the associator of~$\mathcal S$, see \cite[Prop.\,4.2]{CRS3} for details. 
		As shown in \cite[Thm.\,4.5]{CRS3} the orbifold $(\zztriv_3)^\odot_{\A^{\mathcal S}}$ is equivalent to the Turaev--Viro--Barrett--Westbury TQFT \cite{TVmodel, bwTV1} for~$\mathcal S$. 
		Its evaluation on bordisms may be expressed as a \textsl{sum} of \textsl{states}, where now the latter involve both a basis of~$\A^{\mathcal S}_1$ and the simple objects in~$I$. 
		\item 
		\label{item:4dSSM}
		In dimension $n=4$, orbifold data~$\A^{\mathfrak S}$ for the Euler completion $(\zztriv_4)^\odot$ can be extracted from spherical fusion 2-categories~$\mathfrak S$ in a way analogous to the above 3-dimensional case (see Remark~\ref{rem:HigherSSMs} for more on~$\zztriv_4$). 
		As shown in \cite{LukasNilsVincentas}, the state sum model $(\zztriv_4)^\odot_{\A^{\mathfrak S}}$ precisely reproduces the Douglas--Reutter invariants of closed 4-manifolds \cite{DouglasReutter2018}, and lifts them to a TQFT. 
	\end{enumerate}
	\label{exa:SSMTQFT}
\end{example}

\begin{example}
	\label{exa:OrbifoldSigmaModels}
	Orbifolds from group actions are eponymous for the (generalised) orbifold construction: if~$\zz$ is a twisted sigma model whose target manifold~$Y$ comes with a $G$-action, one may consider the corresponding sigma model~$\zz^G$ whose target is the orbifold stack $Y/\!\!/G$, see e.g.\ \cite{OrbifoldsStringTopologyBook}. 
	Alternatively, the $G$-action on~$Y$ may lift to one on~$\zz$, $\Bar G\lra \mathcal D_\zz$, giving a candidate orbifold datum~$\A_G$. 
	If and only if~$\A_G$ is indeed an orbifold datum, we say that the $G$-action \textsl{can be gauged} (without anomaly). 
	In this case the orbifold TQFT~$\zz_{\A_G}$ is expected to be equivalent to~$\zz^G$. 
	
	In dimension $n=2$, this expectation has been verified for many twisted sigma models and Landau--Ginzburg models, see e.g.\ \cite{BCP, BCP1}. 
	In particular, the state space $\zz^G(S^1)$ is naturally recovered as the endomorphisms of~$\A_G$ viewed as a bimodule over itself, thus effortlessly including all ``twisted sectors'', cf.\ Example~\ref{exa:Gequiv} below.  
\end{example}

We end this section with some results and examples that are specific to dimensions~2 and~3, most of which are however expected to generalise to higher dimensions. 
A key tool behind the scenes here is the theory of ``orbifold completion'' discussed in Section~\ref{subsec:OrbifoldCompletion} below. 
In particular, orbifolding with~$\A$ can be undone by orbifolding with a ``quantum symmetry defect''~$\widetilde\A$, at least for $n=2$: 

\begin{theorem}[\cite{cr1210.6363, BCP2}]
	Let~$\A$ be an orbifold datum for a 2-dimensional defect TQFT~$\zz$. 
	The orbifold~$\zz_\A$ naturally lifts to a defect TQFT, and there exists an orbifold datum~$\widetilde\A$ such that $(\zz_\A)_{\widetilde\A} \cong \zz$. 
\end{theorem}

\begin{example}[Orbifolds from invertible quantum dimensions]
	\label{exa:OrbFromInvertibleQuantumDimension}
	In Example~\ref{exa:OrbFromInvertibleSpheres} we constructed an orbifold datum~$\A^X$ for a 2-dimensional defect TQFT~$\zz$ for every $X\in\mathcal D_\zz(a,b)$ with invertible quantum dimensions. 
	This has been applied to the case of the Landau--Ginzburg 2-category $\mathcal{LG}_\C$ (cf.\ Example~\ref{exa:Pivotal2categories}\ref{item:LG}), where checking the invertibility condition reduces to computations with square matrices with polynomial entries. 
	
	Recall that simple isolated singularities over~$\C$ admit an ADE classification (see e.g.\ \cite[Prop.\,8.5]{Yoshinobook}), with examples such as: 
	\be 
	\hspace{-0.45cm}
	\begin{array}{ccc}
		\begin{minipage}{3.4cm}
			\includegraphics[scale=0.2]{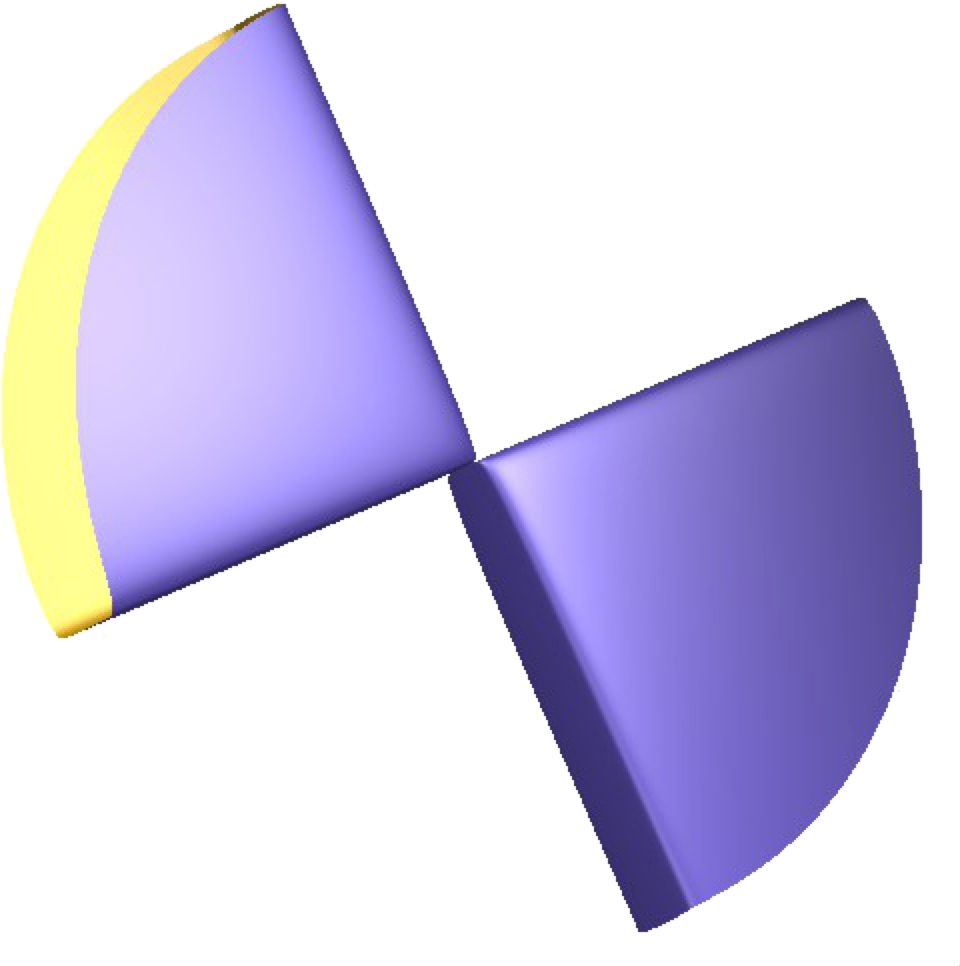}
		\end{minipage} 
		& 
		\begin{minipage}{3cm}
			\includegraphics[scale=0.2]{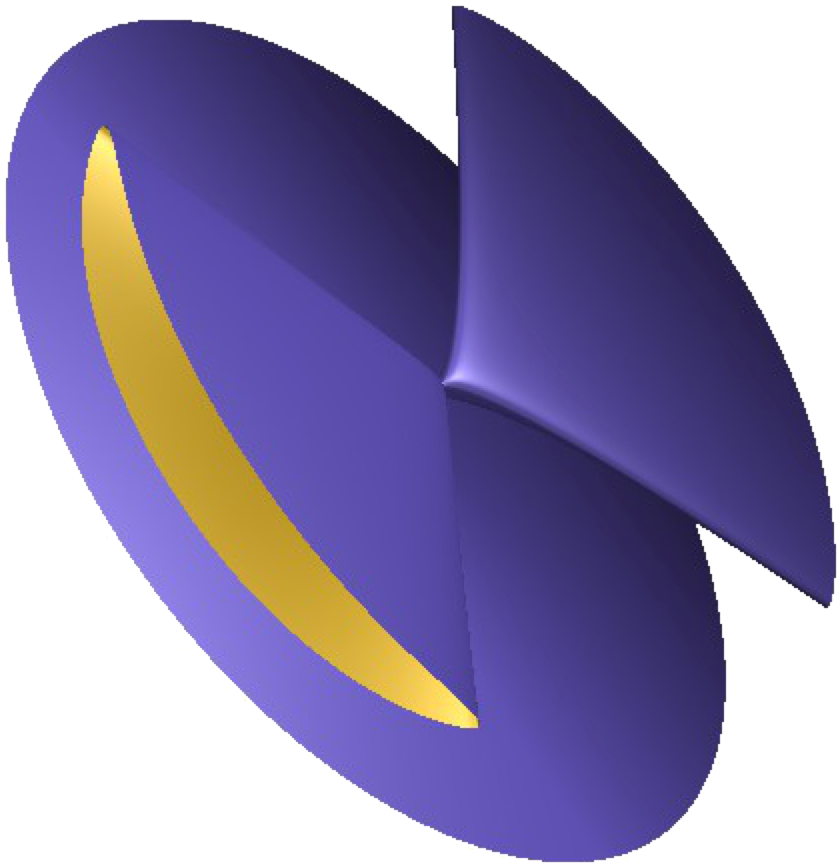}
		\end{minipage} 
		& 
		\begin{minipage}{2.8cm}
			\includegraphics[scale=0.2]{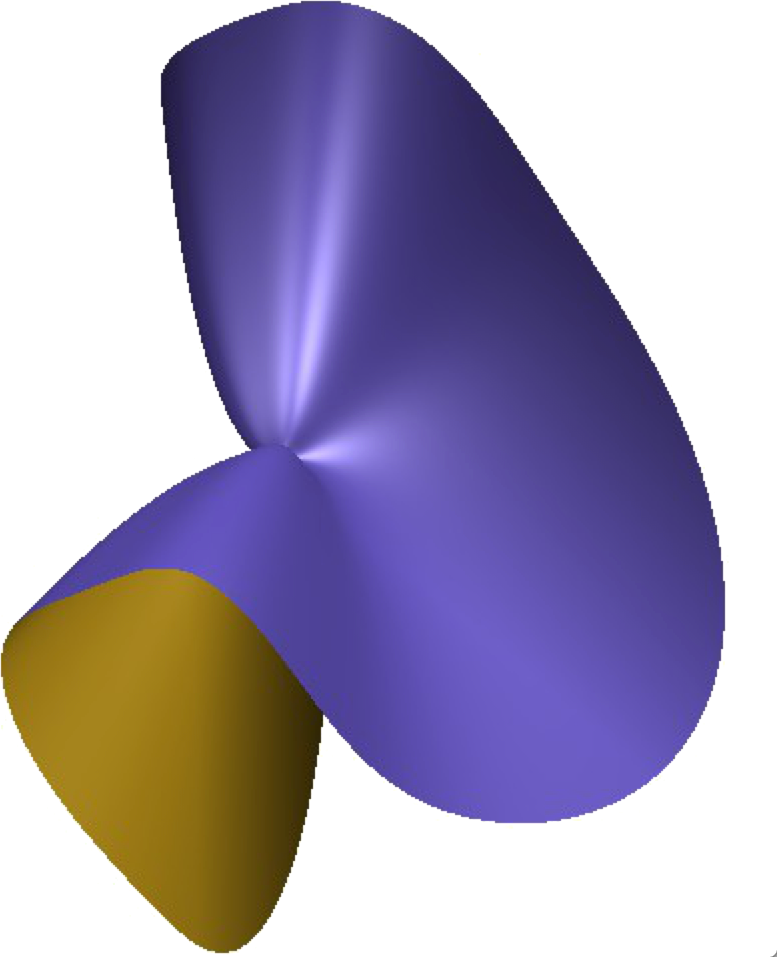}
		\end{minipage} 
		\\
		\big\{ W^{\textrm{A}_{d-1}} \!= x^d + y^2 = 0 \big\} 
		& 
		\big\{ W^{\textrm{D}_{10}} \!= u^9 + uv^2 = 0 \big\}
		& 
		\big\{ W^{\textrm{E}_{6}} \!= a^3 + b^4 = 0 \big\}
	\end{array}
	\ee 
	Viewed as objects in $\mathcal{LG}_\C$, these polynomials are far from equivalent. 
	But as shown in \cite{CRCR} there is $X\in\mathcal{LG}_\C(W^{\textrm{A}_{11}},W^{\textrm{E}_{6}})$, i.e.\ a matrix factorisation of $W^{\textrm{E}_{6}} - W^{\textrm{A}_{11}}$, that induces an equivalence $(\zz_{W^{\textrm{A}_{11}}})_{\A^X} \cong \zz_{W^{\textrm{E}_{6}}}$ between the TQFTs, and analogously for $W^{\textrm{A}_{17}} \sim  W^{\textrm{D}_{10}} \sim  W^{\textrm{E}_7}$ and $W^{\textrm{A}_{29}} \sim W^{\textrm{D}_{16}} \sim W^{\textrm{E}_8}$. 
	
	While the above examples were already expected from the ``CFT/LG correspondence'' combined with \cite{o0111139}, Recknagel--Weinreb \cite{OEReck} have algorithmically constructed several entirely novel examples between non-simple isolated singularities, e.g.\ $W^{\textrm{E}_{13}} \sim W^{\textrm{Z}_{11}}$ and $W^{\textrm{E}_{18}} \sim W^{\textrm{Q}_{12}}$. 
	These give new relations in singularity theory, and between Landau--Ginzburg models. 
\end{example}

\medskip 

Orbifolds in dimension~3 have been studied most extensively for Reshetikhin--Turaev models. 
A key technical result obtained in \cite{MuleRunk} is that to any simple orbifold datum~$\A$ for $\zz^{\textrm{RT}}_{\mathcal M}$ (recall Example~\ref{ex:3dDefectTQFT}\ref{item:RT}) one naturally associates another modular fusion category~$\mathcal M_\A$. 
With this it can be made precise that ``Reshetikhin--Turaev TQFTs close under generalised orbifolds'': 

\begin{theorem}[\cite{CMRSS2, CM3dOrbifoldCompletion}]
	Let~$\mathcal M$ be a modular fusion category, and let~$\A$ be an orbifold datum for $\zz^{\textrm{RT}}_{\mathcal M}$. 
	The orbifold $(\zz^{\textrm{RT}}_{\mathcal M})_\A$ naturally lifts to a defect TQFT, and $(\zz^{\textrm{RT}}_{\mathcal M})_\A \cong \zz^{\textrm{RT}}_{\mathcal M_\A}$. 
\end{theorem}

In addition to those related to state sum models (cf.\ Examples~\ref{exa:OrbForSSM}\ref{item:SSMOrb3} and~\ref{exa:SSMTQFT}\ref{item:3dSSM}) and those coming from group actions (cf.\ Examples~\ref{exa:OrbifoldDataFromGroupActions} and~\ref{exa:OrbFromInvertibleQuantumDimension}), Reshetikhin--Turaev models admit orbifold data~$\A^B$ obtained from \textsl{condensable algebras}, i.e.\ commutative haploid separable symmetric Frobenius algebras~$B$ in~$\mathcal M$, see \cite[Sect.\,3.4]{CRS3}. 
The associated \textsl{condensation} $\mathcal M_{\A^B}$ is equivalent to the category of local $B$-modules, and at least these types of orbifolds can be inverted: 

\begin{theorem}[\cite{Mule1}]
	\label{thm:InvertingCondensations}
	Let~$\mathcal M$ be a modular fusion category, and let~$B$ be a condensable algebra in~$\mathcal M$. 
	There exists an orbifold datum~$\widetilde\A$ for $\zz^{\textrm{RT}}_{\mathcal M_{\A^B}}$ such that $(\mathcal M_{\A^B})_{\widetilde{\A}} \cong \mathcal M$ as ribbon categories. 
\end{theorem}

\begin{example}
	There are several modular fusion categories~$\mathcal I$ of ``Ising type'', with precisely three isomorphism classes of simple objects $\one, \sigma, \varepsilon$ and fusion rules $\varepsilon\otimes \varepsilon \cong \one$ and $\sigma\otimes\sigma \cong \one \oplus \varepsilon$. 
	Via an algorithmic search, explicit orbifold data~$\widetilde\A$ of ``Fibonacci type'' were found in \cite{MuleRunk2}, such that~$\mathcal I_{\widetilde\A}$ is a condensation inversion for an orbifold of the modular fusion category associated to $\mathfrak{sl}(2)$ at level $k=10$. 
	The fact that~$\widetilde\A$ comes neither from condensable algebras nor from group actions again illustrates the usefulness of the general orbifold theory. 
\end{example}

Building on Theorem~\ref{thm:InvertingCondensations}, one obtains an equivalent characterisation of orbifold data for Reshetikhin--Turaev TQFTs in terms of Witt equivalence. 
Recall that two modular fusion categories $\mathcal M, \mathcal M'$ are \textsl{Witt equivalent} if there exists a spherical fusion category together with a ribbon equivalence between its Drinfeld centre and $\mathcal M' \boxtimes \mathcal M^{\textrm{rev}}$, where $(-)^{\textrm{rev}}$ denotes the reversed braiding and twist. 

\begin{theorem}[\cite{Mule1}]
	\label{thm:WittEquivalence}
	Two modular fusion categories $\mathcal M, \mathcal M'$ are Witt equivalent if and only if there exists an orbifold datum~$\A$ for $\zz^{\textrm{RT}}_{\mathcal M}$ such that $\mathcal M' \cong \mathcal M_\A$ as ribbon categories. 
\end{theorem}

\subsection{Orbifold completion}
\label{subsec:OrbifoldCompletion}

An $n$-dimensional orbifold datum for~$\zz$ is a type of algebra internal to the $n$-category~$\mathcal D_\zz$. 
It is then natural to consider the $n$-category whose objects are \textsl{all} orbifold data, and whose (higher) morphisms capture their (higher) representation theory. 
It is also precisely this higher Morita category 
$(\mathcal D_\zz)_{\textrm{orb}}$ 
which allows us to lift the output of the orbifold construction from mere \textsl{closed} TQFTs~$\zz_\A$ to a proper \textsl{defect} TQFT $\zz_{\textrm{orb}}$, whose defects are the morphisms in $(\mathcal D_\zz)_{\textrm{orb}}$. 

\medskip 

While expected to hold in general, the representation theory of orbifold data has so far been rigorously developed only in dimension $n\leqslant 3$. 
The case $n=1$ is trivial, so we start with $n=2$. 

\begin{definition}[\cite{cr1210.6363}]
	\label{def:OrbifoldCompletion}
	Let~$\B$ be a pivotal 2-category with idempotent complete Hom categories. 
	The \textsl{orbifold completion}~$\Borb$ of~$\B$ is the 2-category whose
	\begin{itemize}[leftmargin=*]
		\item 
		objects are orbifold data $\A = (\A_2,\A_1,\A_0^\pm)$ in~$\B$, i.e.\ $\Delta$-separable symmetric Frobenius algebras, 
		\item 
		1-morphisms $\A\lra\A'$ in $\Borb$ are 1-morphisms $\A_2\lra\A'_2$ in~$\B$ together with an $\A'$-$\A$-bimodule structure, 
		\item 
		horizontal composition of $X\colon\A\lra\A'$ and $Y\colon\A'\lra\A''$ is the relative tensor product $Y\otimes_{\A'}X$ in~$\B$ (that exists as Hom categories are idempotent complete),
		\item 
		the identity 1-morphism on $\A\in\Borb$ is~$\A$ viewed as a bimodule over itself, 
		\item 
		2-morphisms in~$\Borb$ are bimodule maps in~$\B$. 
	\end{itemize}
\end{definition}

\begin{theorem}[\cite{cr1210.6363}]
	\label{thm:Borb}
	The pivotal structure on~$\B$ induces a pivotal structure on~$\Borb$, and there is a pivotal equivalence $(\Borb)_{\textrm{orb}} \cong \Borb$.
\end{theorem}

Of course we want to apply this to the case $\B=\mathcal D_\zz$ for some 2-dimensional defect TQFT $\zz\colon\Borddefblank_{2,1}(\mathds{D}) \lra \mathcal C$. 
Then $(\mathcal D_\zz)_{\textrm{orb}}$ is naturally $\mathcal C$-enriched, and we obtain a larger set of defect data~$\mathds{D}^{\textrm{orb}}$ whose label sets~$D_j^{\textrm{orb}}$ are defined to consist of the $(2-j)$-cells of $(\mathcal D_\zz)_{\textrm{orb}}$. 
As shown in \cite[Sect.\,3.4]{cr1210.6363}, this allows us to lift the orbifold construction $\zz\lmt\zz_\A$ from closed to defect TQFTs: 

\begin{definition}
	\label{def:2dOrbifoldDefectTQFT}
	The \textsl{orbifold defect TQFT} 
	$  
	\zz_{\textrm{orb}} \colon \Borddefblank_{2,1}(\mathds{D}^{\textrm{orb}}) \lra \mathcal C
	$ 
	is given on morphisms by replacing $\A$-labelled 2-strata~$\sigma$ by $\A$-labelled substratifications $\sigma^{t,\A}$, connecting $\A_1$-labelled 1-substrata to adjacent ${D}_1^{\textrm{orb}}$-labelled 1-strata via the corresponding bimodule structure morphisms, evaluating with~$\zz$, and taking the colimit over all substratifications. 
\end{definition}

Hence for $X,Y\in(\mathcal D_\zz)_{\textrm{orb}}(\A',\A)$ and $\varphi\colon X\lra Y$, locally the evaluation of $\zz_{\textrm{orb}}$ near a $\varphi$-labelled 0-stratum is (suppressing some labels on the right)
\be 
\mathcal Z_{\textrm{orb}}\Bigg(
%%%%%%%%%%%%%%%%%%%%%%
\begin{tikzpicture}[very thick,scale=0.363,color=green!50!black, baseline, rounded corners=0.1pt]
\clip (0,0) ellipse (6cm and 3cm);
\fill [green!30] (0,-3) -- (0,3) -- (6,3) -- (6,-3);
\nicedashedpalecolourschemegreenedition[opacity=0.7] (0,0) ellipse (6 and 3);
\coordinate (phi) at (0,1);
%
% X/Y-lines: 
\draw[ultra thick, color=blue!70!black] (0,-3) -- (phi);
\draw[blue!70!black] (0.55,-1) node {{\small$X$}};
\draw[ultra thick, color=blue!50!red] (0,3) -- (phi);
\draw[blue!50!red] (0.55,2.1) node {{\small$Y$}};
%
% labels: 
\fill[color=black] (phi) circle (6pt) node[left] (0up) { {\small$\varphi$} };
%
% A-label: 
\draw (-3,0) node {{\small$\mathcal A\vphantom{\mathcal A'}$}};
\draw (+3,0) node {{\small$\mathcal A'$}};
\end{tikzpicture}
%%%%%%%%%%%%%%%%%%%%%%
\Bigg)
= 
\mathcal Z\Bigg(
%%%%%%%%%%%%%%%%%%%%%%
\begin{tikzpicture}[very thick,scale=0.363,color=green!50!black, baseline, rounded corners=0.1pt]
\clip (0,0) ellipse (6cm and 3cm);
\fill [orange!30] (0,-3) -- (0,3) -- (6,3) -- (6,-3);
\nicedashedpalecolourscheme[opacity=0.7] (0,0) ellipse (6 and 3);
\coordinate (phi) at (0,1);
\coordinate (r1) at (0,-1.5);
\coordinate (r2) at (0,2);
\coordinate (l1) at (0,0);
\coordinate (Al1) at (-2,-3);
\coordinate (Al2) at (-1,-3);
\coordinate (Ar1) at (2,-3);
\coordinate (Ar2) at (5,-3);
%
% X/Y-lines: 
\draw[ultra thick, color=blue!70!black] (0,-3) -- (phi);
\draw[blue!70!black] (0.4,-0.7) node {{\tiny$X$}};
\draw[ultra thick, color=blue!50!red] (0,3) -- (phi);
\draw[blue!50!red] (0.4,2.5) node {{\tiny$Y$}};
%
% A-lines: 
\draw[-dot-] (Al1) .. controls +(0,1) and +(0,1) .. (Al2);
\draw (-1.5,-2.2) .. controls +(0,0.75) and +(0,0) .. (l1);
\fill (-1.4,-2.2) circle (0pt) node[left] (0up) { {\tiny$\mathcal A_0^+$} };
\fill (-0.9,-1) circle (5.5pt) node[left] (0up) { };
\fill (-0.8,-1) circle (0pt) node[left] (0up) { {\tiny$\mathcal A_0^-$} };
\draw (-0.9,-1) .. controls +(-0.5,0.75) and +(0,-1) .. (-5,3);
\draw (Ar1) .. controls +(0,0.75) and +(0,0) .. (r1);
\draw (Ar2) .. controls +(0,3) and +(0,0) .. (r2);
%
% labels: 
\fill[color=black] (phi) circle (6pt) node[left] (0up) { {\small$\varphi$} };
\fill[color=blue!70!black] (r1) circle (5.5pt) node[left] (0up) { };
\fill[color=blue!50!red] (r2) circle (5.5pt) node[left] (0up) { };
\fill[color=blue!70!black] (l1) circle (5.5pt) node[left] (0up) { };
%
% A-label: 
\draw (-3.9,-0.6) node {{\tiny$\mathcal A_2$}};
\draw (-1.7,1.9) node {{\tiny$\mathcal A_2$}};
\draw (2,-0.6) node {{\tiny$\mathcal A'_2$}};
\draw (5,0.2) node {{\tiny$\mathcal A'_2$}};
\draw (-3.4,0.6) node {{\tiny$\mathcal A_1$}};
\draw (3.3,0.6) node {{\tiny$\mathcal A'_1$}};
\draw (1.6,-1.8) node {{\tiny$\mathcal A'_1$}};
\end{tikzpicture}
%%%%%%%%%%%%%%%%%%%%%%
\Bigg)
\, . 
\ee 
The fact that this is independent of the choice of substratification near $X,Y,\varphi$ is precisely due to the defining properties of bimodules and bimodule maps. 
Hence $\zz_{\textrm{orb}}$ is well-defined by construction of the orbifold completion $(\mathcal D_\zz)_{\textrm{orb}}$. 
Moreover, by design we have $(\mathcal D_\zz)_{\textrm{orb}} \cong \mathcal D_{\zz_{\textrm{orb}}}$. 

\begin{example}
	\label{exa:2dDefectSSMs}
	Recall from Example~\ref{exa:Pivotal2categories}\ref{item:BC} that the 2-category associated to the $\mathcal C$-valued trivial defect TQFT $\zztriv_2$ is $\mathcal D_{\zztriv_2} = \Bar\mathcal C^{\textrm{d}}$. 
	Hence it directly follows from Definition~\ref{def:OrbifoldCompletion} that $(\mathcal D_{\zztriv_2})_{\textrm{orb}} = \Delta\!\ssFrob(\mathcal C)$. 
	Combining this with Examples~\ref{exa:2dDefectTQFT}\ref{item:2dDefectSSM}, \ref{exa:Pivotal2categories}\ref{item:ssFrob} and~\ref{exa:SSMTQFT}\ref{item:2dSSM}, we realise that the \textsl{defect state sum model is the (Euler completed) orbifold of the trivial defect TQFT}: $\zzss_2 = (\zztriv_2)^\odot_{\textrm{orb}}$. 
\end{example}

The completion property $(\Borb)_{\textrm{orb}} \cong \Borb$ can be viewed as an ``oriented'' categorification of idempotent completion of 1-categories (replacing \textsl{identities} $e\circ e = e$ by \textsl{2-morphisms} $\A_0^+\colon \A_1 \circ \A_1 \lra \A_1$) -- as opposed to the slightly different categorification in \cite{DouglasReutter2018, GaiottoJohnsonFreyd} inspired by \textsl{framed} TQFTs. 
Intuitively, this property should follow from the fact that making a given choice of substratification finer does not affect the orbifold construction (because of the invariance condition). 
One way to rigorously establish the completion property is to use the universal property of $\Borb$. 
To state it concisely we say that for $a\in\B$, an \textsl{orbifold condensation of~$a$ (onto $b\in\B$)} is $X\in\B(a,b)$ such that $\tev_X\circ\coev_X = 1_{1_b}$, and that an orbifold datum $\A\in\Borb$ \textsl{splits} if there exists an orbifold condensation~$X$ of~$\A_2$ such that $\Xd\circ X \cong \A$ as Frobenius algebras (recall Example~\ref{exa:OrbFromInvertibleSpheres}). 

\begin{proposition}[\cite{CM3dOrbifoldCompletion}]
	\label{prop:UniversalPropertyBorb}
	The inclusion  $\mathcal B \longhookrightarrow \Borb$, $a\lmt (a,1_a,\lambda_{1_a}^{\pm 1})$, satisfies the universal property that for every pivotal 2-functor $\mathcal B \lra \overline{\mathcal D}$ in whose codomain every orbifold condensation splits, there exists an essentially unique pivotal 2-functor $\Borb \lra 
	\overline{\mathcal D}$ \mbox{such that the following commutes up to equivalence:}%
	\be 
	\begin{tikzcd}[column sep=3em, row sep=3em]
	\mathcal B
	\ar[r] 
	\ar[d, hook] 
	& 
	\overline{\mathcal D}
	\\ 
	\Borb
	\ar[ur, dashed] 
	& 
	\end{tikzcd}
	\ee  
\end{proposition}

\begin{example}[Orbifold equivalence]
	The orbifold datum~$\A^X$ constructed from $X\in\B(a,b)$ with invertible quantum dimensions (recall Examples~\ref{exa:OrbFromInvertibleSpheres} and~\ref{exa:OrbFromInvertibleQuantumDimension}) is equivalent to the image of $b\in\B$ under $\mathcal B \longhookrightarrow \Borb$, as shown in \cite[Thm.\,4.8]{cr1210.6363}. 
	We denote this \textsl{orbifold equivalence} $\A^X \cong (b,1_b,\lambda_{1_b}^{\pm 1}) \equiv 1_b$ as $a\sim b$. 
	It immediately follows that everything about $b\in\B$ can be described in terms of the algebra~$\A^X$ on~$a$; in particular we have equivalences 
	$
		\B(b,c) 
			\cong 
			\Borb(\A^X,1_c)
			= 
			\textrm{mod}_{\B(a,c)}(\A^X)
	$ 
	for all $c\in\B$. 
	Applying this to, say, $X\in\mathcal{LG}_\C(W^{\textrm{A}_{11}}, W^{\textrm{E}_{6}})$ of Example~\ref{exa:OrbFromInvertibleQuantumDimension}, the orbifold equivalence $W^{\textrm{A}_{11}} \sim W^{\textrm{E}_{6}}$ implies (by choosing $c=0\in\mathcal{LG}_\C$) that the category of matrix factorisations of $W^{\textrm{E}_{6}}$ is equivalent to the category of right $\A^X$-modules internal to the category of matrix factorisations of $W^{\textrm{A}_{11}}$. 
\end{example}

\begin{example}[$G$-equivariantisation]
	\label{exa:Gequiv}
	Let~$\zz$ be a 2-dimensional defect TQFT with a $G$-action $\rho\colon\Bar G\lra \mathcal D_\zz$ that gives rise to an orbifold datum~$\A_G$ as in Example~\ref{exa:OrbifoldSigmaModels}. 
	Since the ``bulk state space is given by endomorphisms of the identity defect'' (as reviewed e.g.\ in \cite[Sect.\,3.1]{2dDefectTQFTLectureNotes}), we have that 
	$
	\zz^G(S^1) 
		= 
		\zz_{\A_G}(S^1) 
		= 
		\End_{(\mathcal D_\zz)_{\textrm{orb}}}(1_{\A_G}) 
		= 
		\End_{\A_G,\A_G}(\A_G)
	$. 
	Using that by definition $(\A_G)_1 = \bigoplus_{g\in G} \rho(g)$, we find that ``twisted sectors'' are automatically included in the formalism, namely as the summands corresponding to $g\neq e$. 
	
	The $G$-action $\rho\colon\Bar G\lra \mathcal D_\zz$ induces a $G$-action on the Hom categories $\mathcal D_{\zz}(\rho(*),c)$ for all $c\in\mathcal D_\zz$, namely by horizontal composition (from the right) with $\rho(g)$, $g\in G$. 
	Then one finds that the $G$-equivariantisation of these Hom categories is equivalent to the categories of right $\A_G$-modules, $\mathcal D_{\zz}(\rho(*),c)^G \cong \textrm{mod}_{\mathcal D_{\zz}(\rho(*),c)}(\A_G)$. 
	This is explained in detail for the case of Landau--Ginzburg models in \cite[Sect.\,7.1]{cr1210.6363}, where it is also shown that~$\A_G$ is an orbifold datum if the quantum dimensions of $\rho(g)$ are all identities. 
\end{example}

\begin{example}[McKay correspondence and orbifold equivalence]
	By enlarging the 2-category $\mathcal{LG}_\C$ to include more general maps than polynomials as objects, it is shown in \cite{IonovMcKay} that the McKay correspondence gives rise to many (new) orbifold equivalences in~$\mathcal{LG}_\C$. 
	Recall that if a finite group~$G$ acts on a variety~$V$ such that $V/G$ has a crepant resolution~$Y$, then the McKay correspondence states that (under certain technical assumptions) the bounded derived category of~$Y$ is equivalent to the $G$-equivariantisation of the derived category of~$V$. 
	On the other hand, it is known that quotienting the derived category of a variety $\{W=0\}$ by the subcategory of perfect complexes is equivalent to the homotopy category of matrix factorisations of~$W$. 
	Hence one may expect that for a $G$-equivariant map $f\colon V \lra \C$ the McKay correspondence induces a relation between the homotopy categories of $G$-equivariant matrix factorisations of~$f$ and of matrix factorisations of the function~$\hat f$ on~$Y$ induced by~$f$. 
	And indeed, $f\sim \hat f$ as shown in \cite[Thm.\,3.7]{IonovMcKay}. 
\end{example}

We now turn to dimension $3$. 
Recall from Theorem~\ref{thm:CMS} that a 3-dimensional defect TQFT~$\zz$ gives rise to a Gray category with duals~$\mathcal D_\zz$. 
An orbifold datum~$\A$ for~$\zz$ is in particular an $E_1$-algebra in the monoidal 2-category $\mathcal D_\zz(\A_3,\A_3)$. 
It is hence natural to consider, for any Gray category with duals~$\mathcal T$ (with appropriate conditions on certain colimits), the Morita 3-category of \cite{TheoClaudia} of such algebras in~$\mathcal T$, as spelled out in \cite[Sect.\,3]{CM3dOrbifoldCompletion}. 
Contrary to the 2-dimensional case in Definition~\ref{def:OrbifoldCompletion}, we need to impose additional constraints on the 1- and 2-morphisms of this 3-category to ensure invariance under Pachner moves in the construction of the orbifold \textsl{defect} TQFT $\zz_{\textrm{orb}}$ below. 
These constraints are identified in \cite[Sect.\,4]{CM3dOrbifoldCompletion}, to which we refer for the precise definition of the \textsl{orbifold completion} 3-category $\mathcal T_{\textrm{orb}}$ of~$\mathcal T$. 
Below we only give a broad sketch. 

Using the graphical calculus of \cite{BMS} (with the conventions of \cite{CMS}), an orbifold datum~$\A$ in~$\mathcal T$ consists of 1-, 2- and 3-morphisms 
\be 
%%%%%%%%%%%%%%%%%%%%%% 
\tikzzbox{
}%%popende
%%%%%%%%%%%%%%%%%%%%%% 
\, . 
\ee 
Such data, subject to the constraints of \cite[Def.\,3.13]{CRS1} (see also \cite[Fig.\,4.1]{CM3dOrbifoldCompletion}), are the objects of $\mathcal T_{\textrm{orb}}$. 
A 1-morphisms $\A\lra\A'$ in $\mathcal T_{\textrm{orb}}$ is a 1-morphism $M\colon\A_3\lra\A'_3$ in~$\mathcal T$ together with 2- and 3-morphisms
\be 
%%%%%%%%%%%%%%%%%%%%%% 
\tikzzbox{%
}%%popende
%%%%%%%%%%%%%%%%%%%%%%
\ee
etc., and 3-morphisms in $\mathcal T_{\textrm{orb}}$ are those in~$\mathcal T$ which are compatible with the above. 

With all details about $\mathcal T_{\textrm{orb}}$ as laid out in \cite{CM3dOrbifoldCompletion}, one finds that the first part of Theorem~\ref{thm:Borb} generalises to 

\begin{theorem}[\cite{CM3dOrbifoldCompletion}]
	%The orbifold completion 
	$\mathcal T_{\textrm{orb}}$ admits adjoints for all 1- and 2-morphisms. 
\end{theorem}

The completion property $(\mathcal T_{\textrm{orb}})_{\textrm{orb}} \cong \mathcal T_{\textrm{orb}}$ is expected to follow from a universal property analogous to that in Proposition~\ref{prop:UniversalPropertyBorb}. 

\begin{example}
	Recall from Examples~\ref{exa:2dDefectSSMs} and~\ref{exa:GrayCatDual}\ref{item:BssFrob} that $(\mathcal D_{(\zztriv_2)^\odot_{\textrm{orb}}}) \cong \ssFrob(\Vectk)$ and $\mathcal D_{\zztriv_3} = \Bar\ssFrob(\Vectk)$. 
	Taking the (Euler completion of the) orbifold completion $(\mathcal D_{\zztriv_3})_{\textrm{orb}}$ recovers the 3-category (with duals) of spherical fusion categories, bimodule categories with trace, bimodule functors, and bimodule natural transformations defined in \cite{Bimodtrace}. 
\end{example}

By design, the orbifold completion $(\mathcal D_\zz)_{\textrm{orb}}$ allows us to construct the \textsl{3-dimensional orbifold defect TQFT} $\zz_{\textrm{orb}}\colon \Borddefblank_{3,2}(\mathds{D}^{\textrm{orb}}) \lra \mathcal C$ in close analogy to Definition~\ref{def:2dOrbifoldDefectTQFT}. 
Defects of dimension~$j$ are $(3-j)$-cells in $(\mathcal D_\zz)_{\textrm{orb}}$, whose defining conditions ensure well-definedness of $\zz_{\textrm{orb}}$ when taking the colimit over all triangulations; the details, including compatibility with given stratifications, are in \cite[Sect.\,6.2]{CM3dOrbifoldCompletion}. 
Again one has $(\mathcal D_\zz)_{\textrm{orb}} \cong \mathcal D_{\zz_{\textrm{orb}}}$. 

\begin{example}
	\label{exa:3dDefectSSM}
	The 3-dimensional \textsl{defect state sum model is the (Euler completed) orbifold of the trivial defect TQFT}: $\zzss_3 = (\zztriv_3)^\odot_{\textrm{orb}}$. 
	In particular, surface defects between Turaev--Viro--Barrett--Westbury models are given by bimodule categories with trace, and line defects at which an arbitrary number of surface defects meet are given by bimodule functors between appropriate relative Deligne 
	%arXiv_v2: 
		%products. 
		 products, in line with \cite{KK1104.5047}. 
	The case of line defects between precisely two surface defects is studied in detail in \cite{Meusburger3dDefectStateSumModels}, which also provides explicit examples for the special case of Dijkgraaf--Witten models. 
\end{example}

\begin{remark}
	\label{rem:HigherSSMs}
	The theme of Examples~\ref{exa:2dDefectSSMs} and~\ref{exa:3dDefectSSM} is expected to continue in higher dimensions: 
	the \textsl{$n$-dimensional trivial defect TQFT} is obtained from the delooping $\Bar\mathcal D_{\zzss_{n-1}}$, and the \textsl{$n$-dimensional defect state sum model} is the (Euler completed) orbifold of the trivial defect TQFT, 
	\be 
	\zzss_n = (\zztriv_n)^\odot_{\textrm{orb}} \, . 
	\ee 
	For $n=4$, this is explained in detail for closed state sum models in \cite{DouglasReutter2018, LukasNilsVincentas}. 
\end{remark}

\begin{example}
	Let~$\mathcal M$ be a modular fusion category. 
	The Reshetikhin--Turaev defect TQFT $\zz^{\textrm{RT}}_{\mathcal M}$ of \cite{KMRS} mentioned in Example~\ref{ex:3dDefectTQFT}\ref{item:RT} is the defect TQFT obtained from the orbifold completion $(\Bar\Delta\!\ssFrob(\mathcal M))_{\textrm{orb}}$, as shown in \cite[Sect.\,6.2]{CM3dOrbifoldCompletion}. 
\end{example}


\begin{thebibliography}{GHNPPS}

\bibitem[ALR]{OrbifoldsStringTopologyBook}
A.~Adem, J.~Leida, and Y.~Ruan,  
\textsl{Orbifolds and Stringy Topology}, 
\doi{10.1017/CBO9780511543081}{\textsl{Cambridge Tracts in Mathematics} \textbf{171}, Cambridge University Press, 2007}. 

\bibitem[At]{AtiyahTQFT}
M.~Atiyah, 
\textsl{Topological quantum field theories}, 
\href{http://www.numdam.org/item?id=PMIHES_1988__68__175_0}{Inst. Hautes \'{E}tudes Sci. Publ. Math. \textbf{68} (1988), 175--186}.

\bibitem[Ba]{BalsamTQFT2}
B.~Balsam, 
\textsl{Turaev-Viro invariants as an extended TQFT II},
\href{https://arxiv.org/abs/1010.1222}{arXiv:1010.1222 [math.QA]}.

%\bibitem[Bar]{BartlettExtendedTQFTs}
%B.~Bartlett, 
%\textsl{Extended and fully extended TQFTs}, 
%in preparation. 
%%\href{https://arxiv.org/abs/1010.1222}{arXiv:1010.1222 [math.QA]}.

\bibitem[BCFR]{BCFR}
I.~Brunner, N.~Carqueville, P.~Fragkos, and D.~Roggenkamp, 
\textsl{Truncated affine Rozansky--Witten models as extended defect TQFTs},
\href{https://arxiv.org/abs/2307.06284}{arXiv:2307.06284 [math-ph]}.

\bibitem[BCP1]{BCP}
I.~Brunner, N.~Carqueville, and D.~Plencner, 
\textsl{Orbifolds and topological defects}, 
\doi{10.1007/s00220-014-2056-3}{Comm.~Math.~Phys. \textbf{315} (2012) 739--769}, 
\href{https://arxiv.org/abs/1307.3141}{arXiv:1307.3141 [hep-th]}. 

\bibitem[BCP2]{BCP1}
I.~Brunner, N.~Carqueville, and D.~Plencner, 
\textsl{A quick guide to defect orbifolds}, 
\href{https://bookstore.ams.org/pspum-88}{Proc.~of Symp.~in Pure Math.~\textbf{88} (2014), 231--241}, 
\href{https://arxiv.org/abs/1310.0062}{arXiv:1310.0062 [hep-th]}.

\bibitem[BCP3]{BCP2}
I.~Brunner, N.~Carqueville, and D.~Plencner, 
\textsl{Discrete torsion defects}, 
\doi{10.1007/s00220-015-2297-9}{Comm. Math. Phys. \textbf{337} (2015), 429--453}, 
\href{https://arxiv.org/abs/1404.7497}{arXiv:1404.7497 [hep-th]}.

\bibitem[BCR]{BCR}
I.~Brunner, N.~Carqueville, and D.~Roggenkamp, 
\textsl{Truncated affine Rozansky--Witten models as extended TQFTs},
\doi{10.1007/s00220-022-04614-4}{Comm.\ Math.\ Phys.\ \textbf{400} (2023), 371--415}, 
\href{https://arxiv.org/abs/2201.03284}{arXiv:2201.03284 [math-ph]}.

\bibitem[BMS]{BMS}
J.~Barrett, C.~Meusburger, and G.~Schaumann, 
\textsl{Gray categories with duals and their diagrams},
\href{https://arxiv.org/abs/1211.0529}{arXiv:1211.0529 [math.QA]}.

\bibitem[BP]{bp9205031}
C.~Bachas and M.~Petropoulos, 
\textsl{Topological Models on the Lattice and a Remark on String Theory Cloning}, 
\doi{10.1007/BF02097063}{Commun.~Math.~Phys.~\textbf{152} (1993), 191--202}, 
\href{https://www.arxiv.org/abs/hep-th/9205031}{arXiv:hep-th/9205031}.

\bibitem[BW]{bwTV1}
J.~Barrett and B.~Westbury, 
\textsl{Invariants of piecewise-linear 3-manifolds}, 
\doi{10.1090/S0002-9947-96-01660-1}{Trans. Amer. Math. Soc. \textbf{348} (1996), 3997--4022}. 

\bibitem[Ca]{2dDefectTQFTLectureNotes}
N.~Carqueville, 
\textsl{Lecture notes on 2-dimensional defect TQFT}, 
\doi{10.4064/bc114-2}{Banach Center Publications \textbf{114} (2018), 49--84}, 
\href{https://arxiv.org/abs/1607.05747}{\mbox{arXiv:}1607.05747 [math.QA]}. 

\bibitem[CMoMo]{CMM}
N.~Carqueville and F.~Montiel Montoya, 
\textsl{Extending Landau-Ginzburg models to the point},
\doi{10.1007/s00220-020-03871-5}{Comm.\ Math.\ Phys.\ \textbf{379} (2020), 955--977}, 
\href{https://arxiv.org/abs/1809.10965}{arXiv:1809.10965 [math.QA]}.

\bibitem[CMRSS1]{CMRSS1}
N.~Carqueville, V.~Mulevi\v{c}ius, I.~Runkel, D.~Scherl, and G.~Schaumann, 
\textsl{Orbifold graph TQFTs}, 
\href{https://arxiv.org/abs/2101.02482}{arXiv:2101.02482 [math.QA]}.

\bibitem[CMRSS2]{CMRSS2}
N.~Carqueville, V.~Mulevi\v{c}ius, I.~Runkel, D.~Scherl, and G.~Schaumann, 
\textsl{Reshetikhin--Turaev TQFTs close under generalised orbifolds}, 
\href{https://arxiv.org/abs/2109.04754}{arXiv:2109.04754 [math.QA]}.

\bibitem[CMS]{CMS}
N.~Carqueville, C.~Meusburger, and G.~Schaumann, 
\textsl{3-dimensional defect TQFTs and their tricategories}, 
\doi{10.1016/j.aim.2020.107024}{Adv. Math. \textbf{364} (2020) 107024},
\href{https://arxiv.org/abs/1603.01171}{arXiv:1603.01171 [math.QA]}.

\bibitem[CMu]{cm1208.1481}
N.~Carqueville and D.~Murfet, 
\textsl{Adjunctions and defects in Landau-Ginzburg models}, 
\doi{10.1016/j.aim.2015.03.033}{Adv. Math. \textbf{289} (2016), 480--566}, 
\href{https://arxiv.org/abs/1208.1481}{arXiv:1208.1481 [math.AG]}. 

\bibitem[CMuMü]{LukasNilsVincentas}
N.~Carqueville, V.~Mulevi\v{c}ius, and L.~Müller, 
\textsl{in preparation}.

\bibitem[CMü]{CM3dOrbifoldCompletion}
N.~Carqueville and L.~Müller, 
\textsl{Orbifold completion of 3-categories},
\href{https://arxiv.org/abs/2307.06485}{arXiv:2307.06485 [math.QA]}.

\bibitem[CR]{cr1210.6363}
N.~Carqueville and I.~Runkel, 
\textsl{Orbifold completion of defect bicategories}, 
\doi{10.4171/QT/76}{Quantum Topology \textbf{7}:2 (2016) 203--279}, 
\href{https://arxiv.org/abs/1210.6363}{arXiv:1210.6363 [math.QA]}.

\bibitem[CRCR]{CRCR}
N.~Carqueville, A.~Ros Camacho, and I.~Runkel, 
\textsl{Orbifold equivalent potentials}, 
\doi{10.1016/j.jpaa.2015.07.015}{Journal of Pure and Applied Algebra \textbf{220} (2016), 759--781}, 
\arxiv{1311.3354}{arXiv:1311.3354 [math.QA]}.

\bibitem[CRS1]{CRS1}
N.~Carqueville, I.~Runkel, and G.~Schaumann, 
\textsl{Orbifolds of $n$-dimensional defect TQFTs}, 
\doi{10.2140/gt.2019.23.781}{Geometry \& Topology \textbf{23} (2019), 781--864},  
\href{https://arxiv.org/abs/1705.06085}{arXiv:1705.06085 [math.QA]}.

\bibitem[CRS2]{CRS2}
N.~Carqueville, I.~Runkel, and G.~Schaumann, 
\textsl{Line and surface defects in Reshetikhin--Turaev TQFT},  
\doi{10.4171/QT/121}{Quantum Topology \textbf{10} (2019), 399--439}, 
\href{https://arxiv.org/abs/1710.10214}{arXiv:1710.10214 [math.QA]}.

\bibitem[CRS3]{CRS3}
N.~Carqueville, I.~Runkel, and G.~Schaumann, 
\textsl{Orbifolds of Reshetikhin--Turaev TQFTs}, 
\href{http://www.tac.mta.ca/tac/volumes/35/15/35-15abs.html}{Theory and Applications of Categories \textbf{35} (2020), 513--561}
\href{https://arxiv.org/abs/1809.01483}{arXiv:1809.01483 [math.QA]}.

\bibitem[CS]{RSpinLorantNils}
N.~Carqueville and L.~Szegedy, 
\textsl{Fully extended $r$-spin TQFTs}, 
\href{https://arxiv.org/abs/2107.02046}{arXiv:2107.02046 [math.QA]}.

\bibitem[CW]{cw1007.2679}
A.~{C\u ald\u araru} and S.~Willerton, 
\textsl{The Mukai pairing, I: a categorical approach},
\href{https://nyjm.albany.edu/j/2010/16-6.html}{New York Journal of Mathematics \textbf{16} (2010), 61--98}, 
\href{https://arxiv.org/abs/0707.2052}{arXiv:0707.2052 [math.AG]}.

\bibitem[DKR]{dkr1107.0495}
A.~Davydov, L.~Kong, and I.~Runkel, 
\textsl{Field theories with defects and the centre functor}, 
\href{http://www.ams.org/bookstore?fn=20&arg1=pspumseries&ikey=PSPUM-83}{Mathematical Foundations of Quantum Field Theory and Perturbative String Theory, Proceedings of Symposia in Pure Mathematics, AMS, 2011}, 
\href{https://arxiv.org/abs/1107.0495}{arXiv:1107.0495 [math.QA]}. 

\bibitem[DR]{DouglasReutter2018}
C.~Douglas and D.~Reutter, 
\textsl{Fusion 2-categories and a state-sum invariant for 4-manifolds}, 
\href{https://arxiv.org/abs/1812.11933}{\mbox{arXiv:}1812.11933 [math.QA]}.

\bibitem[DW]{DijkgraafWitten1990}
R.~Dijkgraaf and E.~Witten, 
\textsl{Topological gauge theories and group cohomology}, 
\doi{10.1007/BF02096988}{Comm.\ Math.\ Phys.\ \textbf{129} (1990), 393--429}.

\bibitem[FFRS]{ffrs0909.5013}
J.~Fr\"ohlich, J.~Fuchs, I.~Runkel, and C.~Schweigert,
\textsl{Defect lines, dualities, and generalised orbifolds}, 
\doi{10.1142/9789814304634_0056}{Proceedings of the XVI International Congress on Mathematical Physics, Prague, August 3--8, 2009}, \href{https://arxiv.org/abs/0909.5013}{arXiv:0909.5013 [math-ph]}.

\bibitem[FH]{FreedHopkinsRefelctionPositivity2016}
D.~S.~Freed and M.~J.~Hopkins, 
\textsl{Reflection positivity and invertible topological phases}, 
\doi{10.2140/gt.2021.25.1165}{Geom. Topol.~\textbf{25} (2021), 1165--1330},
\href{https://arxiv.org/abs/1604.06527}{arXiv:1604.06527 [hep-th]}. 

\bibitem[FHK]{FHK}
M.~Fukuma, S.~Hosono, and H.~Kawai, 
\textsl{Lattice Topological Field Theory in Two Dimensions}, 
\doi{10.1007/BF02099416}{Comm. Math. Phys.~\textbf{161} (1994), 157--176},
\href{https://www.arxiv.org/abs/hep-th/9212154}{arXiv:hep-th/9212154}.

\bibitem[FMT]{FreedMooreTeleman2022}
D.~S.~Freed, G.~W.~Moore, and C.~Teleman, 
\textsl{Topological symmetry in quantum field theory},
\href{https://arxiv.org/abs/2209.07471}{arXiv:2209.07471 [hep-th]}.

\bibitem[FQ]{FreedQuinn1993}
D.~S.~Freed and F.~Quinn, 
\textsl{Chern--Simons theory with finite gauge group}, 
\doi{10.1007/BF02096860}{Comm.\ Math.\ Phys.\ \textbf{156} (1993), 435--472}.

\bibitem[FS]{FuchsStignerFrobeniusAlgebras}
J.~{Fuchs} and C.~{Stigner}, 
\textsl{On Frobenius algebras in rigid monoidal categories}.
\newblock Arab. J. Sci. Eng. \textbf{33-2C} (2008) 175--191,
\href{https://arxiv.org/abs/0901.4886}{arXiv:0901.4886 [math.CT]}.

\bibitem[FSV]{fsv1203.4568}
J.~Fuchs, C.~Schweigert, and A.~Valentino, 
\textsl{Bicategories for boundary conditions and for surface defects in 3-d TFT}, 
\doi{10.1007/s00220-013-1723-0}{Communications in Mathematical Physics \textbf{321}:2 (2013), 543--575}, 
\href{http://arxiv.org/abs/1203.4568}{arXiv:1203.4568 [hep-th]}.

\bibitem[GJF]{GaiottoJohnsonFreyd}
D.~Gaiotto and T.~Johnson-Freyd,
\textsl{Condensations in higher categories},
\arxiv{1905.09566}{arXiv:1905.09566 [math.CT]}.

\bibitem[GPS]{GPS}
R.~Gordon, A.~J.~Power, and R.~Street, 
\textsl{Coherence for Tricategories}, 
Memoirs of the American Mathematical Society \textbf{117}, 
American Mathematical Society, 1995.

\bibitem[Gu]{Gurskibook}
N.~Gurski, 
\textsl{\doi{10.1017/CBO9781139542333}{Coherence in Three-Dimensional Category Theory}}, 
\textsl{Cambridge Tracts in Mathematics} \textbf{201}, Cambridge University Press, 2013.

\bibitem[HKK+]{mirrorbook}
K.~Hori, S.~Katz, A.~Klemm, R.~Pandharipande, R.~Thomas, C.~Vafa, R.~Vakil, and E.~Zaslow, 
\textsl{Mirror symmetry}, 
\href{http://claymath.org/library/monographs/cmim01c.pdf}{Clay Mathematics Monographs \textbf{1} American Mathematical Society, 2003}.

\bibitem[Io]{IonovMcKay}
A.~Ionov, 
\textsl{McKay correspondence and orbifold equivalence}, 
\doi{10.1016/j.jpaa.2022.107297}{Journal of Pure and Applied Algebra \textbf{227}:5 (2023), 107297}, 
\arxiv{2202.12135}{arXiv:2202.12135 [math.AG]}.

\bibitem[JFS]{TheoClaudia}
T.~Johnson-Freyd and C.~Scheimbauer, 
\textsl{(Op)lax natural transformations, twisted quantum field theories, and ``even higher'' Morita categories}, 
\doi{10.1016/j.aim.2016.11.014}{Advances in Mathematics \textbf{307} (2017), 147--223},  
\href{https://arxiv.org/abs/1502.06526}{arXiv:1502.06526 [math.CT]}.

\bibitem[Ju]{Juhasz2014}
A.~Juh\'asz,
\textsl{Defining and classifying TQFTs via surgery},
\doi{10.4171/QT/108}{Quantum Topology \textbf{9} (2018), 229--321}, 
\href{https://arxiv.org/abs/1408.0668}{arXiv:1408.0668 [math.GT]}.

\bibitem[Ka]{KapustinICM2010}
A.~Kapustin, 
\textsl{Topological Field Theory, Higher Categories, and Their Applications}, 
\href{https://www.mathunion.org/fileadmin/ICM/Proceedings/ICM2010.3/ICM2010.3.pdf}{Proceedings of the International Congress of Mathematicians 2010, Volume 3, 2021--2043}, 
\href{https://arxiv.org/abs/1004.2307}{arXiv:1004.2307 [hep-th]}.

\bibitem[KK]{KK1104.5047}
A.~Kitaev and L.~Kong, 
\textsl{Models for gapped boundaries and domain walls},
\doi{10.1007/s00220-012-1500-5}{Commun.\ Math.\ Phys.\ \textbf{313} (2012) 351--373},
\href{https://arxiv.org/abs/1104.5047}{arXiv:1104.5047 [cond-mat.str-el]}.

\bibitem[KMRS]{KMRS}
V.~Koppen, V.~Mulevi\v{c}ius, I.~Runkel, and C.~Schweigert, 
\textsl{Domain walls between 3d phases of Reshetikhin-Turaev TQFTs}, 
\doi{10.1007/s00220-022-04489-5}{Communications in Mathematical Physics \textbf{396} (2022), 1187--1220}, 
\href{https://arxiv.org/abs/2105.04613}{arXiv:2105.04613 [hep-th]}.

\bibitem[Ko]{Kockbook}
J.~Kock, 
\textsl{Frobenius algebras and 2D topological quantum field theories}, 
\textsl{London Mathematical Society Student Texts} \textbf{59}, Cambridge University Press, 2003. 

\bibitem[KR]{KR0909.3643}
A.~Kapustin and L.~Rozansky, 
\textsl{Three-dimensional topological field theory and symplectic algebraic geometry II}, 
\doi{10.4310/CNTP.2010.v4.n3.a1}{Communications of Number Theory and Physics \textbf{4} (2010), 463--549}, 
\href{https://arxiv.org/abs/0909.3643}{arXiv:0909.3643 [math.AG]}.

\bibitem[KRS]{KRS}
A.~Kapustin, L.~Rozansky, and N.~Saulina, 
\textsl{Three-dimensional topological field theory and symplectic algebraic geometry I}, 
\doi{10.1016/j.nuclphysb.2009.01.027}{Nuclear Physics B \textbf{816} (2009), 295--355}, 
\href{https://arxiv.org/abs/0810.5415}{arXiv:0810.5415 [hep-th]}.

\bibitem[KS]{ks1012.0911}
A.~Kapustin and N.~Saulina, 
\textsl{Surface operators in 3d Topological Field Theory and 2d Rational Conformal Field Theory}, 
Mathematical Foundations of Quantum Field Theory and Perturbative String Theory, 
Proceedings of Symposia in Pure Mathematics \textbf{83}, 175--198, 
American Mathematical Society, 2011, 
\href{https://arxiv.org/abs/1012.0911}{arXiv:1012.0911 [hep-th]}.

\bibitem[KW]{kw0604151}
A.~Kapustin and E.~Witten, 
\textsl{Electric-Magnetic Duality And The Geometric Langlands Program}, 
\doi{10.4310/CNTP.2007.v1.n1.a1}{Communications in Number Theory and Physics \textbf{1} (2007), 1--236}, 
\href{https://arxiv.org/abs/hep-th/0604151}{arXiv:hep-th/0604151}. 

\bibitem[LP]{lp0602047}
A.~D.~Lauda and H.~Pfeiffer, 
\textsl{State sum construction of two-dimensional open-closed Topological Quantum Field Theories}, 
\doi{10.1142/S0218216507005725}{J. Knot Theor. Ramifications \textbf{16} (2007), 1121--1163}, 
\href{https://arxiv.org/abs/math/0602047}{arXiv:math/0602047 [math.QA]}.

\bibitem[Lu]{l0905.0465}
J.~Lurie, 
\textsl{On the Classification of Topological Field Theories},
\href{https://projecteuclid.org/euclid.cdm/1254748657}{Current Developments in Mathematics \textbf{2008} (2009), 129--280}, 
\arxiv{0905.0465}{arXiv:0905.0465 [math.CT]}.

\bibitem[Me]{Meusburger3dDefectStateSumModels}
C.~Meusburger, 
\textsl{State sum models with defects based on spherical fusion categories}, 
%arXiv_v2: 
	\doi{10.1016/j.aim.2023.109177}{Advances in Mathematics \textbf{429} (2023), 109177},  
\href{https://arxiv.org/abs/2205.06874}{arXiv:2205.06874 [math.QA]}.

\bibitem[MR1]{MuleRunk}
V.~Mulevi\v{c}ius and I.~Runkel, 
\textsl{Constructing modular categories from orbifold data}, 
\doi{10.4171/QT/170}{Quantum Topology \textbf{13}:3 (2022) 459--523}, 
\href{https://arxiv.org/abs/2002.00663}{arXiv:2002.00663 [math.QA]}.

\bibitem[MR2]{MuleRunk2}
V.~Mulevi\v{c}ius and I.~Runkel, 
\textsl{Fibonacci-type orbifold data in Ising modular categories}, 
\doi{10.1016/j.jpaa.2022.107301}{Journal of Pure and Applied Algebra \textbf{227}:6 (2023), 107301}, 
\href{https://arxiv.org/abs/2010.00932}{arXiv:2010.00932 [math.QA]}.

\bibitem[Mul]{Mule1}
V.~Mulevi\v{c}ius, 
\textsl{Condensation inversion and Witt equivalence via generalised orbifolds}, 
\href{https://arxiv.org/abs/2206.02611}{arXiv:2206.02611 [math.QA]}.

\bibitem[Mun]{Munkres} 
J.~R.~Munkres, 
\textsl{Elementary Differential Topology}, 
Annals of Mathematics Studies~\textbf{54}, 
Princeton University Press, 
1967.

\bibitem[NR]{NovakRunkel}
S.~Novak and I.~Runkel, 
\textsl{State sum construction of two-dimensional topological quantum field theories on spin surfaces}, 
\doi{10.1142/S0218216515500285}{J.\ Knot Theory Ramifications  \textbf{24}:5 (2015), 1550028}, 
\href{https://arxiv.org/abs/1402.2839}{arXiv:1402.2839 [math.QA]}.

\bibitem[Os]{o0111139}
V.~Ostrik, 
\textsl{Module categories, weak Hopf algebras and modular invariants}, 
\doi{10.1007/s00031-003-0515-6}{Transform. Groups \textbf{8} (2003), 177--206}, 
\href{https://www.arxiv.org/abs/math/0111139}{arXiv:math/0503632 [math.AG]}.

\bibitem[Pa]{Pachpaper}
U.~Pachner, 
\textsl{P.L.~Homeomorphic Manifolds are Equivalent by Elementary Shellings}, 
\doi{10.1016/S0195-6698(13)80080-7}{European Journal of Combinatorics \textbf{12}:2 (1991), 129--145}. 

%\bibitem[Pl]{PlavnikMFC}
%J.~Plavnik, 
%\textsl{Modular tensor categories}, 
%in preparation. 
%%\href{https://arxiv.org/abs/1010.1222}{arXiv:1010.1222 [math.QA]}.

\bibitem[Po]{PolishchukKernelAlgebras}
A.~Polishchuk, 
\textsl{Kernel algebras and generalized Fourier--Mukai transforms}, 
\doi{10.4171/JNCG/73}{J.\ Noncommut.\ Geom.\ \textbf{5}:2 (2011), 153--251}, 
\href{https://arxiv.org/abs/0810.1542}{arXiv:0810.1542 [math.AG]}. 

\bibitem[Qu]{Quinnlectures}
F.~Quinn, 
\textsl{Lectures on axiomatic topological quantum field theory}, 
IAS/Park City Mathematics Series \textbf{1} (1995), 325--433.

\bibitem[ReW]{OEReck}
A.~Recknagel and P.~Weinreb, 
\textsl{Orbifold equivalence: structure and new examples}, 
\doi{10.5427/jsing.2018.17j}{Journal of Singularities \textbf{17} (2018), 216--244}, 
\href{https://arxiv.org/abs/1708.08359}{arXiv:1708.08359 [math.QA]}.

\bibitem[RoW]{RW1996}
L.~Rozansky and E.~Witten, 
\textsl{Hyper-K\"ahler geometry and invariants of three-manifolds}, 
\doi{10.1007/s000290050016}{Selecta Math.\ \textbf{3} (1997), 401--458}, 
\arxiv{hep-th/9612216}{arXiv:hep-th/9612216}. 

\bibitem[RS]{RunkelSuszekDefects}
I.~Runkel and R.~R.~Suszek, 
\textsl{Gerbe-holonomy for surfaces with defect networks}, 
\href{https://projecteuclid.org/journals/advances-in-theoretical-and-mathematical-physics/volume-13/issue-4/Gerbe-holonomy-for-surfaces-with-defect-networks/atmp/1278423132.full}{Adv.\ Theor.\ Math.\ Phys.\ \textbf{13}:4 (2009), 1137--1219}, 
\href{https://arxiv.org/abs/0808.1419}{arXiv:0808.1419 [hep-th]}.

\bibitem[RSW]{RunkelSzegedyWatts}
I.~Runkel, L.~Szegedy, and G.~M.~T.~Watts,  
\textsl{Parity and Spin CFT with boundaries and defects}, 
\href{https://arxiv.org/abs/2210.01057}{arXiv:2210.01057 [hep-th]}.

\bibitem[RT]{retu2}
N.~Reshetikhin and V.~G.~Turaev,
\textsl{Invariants of 3-manifolds via link polynomials and quantum groups},
\doi{10.1007/BF01239527}{Inv.\ Math.\ \textbf{103} (1991), 547--597}.

%\bibitem[Ru]{RunkelTopologicalDefects}
%I.~Runkel, 
%\textsl{Topological defects}, 
%in preparation. 
%%\href{https://arxiv.org/abs/1010.1222}{arXiv:1010.1222 [math.QA]}.

\bibitem[Sc]{Bimodtrace}
G.~Schaumann, 
\textsl{Traces on module categories over fusion categories},
\doi{10.1016/j.jalgebra.2013.01.013}{Journal of Algebra \textbf{379} (2013), 382--425}, 
\href{https://arxiv.org/abs/1206.5716}{arXiv:1206.5716 [math.QA]}.

\bibitem[SW]{SchweigertWoike}
C.~Schweigert and L.~Woike,
\textsl{Orbifold Construction for Topological Field Theories}, 
\doi{10.1016/j.jpaa.2018.05.020}{Journal of Pure and Applied Algebra \textbf{223}:3 (2019), 1167--1192}, 
\href{https://arxiv.org/abs/1705.05171}{arXiv:1705.05171 [math.QA]}.

\bibitem[Tu]{tur} 
V.~G.~Turaev, 
\textsl{Quantum Invariants of Knots and 3-Manifolds},  
de Gruyter, New York, 1991.

\bibitem[TVire1]{TVire1}
V.~Turaev and O.~Virelizier, 
\textsl{On two approaches to 3-dimensional TQFTs}, 
\href{https://arxiv.org/abs/1006.3501}{arXiv:1006.3501 [math.GT]}.  

\bibitem[TVire2]{TVireBook}
V.~Turaev and A.~Virelizier, 
\textsl{Monoidal Categories and Topological Field Theories}, 
\doi{10.1007/978-3-319-49834-8}{Progress in Mathematics \textbf{322}, Birkh\"auser, 2017}.

\bibitem[TViro]{TVmodel}
V.~Turaev and O.~Viro, 
\textsl{State sum invariants of 3-manifolds and quantum $6j$-symbols}, 
\doi{10.1016/0040-9383(92)90015-A}{Topology \textbf{31}:4 (1992), 865--902}.

%\bibitem[Vi]{Virelizier3d}
%A.~Virelizier, 
%\textsl{3d TQFTs and 3-manifold invariants}, 
%in preparation. 
%%\href{https://arxiv.org/abs/1010.1222}{arXiv:1010.1222 [math.QA]}.

\bibitem[Yon]{Yonekura2018}
K.~Yonekura, 
\textsl{On the Cobordism Classification of Symmetry Protected Topological Phases},
\doi{10.1007/s00220-019-03439-y}{Comm.\ Math.\ Phys.\ \textbf{368} (2019), 1121--1173}, 
\href{https://arxiv.org/abs/1803.10796}{arXiv:1803.10796 [hep-th]}.

\bibitem[Yos]{Yoshinobook}
Y.~Yoshino, 
\textsl{Maximal Cohen--Macaulay Modules Over Cohen--Macaulay Rings}, 
London Mathematical Society Lecture Note Series \textbf{146}, 
Cambridge University Press, 
1990. 


\end{thebibliography}
\end{document}